\documentclass[twocolumn,english]{revtex4-1}
\usepackage[T1]{fontenc}
\usepackage{CJKutf8}
\usepackage{mathrsfs}
\usepackage{mathtools}
\usepackage{amsmath}
\usepackage{amssymb}
\usepackage{graphicx}

\makeatletter

\newcommand*\LyXThinSpace{\,\hspace{0pt}}
\providecommand{\tabularnewline}{\\}

\usepackage{babel}

\makeatother

\usepackage{babel}
\begin{document}
\title{Extending the Gutzwiller approximation to intersite interactions}
\author{Garry Goldstein$^{1}$, Nicola Lanata$^{2}$, Gabriel Kotliar$^{1,3}$}
\address{$^{1}$Physics and Astronomy Department, Rutgers University, Piscataway,
NJ 08854, USA}
\address{$^{2}$Department of physics and astronomy, Aarhus University, 800
Arhus C, Denmark}
\address{$^{3}$Condensed Matter Physics and Materials Science Division, Brookhaven
National Laboratory, Upton, NY 11973-5000, USA}
\begin{abstract}
We develop an extension of the Gutzwiller Approximation (GA) formalism
that includes the effects of Coulomb interactions of arbitrary range
(including density density, exchange, pair hopping and Coulomb assisted
hopping terms). This formalism reduces to the ordinary GA formalism
for the multi-band Hubbard models in the presence of only local interactions.
This is accomplished by combining the $1/z$ expansion ---where $z$
is the coordination number and only the leading order terms contribute
in the limit of infinite dimensions--- with a $P_{R}^{\dagger}P_{R}-I$
expansion, where $P_{R}$ is the Gutzwiller projector on the site
$R$. The method is conveniently formulated in terms of a Gutzwiller
Lagrange function. We apply our theory to the extended single band
Hubbard model. Similarly to the usual Brinkman-Rice mechanism we find
a Mott transition. A valence skipping transition is observed, where
the occupation of the empty and doubly occupied states for the Gutzwiller
wavefunction is enhanced with respect to the uncorrelated Slater determinant
wavefunction. 
\end{abstract}
\maketitle

\section{\label{sec:Introduction}Introduction }

Over the last three decades there has been renewed interest and substantial
progress in the development of methods for treating strongly correlated
electron systems. Various approximations to the Density Functional
Theory (DFT) such as the Local Density Approximation (LDA) proved
to be a good starting point for combinations with more advanced methodologies~\citep{Hybertsen_1989}
to study strongly correlated systems. In this regard, of particular
interest are quantum embedding methods, such as the Dynamical Mean
Field Theory (DMFT) \citep{Georges_1996}, Density Matrix Embedding
Theory (DMET) \citep{Wouters_2016}, the GA \citep{Bunemann1998,Brinkman_1970,fabrizio2012,Schiro_2010,Metzner_1989,Fazekas_1999,Lanata_2015,Ferrero_2009,Bunemann_2007,Gebrhard_2017,Gebhard_1990,Ayral_2017,Gutzwiller_1963,Gutzwiller_1964,Gutzwiller_1965,Lanata_2008,Bunemann_2003,Oelsen_2011,Oelsen_2011(2),Bunemann_2013,Munster_2016}
and the slave particles methods \citep{Kotliar_1986,Lechermann2007,Bunemann_2007,Ferrero_2009,Li_1989},
which share many common elements \citep{Ayral_2017,Lanata_2015,Lee_2019,Bunemann_2007,Kotliar_1986,Lanata_2008}.
In this work we focus on the GA, which has been actively developed
in recent years. Combining these embedding methods with density functional
theory gives rise to (LDA+DMFT) \citep{Kotliar_2006,Anisimov_1997}
and LDA in combination with the GA (LDA+GA) \citep{Lanata_2015,Deng_2009,Bunemann_2012,Schickling_2014}.
Furthermore these methods can be cast in a framework of functionals
of multiple observables, making them convenient for ab-initio numerical
simulations \citep{Savrasov_2004,Lanata_2015}.

In many currently available theoretical frameworks to study strongly
correlated systems, the non-local components of the Coulomb interaction
have been treated at the mean field level. On the other hand, this
may not be sufficient in many cases. For example, the non-local Coulomb
interactions can be as important as the local contributions in organic
materials, where even the electrons of s and p orbitals can induce
strong-correlation effects \citep{McKenzie_1998}. More generally,
in many materials the bare nearest neighbor Coulomb matrix elements
are the same order of magnitude as the hopping matrix elements \citep{Fazekas_1999,Hubbard_1963},
suggesting that it is necessary to take them into account more accurately.

Many techniques to treat short-ranged non-local interactions have
been developed in the context of model Hamiltonians. For extensions
of DMFT to treat this problem see \citep{Ayral_2013,Terletska_2017,Rohringer_2018,Schiller_1995,Stanescu_2004,Lichtenstein_2000,biermann_2003,Si_1996,Chitra_2000,Werner_2010}.
In this work we will focus on extensions of the GA, that is computationally
significantly less expensive than DMFT. A pioneering extension of
the GA to treat the t-J model was introduced by Zhang et. al. \citep{Zhang_1988}.
Ogata et al. made calculations of higher order corrections for the
t-J model within the GA \citep{Ogata_2003}. The effects of different
intersite interactions for the extended t-J model were studied by
Sensarma et. al. within the GA \citep{Sensarma_2007}. An operatorial
approach to the GA, where expectation values of Gutzwiller projected
operators were calculated in a $1/z$ expansion, was proposed in \citep{Li2009}.
Benchmark calculations for hydrogen like systems, including the effects
of intersite interactions within the framework of the extended Hubbard
model, were performed within the GA in Refs. \citep{Yao_2014,Liu_2016}.
The so called ``statistically consistent GA'' for non-local interactions
was studied in \citep{Zegrodnik_2019,Wysokinski_2015,Zegrodnik_2018,Fidrysiak_2019}
and the so called ``diagramattic expansion of the Gutzwiller wavefunction''
with intersite interactions was developed in \citep{Zegodnik_2017,Zegrerodnik_2018,Abram_2016}
for many models. However, GA methodologies able to account systematically
for the effects of non-local interactions in realistic first-principle
calculations, without empirical adjustments, are still not available.
Here we propose a new generalization of the GA, that constitutes a
step towards this ambitious goal. In fact we show that, combining
the ideas underlying the $1/z$ expansion \citep{Lanata2009,Sandri_2014,Metzner_1989}
with the $P_{R}^{\dagger}P_{R}-I$ expansion \citep{Wysokinski_2015,Fidrysiak_2019,Major_2018,Zegodnik_2017,Zegrerodnik_2018},
it is possible to tackle systematically non-local two site interactions
for general multi-orbital Hubbard models. Our work is an extension
of the GA to intersite interactions, in the same spirit as the extended
DMFT \citep{Stanescu_2004,Chitra_2000,Si_1996} and the dual boson
method \citep{Rubtsov_2011,Loon_2014} extend Dynamical Mean Field
Theory \citep{Georges_1996}. To illustrate our method we present
calculations for the extended single band Hubbard model, including
nearest-neighbor hopping, density density, correlated hopping, pair
hopping and exchange interactions. In particular, our calculations
of the single band extended Hubbard model indicate that the nearest-neighbor
Coulomb interactions can induce a phase transition where charge fluctuations
are enhanced rather than suppressed.

The setup of the paper is as follows. In Sec. \ref{sec:Example:-single-band-1}
we present an application of our formalism to the single band extended
Hubbard model. In Sec. \ref{sec:Main-Hamiltonian} we present the
main general Hamiltonian studied throughout the text. In Sec. \ref{sec:Gutzwiller-Approximation-and}
we discuss the simplifications of GA formalism arising from retaining
only the leading order in the $P_{R}^{\dagger}P_{R}-I$ expansion.
In Sec. \ref{sec:Equivalences} we show that by combining the $P_{R}^{\dagger}P_{R}-I$
formalism with the $1/z$ expansion it is possible to express semi-analytically
the variational energy (including the contribution of the non-local
interaction terms) as function of the GA variational parameters. In
Sec. \ref{sec:Gutzwiller-Lagrangian} we conveniently reformulate
our theory in terms of a GA Lagrange function, which reduces to the
result of Ref. \citep{Lanata_2017} for the special case of only-local
interactions. In Sec. \ref{sec:Conclusions} we conclude. The more
technical derivations are relegated to the appendices.

\section{\label{sec:Example:-single-band-1}The extended single-band Hubbard
model}

\subsection{\label{subsec:Main-Hamiltonian-1}Hamiltonian and setup}

As an example of our general formalism, that will be presented in
the following sections, here we consider the single-band extended
Hubbard model \cite{Amaricci_2010,Zhang_1989,Yan_1993,Dogen_1994,Dogen_1994-1,Dogen_1996,Chattopadhyay,Nayak_2002,Aichorn_2004,Onari_2004,Schuler_2013,Ayral_2013,Terletska_2017,Rohringer_2018,Schiller_1995,Stanescu_2004,Lichtenstein_2000,biermann_2003,Si_1996,Chitra_2000,Werner_2010}:
\begin{align}
H= & -t\sum_{\left\langle R,R'\right\rangle ,\sigma=\pm}\left(c_{R\sigma}^{\dagger}c_{R'\sigma}+h.c.\right)+U\sum_{R}n_{R\uparrow}n_{R\downarrow}\nonumber \\
 & +V\sum_{\left\langle R,R'\right\rangle }n_{R}n_{R'}-\mu\sum_{R\sigma}n_{R\sigma},\label{eq:Extended_Hubbard_simple}
\end{align}
where $\left\langle R,R'\right\rangle $ denotes nearest neighbors
$R$ and $R'$. 

As in the classic GA theory, our formalism is based on the following
variational wave function for the ground state of the system: 
\begin{equation}
\left|\Psi\right\rangle =\prod_{R}P_{R}\left|\Psi_{0}\right\rangle ,\label{eq:Gutzwiller-2-1}
\end{equation}
where $P_{R}$ is a bosonic operator acting on a single site $R$
and $\left|\Psi_{0}\right\rangle $ a single band Slater determinant
wave-function. For simplicity, here we will focus on the normal phase,
i.e. we will assume that $\left|\Psi\right\rangle $ does not break
any symmetry of the Hamiltonian $H$. Following Refs. \cite{Lanata2009,Lanata2012,Lanata_2008,Lanata_2015,Lanata_2016,Lanata_2017,Lanata_2017(2)}
we aim to minimize the following energy function:
\begin{equation}
\mathcal{E}=\frac{\left\langle \Psi\right|H\left|\Psi\right\rangle }{\left\langle \Psi\mid\Psi\right\rangle },\label{eq:Energy-3}
\end{equation}
while fulfilling the following subsidiary conditions, known as the
Gutzwiller constraints: \citep{Lanata2009,Lanata2012,Lanata_2015,Lanata_2016,Lanata_2017,Lanata_2017(2),Sandri_2014}:
\begin{align}
\left\langle \Psi_{0}\right|P_{R}^{\dagger}P_{R}\left|\Psi_{0}\right\rangle  & =1\nonumber \\
\left\langle \Psi_{0}\right|P_{R}^{\dagger}P_{R}c_{R\sigma}^{\dagger}c_{R\sigma}\left|\Psi_{0}\right\rangle  & =\left\langle \Psi_{0}\right|c_{R\sigma}^{\dagger}c_{R\sigma}\left|\Psi_{0}\right\rangle .\label{eq:Conditions-4-1-1}
\end{align}
Besides the GA, which is an approximation that becomes exact in the
limit of infinite dimensions, our general theory, to include non-local
interactions, will be based on the $P_{R}^{\dagger}P_{R}-I$ approximation
(see Appendices \ref{sec:Weak-coupling-expansion} and \ref{sec:The--scaling}).

\subsection{\label{sec:Gutzwiller-Lagrangian-1-1}Gutzwiller Lagrange function}

\begin{figure}
\begin{centering}
\includegraphics[width=0.47\columnwidth]{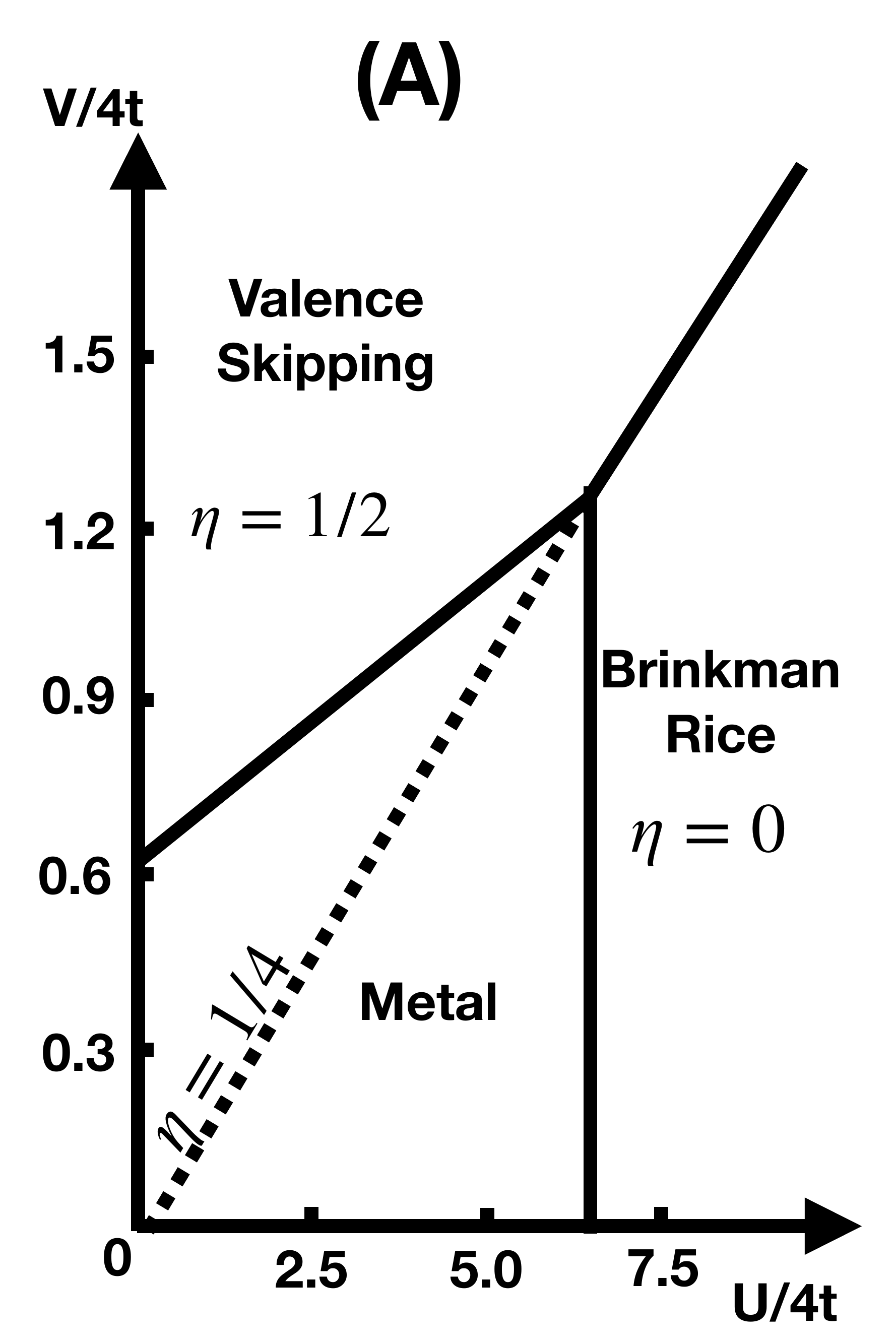}\includegraphics[width=0.47\columnwidth]{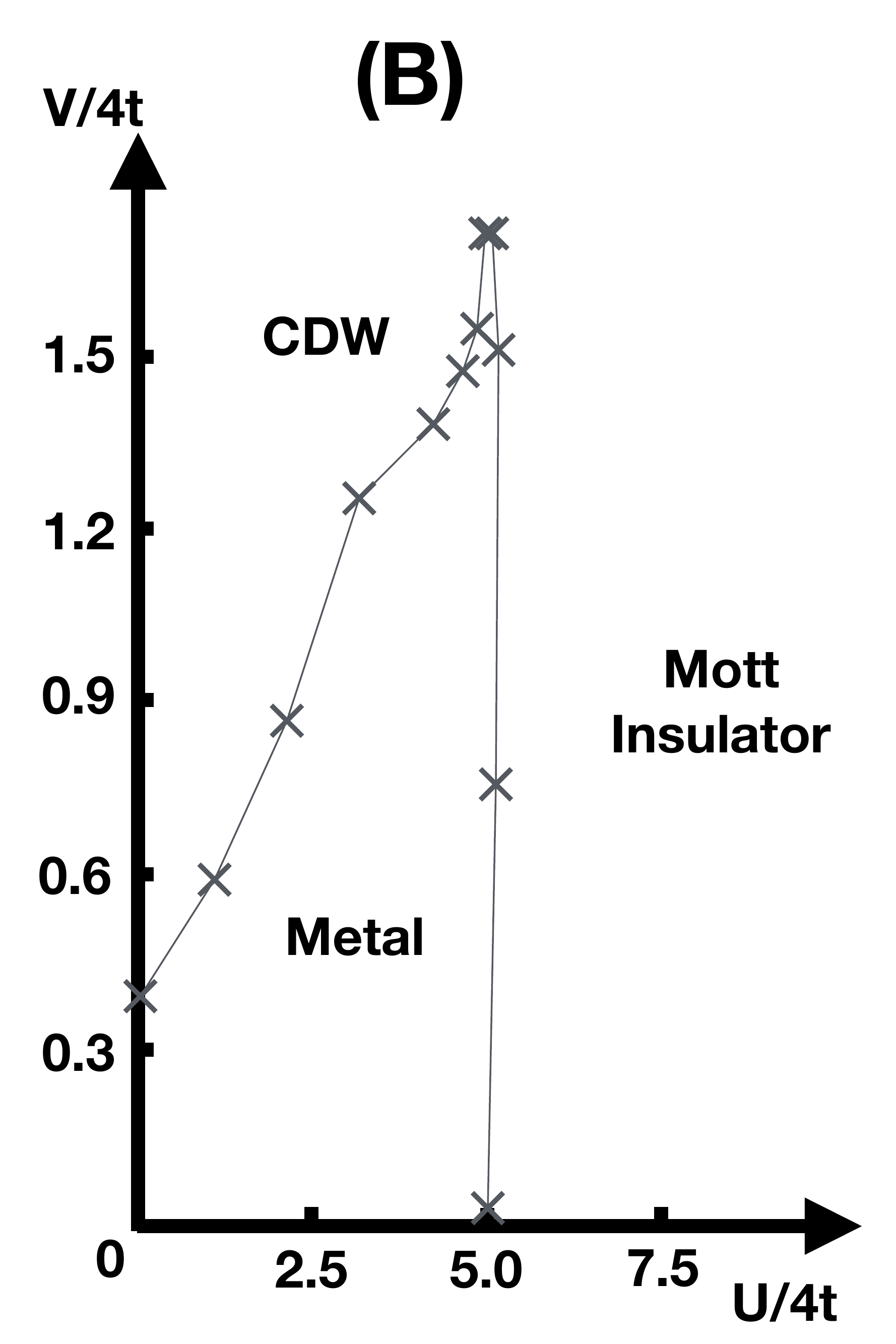}
\par\end{centering}
\caption{\label{fig-Phase-1-1} (A) Phase diagram of the single-band extended
2D Hubbard model for $U$ and $V$. (B) Phase diagram of the single-band
extended Hubbard model from Ref. \cite{Ayral_2013} for $U$ and $V$.}
\end{figure}

\subsubsection{\label{subsec:Regular-Hubbard-model}Hubbard model}

Following Refs. \cite{Lanata_2015,Lanata_2017}, the GA solution of
the energy minimization in Eq. (\ref{eq:Energy-3}), for $V=0$, can
be determined by extremizing the following Lagrange function:
\begin{align}
 & \mathcal{L}_{N}\left(D,\mathcal{R},\lambda,\lambda^{c},\Delta,E^{c},E,\mu,\left|\Psi_{0}\right\rangle ,\left|\Phi\right\rangle \right)\nonumber \\
 & =\frac{1}{\mathcal{N}}\left\langle \Psi_{0}\right|H_{QP}\left|\Psi_{0}\right\rangle +E\left(1-\left\langle \Psi_{0}\mid\Psi_{0}\right\rangle \right)\nonumber \\
 & +\left\langle \Phi\right|H_{Emb}\left|\Phi\right\rangle +E^{c}\left(1-\left\langle \Phi\mid\Phi\right\rangle \right)\nonumber \\
 & +\mathcal{L}_{Mix}\left(D,\mathcal{R},\lambda,\lambda^{c},\Delta\right)+\mu\frac{N}{\mathcal{N}},\label{eq:Lagrnage_function_local}
\end{align}
where: 
\begin{align}
H_{QP} & =-t\sum_{\left\langle RR'\right\rangle \sigma}\left[\mathcal{R}\mathcal{R}^{\ast}f_{R\sigma}^{\dagger}f_{R'\sigma}+h.c.\right]\nonumber \\
 & +\sum_{R\sigma}\lambda f_{R\sigma}^{\dagger}f_{R\sigma}-\mu\sum_{R\sigma}f_{R\sigma}^{\dagger}f_{R\sigma}\label{eq:Quasiparticle_Hamiltonian-1}
\end{align}

\begin{align}
H_{Emb} & =U\hat{c}_{\uparrow}^{\dagger}\hat{c}_{\uparrow}\hat{c}_{\downarrow}^{\dagger}\hat{c}_{\downarrow}+\sum_{\sigma}\lambda^{c}\hat{f}_{\sigma}\hat{f}_{\sigma}^{\dagger}\nonumber \\
 & +\left[\sum_{\sigma}D\hat{c}_{\sigma}^{\dagger}\hat{f}_{\sigma}+h.c.\right],\label{eq:Embedding_local}
\end{align}
\begin{align}
 & \mathcal{L}_{Mix}\left(D,\mathcal{R},\lambda^{c},\lambda,\Delta\right)\nonumber \\
 & =-2\left(\lambda+\lambda^{c}\right)\Delta-4D\left(\mathcal{R}\sqrt{\left(1-\Delta\right)\Delta}\right).\label{eq:Mixing_local}
\end{align}
Here $\mathcal{R}$ and $\lambda$ are the renormalization coefficients
in the quasiparticle Hamiltonian in Eq. (\ref{eq:Quasiparticle_Hamiltonian-1}),
$D$ and $\lambda^{c}$ are parameters of the embedding Hamiltonian
$H_{Emb}$ in Eq. (\ref{eq:Embedding_local}), $\left|\Phi\right\rangle $
is a generic wave function in the Hilbert space of the embedding Hamiltonian
$H_{Emb}$, $E^{c}$ is a Lagrange multiplier enforcing the normalization
of $\left|\Phi\right\rangle $, $E$ is a Lagrange multiplier used
to enforce the normalization of the Slater determinant $\left|\Psi_{0}\right\rangle $,
$\Delta$ is the local density of quasiparticles of the quasiparticle
Hamiltonian in Eq. (\ref{eq:Quasiparticle_Hamiltonian-1}), $N$ is
the total number of electrons and $\mathcal{N}$ is the total number
of sites. 

\subsubsection{\label{subsec:Hartree-Fock-Lagrange}Hartree Fock Lagrange function}

For later convenience, before introducing our extensions of the GA,
here we outline the Hartree Fock formalism for $V\neq0$. The Hartree
Fock solution to Eq. (\ref{eq:Extended_Hubbard_simple}) can be obtained
by extremizing the following Lagrange function: 
\begin{align}
 & \mathcal{L}_{N}\left(\lambda,\lambda^{\left(n.n.\right)},\Delta,\Delta^{\left(n.n.\right)},E,\mu,\left|\Psi_{0}\right\rangle \right)\nonumber \\
 & =\frac{1}{\mathcal{N}}\left\langle \Psi_{0}\right|H_{QP}\left|\Psi_{0}\right\rangle +E\left(1-\left\langle \Psi_{0}\mid\Psi_{0}\right\rangle \right)\nonumber \\
 & +\mathcal{L}_{Mix}\left(\lambda,\lambda^{\left(n.n.\right)},\Delta,\Delta^{\left(n.n.\right)}\right)+\mu\frac{N}{\mathcal{N}},\label{eq:Hartree_Fock_Lagrange_Function}
\end{align}
where: 
\begin{align}
 & H_{QP}=-t\sum_{\left\langle RR'\right\rangle \sigma}\left[f_{R\sigma}^{\dagger}f_{R'\sigma}+h.c.\right]\nonumber \\
 & +\sum_{\left\langle RR'\right\rangle \sigma}\lambda^{\left(n.n.\right)}f_{R\sigma}^{\dagger}f_{R'\sigma}+\sum_{R\sigma}\lambda f_{R\sigma}^{\dagger}f_{R\sigma}-\mu\sum_{R\sigma}f_{R\sigma}^{\dagger}f_{R\sigma},\label{eq:Quasiparticle_Hamiltonian-1-1}
\end{align}

\begin{align}
 & \mathcal{L}_{Mix}\left(\lambda,\lambda^{\left(n.n.\right)},\Delta,\Delta^{\left(n.n.\right)}\right)\nonumber \\
 & =-2\lambda\Delta-2\lambda^{\left(n.n.\right)}\Delta^{\left(n.n.\right)}+\nonumber \\
 & +U\Delta^{2}+2zV\Delta^{2}-zV\left[\Delta^{\left(n.n.\right)}\right]^{2}.\label{eq:Mixing_HF}
\end{align}
Here $\lambda$ and $\lambda^{\left(n.n.\right)}$ are Lagrange multipliers
used to enforce that $\Delta$ and $\Delta^{\left(n.n.\right)}$ are
the local and nearest neighbor density of quasiparticles for the quasiparticle
Hamiltonian in Eq. (\ref{eq:Quasiparticle_Hamiltonian-1-1}), $E$
is a Lagrange multiplier used to enforce the normalization of the
Slater determinant $\left|\Psi_{0}\right\rangle $ in Eq. (\ref{eq:Quasiparticle_Hamiltonian-1-1}).
We note that, because of the non-local interaction in Eq. (\ref{eq:Extended_Hubbard_simple}),
$H_{QP}$ includes the nonlocal term $\lambda^{\left(n.n.\right)}$.

\subsubsection{\label{subsec:The-extended-Hubbard}The extended Hubbard model Gutzwiller
Lagrange Function}

Let us now consider the extended Hubbard model with $V\neq0$ within
the GA. As we are going to show below, it is possible to extend the
classical GA Lagrange function as follows:
\begin{align}
 & \mathcal{L}_{N}\left(D,F,\mathcal{R},\mathcal{T},E^{c}\right.\nonumber \\
 & \left.\lambda,\lambda^{\left(n.n.\right)},\lambda^{c},\lambda^{b},\Delta,\Delta^{\left(n.n.\right)},o,E,\mu,\left|\Psi_{0}\right\rangle ,\left|\Phi\right\rangle \right)\nonumber \\
 & =\frac{1}{\mathcal{N}}\left\langle \Psi_{0}\right|H_{QP}\left|\Psi_{0}\right\rangle +E\left(1-\left\langle \Psi_{0}\mid\Psi_{0}\right\rangle \right)\nonumber \\
 & +\left\langle \Phi\right|H_{Emb}\left|\Phi\right\rangle +E^{c}\left(1-\left\langle \Phi\mid\Phi\right\rangle \right)+\mu\frac{N}{\mathcal{N}}+\nonumber \\
 & \mathcal{L}_{Mix}\left(D,F,\mathcal{R},\mathcal{T},\lambda,\lambda^{\left(n.n.\right)},\lambda^{c},\lambda^{b},\Delta,\Delta^{\left(n.n.\right)},o\right),\label{eq:Lagrange_GA_simple}
\end{align}
where: 
\begin{align}
 & H_{QP}=-t\sum_{\left\langle RR'\right\rangle \sigma}\left[\mathcal{R}\mathcal{R}^{*}f_{R\sigma}^{\dagger}f_{R'\sigma}+h.c.\right]\nonumber \\
 & +\sum_{\left\langle RR'\right\rangle \sigma}\lambda^{\left(n.n.\right)}f_{R\sigma}^{\dagger}f_{R'\sigma}+\sum_{R\sigma}\lambda f_{R\sigma}^{\dagger}f_{R\sigma}-\mu\sum_{R\sigma}f_{R\sigma}^{\dagger}f_{R\sigma},\label{eq:Quasiparticle_Hamiltonian-1-1-1}
\end{align}
\begin{align}
H_{Emb} & =U\hat{c}_{\uparrow}^{\dagger}\hat{c}_{\uparrow}\hat{c}_{\downarrow}^{\dagger}\hat{c}_{\downarrow}+\sum_{\sigma}\lambda_{\sigma}^{c}\hat{f}_{\sigma}\hat{f}_{\sigma}^{\dagger}\nonumber \\
 & +\left[\sum_{\sigma}D\hat{c}_{\sigma}^{\dagger}\hat{f}_{\sigma}+h.c.\right]+\nonumber \\
 & -F\left(n_{c}n_{f}-2n_{c}\left(1-\Delta\right)\right)+\lambda^{b}n_{c},\label{eq:Embedding_Hamiltonian_V}
\end{align}

\begin{align}
 & \mathcal{L}_{Mix}\left(D,F,\mathcal{R},\mathcal{T},\lambda,\lambda^{\left(n.n.\right)},\lambda^{c},\lambda^{b},\Delta,\Delta^{\left(n.n.\right)},o\right)\nonumber \\
 & =-2\left(\lambda+\lambda^{c}\right)\Delta-2\lambda^{\left(n.n.\right)}\Delta^{\left(n.n.\right)}-2\lambda^{b}o-\nonumber \\
 & -4\left[D\left(\mathcal{R}\sqrt{\left(1-\Delta\right)\Delta}\right)\right]\nonumber \\
 & -2F\left(1-\Delta\right)\Delta\mathcal{T}+2zVo^{2}-zV\left[\Delta^{\left(n.n.\right)}\mathcal{T}\right]^{2}.\label{eq:Mixing_GA_simple}
\end{align}
The derivation of the Lagrange function in Eq. (\ref{eq:Lagrange_GA_simple})
is provided below for the general multi-orbital case. Here we focus
on explaining the main physical meaning of the terms appearing in
the Lagrange function in Eq. (\ref{eq:Lagrange_GA_simple}). 

The main differences of the Lagrange function in Eq. (\ref{eq:Lagrange_GA_simple})
with respect to Eq. (\ref{eq:Lagrnage_function_local}) are: (1) $H_{QP}$
now contains a non-local term $\lambda^{\left(n.n.\right)}$, which
is equal to Eq. (\ref{eq:Quasiparticle_Hamiltonian-1-1}) for the
Hartree-Fock case, but also includes the renormalization factors $\mathcal{R}$.
(2) the embedding Hamiltonian $H_{Emb}$ ---which is the reference
system describing the coupling of the impurity to the environment---
now also includes a density density interaction coupling between the
impurity and the bath. (3) $\mathcal{L}_{Mix}$ contains the new Lagrange
multipliers $F$ and $\lambda^{b}$. Furthermore, it includes the
factor $\mathcal{T}$, which is a correlation-induced correction with
respect to the last term of Eq. (\ref{eq:Mixing_HF}). At the saddle
point, the parameter $o$ equals the local electron occupation per
spin.

\subsection{\label{subsec:Phase-diagram}Benchmark calculations}

Here we focus on the 2D Hubbard model on the square lattice. As shown
in Appendix \ref{sec:Example:-single-band}, extremizing Eq.~(\ref{eq:Lagrange_GA_simple})
is equivalent to minimizing the following energy function of the local
double occupancy $\eta$:
\begin{equation}
\mathcal{E}\left(\eta\right)=\eta^{2}\left[32tz\chi-16Vz\chi^{2}\right]+\eta\left[-16tz\chi+U\right],\label{eq:Energy_polynomial}
\end{equation}
where $0\leq\eta\leq\frac{1}{2}$ , $z=4$ is the number of nearest
neighbors per site and
\begin{equation}
\chi=4\int_{0}^{\pi}\frac{dk_{x}}{2\pi}\int_{0}^{\pi-k_{x}}\frac{dk_{y}}{2\pi}\left(\cos\left(k_{x}\right)+\cos\left(k_{y}\right)\right)=\frac{4}{\pi^{2}}.\label{eq:Xi}
\end{equation}
In Fig.~\ref{fig-Phase-1-1}(A) we show the phase diagram of this
system. 

It is insightful to compare our phase diagram with the DMFT+GW study
of Ref.~\citep{Ayral_2013} (the relevant data is reproduced in Fig.
\ref{fig-Phase-1-1}(B)), where CDW and SDW symmetry breaking was
allowed while we considered only the normal phase. Remarkably our
solution, which is completely encoded in Eq. (\ref{eq:Energy_polynomial}),
is in good quantitative agreement with the numerical DMFT+GW data
\citep{Ayral_2013}. 

\subsubsection{\label{subsec:Brinkman-Rice-phase}Brinkman Rice phase}

Consistently with the fact that our theory reduces to the ordinary
GA in the limit of vanishing intersite interactions, at $V=0$ we
recover the Brinkman Rice transition \citep{Brinkman_1970}, where
$\eta=0$ (i.e., the charge fluctuations are frozen) for all $U\geq16tz\chi$.
More generally the Brinkman Rice phase occurs when $U>16tz\chi$ and
$V<\frac{U}{8tz\chi^{2}}$.

\subsubsection{\label{subsec:Goldstein-Kotliar-Corssover-1-1}Metallic phase: enhanced-valence
crossover}

Minimizing the energy function {[}Eq.~(\ref{eq:Energy_polynomial}){]}
it can be readily shown that for $U<16tz\chi$ and $V<V_{c}=\chi^{-1}\left(1+\frac{U}{16tz\chi}\right)$
the system remains metallic and that, in this phase, the double occupancy
is given by: 
\begin{equation}
\eta=\frac{1-\frac{U}{16tz\chi}}{4\left(1-V\chi/2\right)}\,.\label{eq:sSoluton-1}
\end{equation}
Eq.~(\ref{eq:sSoluton-1}) shows that the intersite Coulomb interaction
can enhance dramatically charge fluctuations. In particular, we note
that $\eta$ can even exceed $\frac{1}{4}$ for $V_{c}>V>\frac{U}{8tz\chi^{2}}$,
which is impossible in the half-filled single-band Hubbard model with
only local Hubbard repulsion. The points where $\eta=\frac{1}{4}$,
which here we refer to as the "enhanced-valence crossover," are
marked by a dotted line in Fig.~(\ref{fig-Phase-1-1}).

\subsubsection{\label{subsec:Goldstein-kotliar-Transition-1-1}Valence-skipping
phase}

The non-local Coulomb interaction can induce a phase transition into
a phase with double occupancy $\eta=\frac{1}{2}$, which is stable
for $V>\chi^{-1}\left(1+\frac{U}{16tz\chi}\right)$ and $V>\frac{U}{8tz\chi^{2}}$.
In this work we refer to this region as the "valence-skipping phase"
as the single site expectation values are given by: 
\begin{align}
\left\langle \Psi\mid0\right\rangle \left\langle 0\mid\Psi\right\rangle  & =\left\langle \Psi\mid\uparrow\downarrow\right\rangle \left\langle \uparrow\downarrow\mid\Psi\right\rangle =\frac{1}{2}\nonumber \\
\left\langle \Psi\mid\uparrow\right\rangle \left\langle \uparrow\mid\Psi\right\rangle  & =\left\langle \Psi\mid\downarrow\right\rangle \left\langle \downarrow\mid\Psi\right\rangle =0.\label{eq:Expectation_values}
\end{align}
For $V\chi>2$ there is a first order phase transition between the
Brinkman Rice phase \citep{Brinkman_1970} and the valence skipping
phase while for $V\chi<2$ there is a second order phase transition
between the metallic phase and the valence skipping phase. The order
of the phase transition can be inferred from the continuity or discontinuity
of $\eta$ across these phase transition line (obtained analytically
in the metallic phase in Eq. (\ref{eq:sSoluton-1})). The tricritical
point is at $\left(U,V\right)=\left(16tz\chi,\frac{2}{\chi}\right)$.

\section{\label{sec:Main-Hamiltonian} Extended multi-orbital Hubbard Hamiltonian }
\begin{widetext}
We consider a generic electronic Hamiltonian, which can be represented
in second quantization notation as follows: 
\begin{align}
H & =\sum_{R}\sum_{\alpha,\beta=1}^{N}E_{R}^{\alpha\beta}\big[c_{R\alpha}^{\dagger}c_{R\beta}\big]+\sum_{R_{1}\neq R_{2}}\sum_{\alpha,\beta=1}^{N}t_{R_{1};R_{2}}^{\alpha;\beta}\big[c_{R_{1}\alpha}^{\dagger}\big]\big[c_{R_{2}\beta}\big]+\sum_{R}\sum_{\alpha,\beta,\gamma,\delta=1}^{N}U_{R}^{\alpha\beta\gamma\delta}\big[c_{R\alpha}^{\dagger}c_{R\beta}^{\dagger}c_{R\gamma}c_{R\delta}\big]\nonumber \\
 & +\sum_{R_{1}\neq R_{2}}\sum_{\alpha,\beta,\gamma,\delta=1}^{N}V_{R_{1};R_{2}}^{\alpha\beta;\gamma\delta}\big[c_{R_{1}\alpha}^{\dagger}c_{R_{1}\beta}\big]\big[c_{R_{2}\gamma}^{\dagger}c_{R_{2}\delta}\big]+\sum_{R_{1}\neq R_{2}}\sum_{\alpha,\beta,\gamma,\delta=1}^{N}Y_{R_{1};R_{2}}^{\alpha\beta;\gamma\delta}\big[c_{R_{1}\alpha}^{\dagger}c_{R_{1}\beta}^{\dagger}\big]\big[c_{R_{2}\gamma}c_{R_{2}\delta}\big]\nonumber \\
 & +\left(\sum_{R_{1}\neq R_{2}}\sum_{\alpha,\beta,\gamma,\delta=1}^{N}X_{R_{1};R_{2}}^{\alpha\beta\gamma;\delta}\big[c_{R_{1}\alpha}^{\dagger}c_{R_{1}\beta}^{\dagger}c_{R_{1}\gamma}\big]\big[c_{R_{2}\delta}\big]+h.c.\right)+\sum_{R_{1}\neq R_{2}\neq R_{3}}\sum_{\alpha,\beta,\gamma,\delta=1}^{N}\mathcal{V}_{R_{1};R_{2};R_{3}}^{\alpha\beta;\gamma;\delta}\big[c_{R_{1}\alpha}^{\dagger}c_{R_{1}\beta}\big]\big[c_{R_{2}\gamma}^{\dagger}\big]\big[c_{R_{3}\delta}\big]\nonumber \\
 & +\left(\sum_{R_{1}\neq R_{2}\neq R_{3}}\sum_{\alpha,\beta,\gamma,\delta=1}^{N}\mathcal{Y}_{R_{1};R_{2};R_{3}}^{\alpha\beta;\gamma;\delta}\big[c_{R_{1}\alpha}^{\dagger}c_{R_{1}\beta}^{\dagger}\big]\left[c_{R_{2}\gamma}\right]\left[c_{R_{3}\delta}\right]+h.c.\right)\nonumber \\
 & +\sum_{R_{1}\neq R_{2}\neq R_{3}\neq R_{4}}\sum_{\alpha,\beta,\gamma,\delta=1}^{N}S_{R_{1};R_{2};R_{3};R_{4}}^{\alpha;\beta;\gamma;\delta}\big[c_{R_{1}\alpha}^{\dagger}\big]\big[c_{R_{2}\beta}^{\dagger}\big]\big[c_{R_{3}\gamma}\big]\big[c_{R_{4}\delta}\big].\label{eq:Hamiltonian-1}
\end{align}
Here $\alpha,\beta,\gamma,\delta$ represent both spin and orbital
degrees of freedom per site, of which there are $N$ in total. We
note that this Hamiltonian represents all possible one and two particle
terms that come from the kinetic energy and Coulomb interaction of
a first principles Hamiltonian. In particular it includes the regular
Hubbard Hamiltonian (the first three terms of the first line). The
square brackets are used to mark explicitly the operators acting over
the same site.

For later convenience, we formally express the Hamiltonian in {[}Eq.
(\ref{eq:Hamiltonian-1}){]} also as follows: 
\begin{align}
H & =\sum_{R}H_{R}^{loc}+\sum_{R_{1}\neq R_{2}}\sum_{\mu,\nu=1}^{2^{2N}}J_{R_{1};R_{2}}^{\mu;\nu}O_{R_{1}\mu}O_{R_{2}\nu}+\sum_{R_{1}\neq R_{2}\neq R_{3}}\sum_{\mu,\nu,\eta=1}^{2^{2N}}J_{R_{1};R_{1};R_{3}}^{\mu;\nu;\eta}O_{R_{1}\mu}O_{R_{2}\nu}O_{R_{3}\eta}+\nonumber \\
 & +\sum_{R_{1}\neq R_{2}\neq R_{3}\neq R_{4}}\sum_{\mu,\nu,\eta,\rho=1}^{2^{2N}}J_{R_{1};R_{2};R_{3};R_{4}}^{\mu;\nu;\eta;\rho}O_{R_{1}\mu}O_{R_{2}\nu}O_{R_{3}\eta}O_{R_{4}\rho}\,,\label{eq:Generic_Hamiltonian}
\end{align}
where the $O_{R\mu}\big[\big\{ c_{R\alpha}^{\dagger},c_{R\beta}\big\}\big]$
are a basis of the linear space of local operators (which can be written
in terms of the local creation and annihilation operators $\big\{ c_{R\alpha}^{\dagger},c_{R\beta}\big\}$)
and the $J$ are complex coefficients. 
\end{widetext}

\section{\label{sec:Gutzwiller-Approximation-and}GA + $P_{R}^{\dagger}P_{R}-I$
expansion}

As in the classic GA theory, we consider the following variational
wave function: 
\begin{equation}
\left|\Psi\right\rangle =\prod_{R}P_{R}\left|\Psi_{0}\right\rangle ,\label{eq:Gutzwiller-2}
\end{equation}
where $P_{R}$ is the most general operator acting on a single site
$R$ and $\left|\Psi_{0}\right\rangle $ is any generic multi-band
Slater determinant wave-function. For simplicity, here we consider
the case of no superconductivity, so in particular the projector $P_{R}$
satisfies $\left[P_{R},\hat{N}\right]=0$, where $\hat{N}$ is the
number operator. We introduce the following subsidiary conditions,
known as the Gutzwiller constraints \citep{Lanata2009,Lanata2012,Lanata_2015,Lanata_2016,Lanata_2017,Lanata_2017(2),Sandri_2014}:
\begin{align}
\left\langle \Psi_{0}\right|P_{R}^{\dagger}P_{R}\left|\Psi_{0}\right\rangle  & =1\nonumber \\
\left\langle \Psi_{0}\right|P_{R}^{\dagger}P_{R}c_{R\alpha}^{\dagger}c_{R\beta}\left|\Psi_{0}\right\rangle  & =\left\langle \Psi_{0}\right|c_{R\alpha}^{\dagger}c_{R\beta}\left|\Psi_{0}\right\rangle .\label{eq:Conditions-4-1}
\end{align}
Our goal is to evaluate the expectation value of the Hamiltonian in
Eq. (\ref{eq:Hamiltonian-1}) with respect to the Gutzwiller wavefunction
in Eq. (\ref{eq:Gutzwiller-2}) subject to the Gutzwiller constraints.

The starting point of our approach consists in retaining only the
leading order in the the $P_{R}^{\dagger}P_{R}-I$ expansion, which
was previously introduced in Refs.~\citep{Wysokinski_2015,Zegrodnik_2019,Zegrerodnik_2018}
and is also summarized in Appendix \ref{sec:Weak-coupling-expansion}
for completeness. Within this approximation we have:

\begin{equation}
\frac{\left\langle \Psi\right|H\left|\Psi\right\rangle }{\left\langle \Psi\mid\Psi\right\rangle }\cong\left\langle \Psi_{0}\right|H_{P}\left|\Psi_{0}\right\rangle \,,\label{eq:Equality_ish-1}
\end{equation}
where: 
\begin{widetext}
\begin{align}
H_{P} & =\sum_{R}P_{R}^{\dagger}H_{R}^{loc}P_{R}+\sum_{R_{1}\neq R_{2}}\sum_{\mu\nu}J_{R_{1};R_{2}}^{\mu;\nu}P_{R_{1}}^{\dagger}O_{R_{1}\mu}P_{R_{1}}P_{R_{2}}^{\dagger}O_{R_{2}\nu}P_{R_{2}}+\nonumber \\
 & +\sum_{R_{1}\neq R_{2}\neq R_{3}}\sum_{\mu\nu\eta}J_{R_{1};R_{2};R_{3}}^{\mu;\nu;\eta}P_{R_{1}}^{\dagger}O_{R_{1}\mu}P_{R_{1}}P_{R_{2}}^{\dagger}O_{R_{2}\nu}P_{R_{2}}P_{R_{3}}^{\dagger}O_{R_{3}\eta}P_{R_{3}}+\nonumber \\
 & +\sum_{R_{1}\neq R_{2}\neq R_{3}\neq R_{4}}\sum_{\mu\nu\eta\rho}J_{R_{1};R_{2};R_{3};R_{4}}^{\mu;\nu;\eta;\rho}P_{R_{1}}^{\dagger}O_{R_{1}\mu}P_{R_{1}}P_{R_{2}}^{\dagger}O_{R_{2}\nu}P_{R_{2}}P_{R_{3}}^{\dagger}O_{R_{3}\eta}P_{R_{3}}P_{R_{4}}^{\dagger}O_{R_{4}\rho}P_{R_{4}}\,.\label{eq:H_P}
\end{align}

Note that this is a key simplification, as in all terms of Eq~(\ref{eq:H_P})
only the operators $P_{R}$ acting over sites with operators $O_{R}$
appear. 
\end{widetext}

\section{\label{sec:Equivalences}Equivalences}
\begin{widetext}
For simplicity, in the main text of this work we will explicitly account
only for the two-site contributions to the Hamiltonian, corresponding
to the following terms of Eq.~(\ref{eq:Hamiltonian-1}) and the first
line of Eq. (\ref{eq:H_P}): 
\begin{align}
\tilde{H}= & \sum_{R}\sum_{\alpha,\beta=1}^{N}E_{R}^{\alpha\beta}\big[c_{R\alpha}^{\dagger}c_{R\beta}\big]+\sum_{R_{1}\neq R_{2}}\sum_{\alpha,\beta=1}^{N}t_{R_{1};R_{2}}^{\alpha;\beta}\big[c_{R_{1}\alpha}^{\dagger}\big]\big[c_{R_{2}\beta}\big]+\sum_{R}\sum_{\alpha,\beta,\gamma,\delta=1}^{N}U_{R}^{\alpha\beta\gamma\delta}\big[c_{R\alpha}^{\dagger}c_{R\beta}^{\dagger}c_{R\gamma}c_{R\delta}\big]\nonumber \\
 & +\sum_{R_{1}\neq R_{2}}\sum_{\alpha,\beta,\gamma,\delta=1}^{N}V_{R_{1};R_{2}}^{\alpha\beta;\gamma\delta}\big[c_{R_{1}\alpha}^{\dagger}c_{R_{1}\beta}\big]\big[c_{R_{2}\gamma}^{\dagger}c_{R_{2}\delta}\big]+\sum_{R_{1}\neq R_{2}}\sum_{\alpha,\beta,\gamma,\delta=1}^{N}Y_{R_{1};R_{2}}^{\alpha\beta;\gamma\delta}\big[c_{R_{1}\alpha}^{\dagger}c_{R_{1}\beta}^{\dagger}\big]\big[c_{R_{2}\gamma}c_{R_{2}\delta}\big]\nonumber \\
 & +\left(\sum_{R_{1}\neq R_{2}}\sum_{\alpha,\beta,\gamma,\delta=1}^{N}X_{R_{1};R_{2}}^{\alpha\beta\gamma;\delta}\big[c_{R_{1}\alpha}^{\dagger}c_{R_{1}\beta}^{\dagger}c_{R_{1}\gamma}\big]\big[c_{R_{2}\delta}\big]+h.c.\right).\label{eq:Equality-1}
\end{align}
Therefore, the effective Hamiltonian $\tilde{H}_{P}$ of Eq.~(\ref{eq:H_P})
reduces to: 
\begin{equation}
\tilde{H}_{P}=\sum_{R}P_{R}^{\dagger}H_{R}^{loc}P_{R}+\sum_{R_{1}\neq R_{2}}\sum_{\mu\nu}J_{R_{1};R_{2}}^{\mu;\nu}\,\left[P_{R_{1}}^{\dagger}O_{R_{1}\mu}P_{R_{1}}\right]\left[P_{R_{2}}^{\dagger}O_{R_{2}\nu}P_{R_{2}}\right]\,.\label{eq:H_P_shorter}
\end{equation}
The treatment of the terms of Eq. (\ref{eq:Hamiltonian-1}) involving
three and four sites will be discussed in Appendix \ref{sec:Towards-ab-intio}.

As we are going to demonstrate, at the leading order of the $1/z$
expansion the expression for the total energy simplifies as follows:
\begin{equation}
\left\langle \Psi_{0}\right|\tilde{H}_{P}\left|\Psi_{0}\right\rangle \cong\left\langle \Psi_{0}\right|\tilde{H}_{Eff}\left|\Psi_{0}\right\rangle \,,\label{eq:Equality_ish}
\end{equation}
where 
\begin{equation}
\tilde{H}_{Eff}=\sum_{R}P_{R}^{\dagger}H_{R}^{loc}P_{R}+\sum_{R_{1}\neq R_{2}}\sum_{\mu\nu}J_{R_{1};R_{2}}^{\mu;\nu}\left[\sum_{i}\mathcal{Z}_{R_{1}\mu i}O_{R_{1}i}\right]\left[\sum_{j}\mathcal{Z}_{R_{2}\nu j}O_{R_{2}j}\right],\label{eq:Generic_hamiltonian-1}
\end{equation}
and the $\mathcal{Z}_{R\mu i}$ are complex numbers 
 that can be expressed as a function of the Gutzwiller variational
parameters.

We note that, analogously to previous work~\citep{Lanata2009,Lanata2012,Lanata_2015,Lanata_2016,Lanata_2017,Sandri_2014},
this simplification amounts to formally replace the operators $P_{R}^{\dagger}O_{R\mu}P_{R}$
of Eq.~(\ref{eq:H_P_shorter}) with $\sum_{i}\mathcal{Z}_{R\mu i}O_{Ri}$.
From now on we are going to refer to these formal substitutions as
``Gutzwiller equivalences''. As shown in Appendix \ref{sec:Sanity-Check},
$H_{Eff}$ is Hermitian (consistently with the fact that the total
energy for the Hamiltonians in Eq. (\ref{eq:Equality_ish}) is real). 
\end{widetext}

\subsection{\label{subsec:Definitions}Definitions}

For completeness, here we summarize the definitions of the variational
parameters previously introduced in Refs. \citep{Lanata_2015,Lanata_2016,Lanata_2017},
in terms of which it will be possible to express conveniently the
total energy also for the non-local interactions at the core of the
present theory.

We express the Gutzwiller operators in the so-called ``mixed-basis
representation'' \citep{Lanata_2015,Lanata_2016,Lanata_2017} defined
as follows: 
\begin{equation}
P_{R}=\sum_{\Gamma,n}[\Lambda_{R}]_{\Gamma n}\left|\Gamma_{R}\right\rangle \left\langle n_{R}\right|\,,\label{eq:Projector-1}
\end{equation}
where

\begin{align}
\left|\Gamma_{R}\right\rangle  & =\left(c_{R1}^{\dagger}\right)^{n_{1}\left(\Gamma,R\right)}.....\left(c_{RN}^{\dagger}\right)^{n_{N}\left(\Gamma,R\right)}\left|0,R\right\rangle \nonumber \\
\left|n_{R}\right\rangle  & =\left(f_{R1}^{\dagger}\right)^{n_{1}\left(n,R\right)}.....\left(f_{RN}^{\dagger}\right)^{n_{N}\left(n,R\right)}\left|0,R\right\rangle .\label{eq:Basis}
\end{align}
Here $n_{\alpha}\left(\Gamma,R\right),\,n_{a}\left(n,R\right)\in\left\{ 0,1\right\} $
represent the occupation numbers of the states $\left|n_{R}\right\rangle ,\,\left|\Gamma_{R}\right\rangle $
respectively and the operators $f_{Ra}$ are the so-called ``quasi-particle''
fermionic operators, which are related to the operators $c_{R\alpha}$
through an arbitrary unitary transformation.

We conveniently express the local reduced density matrix of $|\Psi_{0}\rangle$
as follows: 
\begin{align}
\rho_{R}^{0} & \equiv Tr_{R'\neq R}\left|\Psi_{0}\right\rangle \left\langle \Psi_{0}\right|\nonumber \\
 & =\frac{1}{Z}\exp\left(-\sum_{ab}\left[\ln\left(\frac{\mathbb{I}-\Delta_{RR}^{T}}{\Delta_{RR}^{T}}\right)\right]_{ab}f_{Ra}^{\dagger}f_{Rb}\right)\,,\label{eq:Density_matrix-1}
\end{align}
where 
\begin{align}
[\Delta_{R_{1}R_{2}}]_{ab}=\left\langle \Psi_{0}\right|f_{R_{1}a}^{\dagger}f_{R_{2}b}\left|\Psi_{0}\right\rangle \label{deltadef}
\end{align}
$Z$ is a normalization constant insuring that $Tr\left[\rho_{R}^{0}\right]=1$
and the superscript $T$ indicates the transpose.

It is also convenient to introduce the so-called ``matrix of slave-boson
amplitudes'', see \citep{Lanata_2015,Lanata_2016,Lanata_2017,Lanata_2017(2),Lee_2019},
which is defined as follows: 
\begin{align}
\phi_{R}=\Lambda_{R}\sqrt{P_{R}^{0}}\,,
\end{align}
where 
\begin{equation}
[P_{R}^{0}]_{nn'}=\left\langle n_{R}\mid\rho_{R}^{0}\mid n'_{R}\right\rangle \,.\label{eq:precise_for_beginers}
\end{equation}

Following Refs.~\citep{Lanata_2015,Lanata_2016}, we also introduce
the so-called ``embedding mapping'', which relates the matrix $\phi_{R}$
to the states $\left|\Phi_{R}\right\rangle $ belonging to an auxiliary
impurity model with a bath site of size equal to the size of the impurity.
The definition of the embedding mapping is the following: 
\begin{align}
\left|\Phi_{R}\right\rangle  & =\sum_{\Gamma,n}\exp\left(i\frac{\pi}{2}N\left(\hat{n}_{R}\right)\left(N\left(\hat{n}_{R}\right)-1\right)\right)\times\nonumber \\
 & \times[\phi_{R}]_{\Gamma n}\,U_{PH}\left|\hat{\Gamma}_{R}\right\rangle \left|\hat{n}_{R}\right\rangle \,,\label{eq:State}
\end{align}
where

\begin{align}
\left|\hat{\Gamma}_{R}\right\rangle  & =\left(\hat{c}_{R1}^{\dagger}\right)^{n_{1}\left(\hat{\Gamma},R\right)}.....\left(\hat{c}_{RN}^{\dagger}\right)^{n_{N}\left(\hat{\Gamma},R\right)}\left|0,R\right\rangle \nonumber \\
\left|\hat{n}_{R}\right\rangle  & =\left(\hat{f}_{R1}^{\dagger}\right)^{n_{1}\left(n,R\right)}.....\left(\hat{f}_{RN}^{\dagger}\right)^{n_{N}\left(n,R\right)}\left|0,R\right\rangle .\label{eq:Basis-1}
\end{align}
Here $n_{\alpha}\left(\hat{\Gamma},R\right),\,n_{a}\left(\hat{n},R\right)\in\left\{ 0,1\right\} $
represent the occupation numbers of the states $\left|\hat{n}_{R}\right\rangle ,\,\left|\hat{\Gamma}_{R}\right\rangle $
respectively, and 
\begin{align}
N\left(\hat{n}_{R}\right)= & \sum_{a=1}^{N}n_{a}\left(\hat{n},R\right)
\end{align}
is the sum of the occupation numbers of the single particle states
for the bath. Furthermore $U_{PH}$ is defined as follows: 
\begin{align}
U_{PH}^{\dagger}\hat{c}_{R\alpha}^{\dagger}U_{PH} & =\hat{c}_{R\alpha}^{\dagger}\nonumber \\
U_{PH}^{\dagger}\hat{c}_{R\alpha}U_{PH} & =\hat{c}_{R\alpha}\nonumber \\
U_{PH}^{\dagger}\hat{f}_{Ra}^{\dagger}U_{PH} & =\hat{f}_{Ra}\nonumber \\
U_{PH}^{\dagger}\hat{f}_{Ra}U_{PH} & =\hat{f}_{Ra}^{\dagger}.\label{eq:Particle_hole}
\end{align}
\begin{equation}
U_{PH}\left|0\right\rangle =\prod_{a=1}^{N}\hat{f}_{Ra}^{\dagger}\left|0\right\rangle \,.\label{eq:Particle_hole_vacuum}
\end{equation}
Here the fact that $\left[P_{R},\hat{N}\right]=0$ implies that the
state $\left|\Phi_{R}\right\rangle $ is at half filling, i.e.: 
\begin{equation}
\left[\sum_{\alpha}\hat{c}_{R\alpha}^{\dagger}\hat{c}_{R\alpha}+\sum_{a}\hat{f}_{Ra}^{\dagger}\hat{f}_{Ra}\right]=N\,|\Phi_{R}\rangle\,.\label{eq:Number_half_filling}
\end{equation}
Note that, within the definitions above, the Gutzwiller constraints
in Eq. (\ref{eq:Conditions-4-1}) can be written as \citep{Lanata2009,Lanata2012,Lanata_2015,Lanata_2016,Lanata_2017,Sandri_2014}:
\begin{align}
\left\langle \Phi_{R}\mid\Phi_{R}\right\rangle  & =1\nonumber \\
\left\langle \Phi_{R}\right|\hat{f}_{Rb}\hat{f}_{Ra}^{\dagger}\left|\Phi_{R}\right\rangle  & =\left\langle \Psi_{0}\right|f_{Ra}^{\dagger}f_{Rb}\left|\Psi_{0}\right\rangle \label{eq:Conditions-4}
\end{align}

\subsection{\label{sec:Fermionic-Equivalences}Fermionic Equivalences}

Let us derive the equivalence relations for all local operators in
Eq.~(\ref{eq:Equality-1}) that increase the number of electrons
by one, i.e., $c_{R\alpha}^{\dagger}$ and $c_{R\alpha}^{\dagger}c_{R\beta}^{\dagger}c_{R\gamma}$.

As demonstrated in the Appendices \ref{sec:The--scaling} and \ref{subsec:Fermionic-operators},
at the leading order of the $1/z$ expansion the following Gutzwiller
equivalences hold: 
\begin{align}
P_{R}^{\dagger}c_{R\alpha}^{\dagger}P_{R} & \sim\sum_{a}\mathcal{R}_{R\alpha a}f_{Ra}^{\dagger},\nonumber \\
P_{R}^{\dagger}c_{R\alpha}^{\dagger}c_{R\beta}^{\dagger}c_{R\gamma}P_{R} & \sim\sum_{a}\mathcal{S}_{R\alpha\beta\gamma a}f_{Ra}^{\dagger}.\label{eq:fermion_equivalence_example}
\end{align}
Here $\mathcal{R}_{R}$ and $\mathcal{S}_{R}$ are examples of the
coefficients $\mathcal{Z}_{R\mu i}$. The Hermitian conjugate of Eq.
(\ref{eq:fermion_equivalence_example}) also holds. Here the coefficients
$\mathcal{R}_{R\alpha a},\,\mathcal{S}_{R\alpha\beta\gamma a}$ are
determined by the following equations: 
\begin{align}
\left\langle \Psi_{0}\right|P_{R}^{\dagger}c_{\alpha}^{\dagger}P_{R}f_{Ra}\left|\Psi_{0}\right\rangle  & =\left\langle \Psi_{0}\right|\left[\sum_{b}\mathcal{R}_{R\alpha b}f_{Rb}^{\dagger}\right]f_{Ra}\left|\Psi_{0}\right\rangle \nonumber \\
\left\langle \Psi_{0}\right|P_{R}^{\dagger}c_{\alpha}^{\dagger}c_{\beta}^{\dagger}c_{\gamma}P_{R}f_{Ra}\left|\Psi_{0}\right\rangle  & =\left\langle \Psi_{0}\right|\left[\sum_{b}\mathcal{S}_{R\alpha\beta\gamma b}f_{Rb}^{\dagger}\right]f_{Ra}\left|\Psi_{0}\right\rangle .\label{eq:Equivalence_equations_fermions}
\end{align}
Furthermore in Appendix \ref{subsec:Fermionic-Equivalences} we solve
explicitly these equations and show that: 
\begin{align}
\mathcal{R}_{R\alpha a} & =\sum_{b}\left\langle \Phi_{R}\right|\hat{c}_{R\alpha}^{\dagger}\hat{f}_{Rb}\left|\Phi_{R}\right\rangle 
\\
\mathcal{S}_{R\alpha\beta\gamma a} & =\sum_{b}\left\langle \Phi_{R}\right|\hat{c}_{R\alpha}^{\dagger}\hat{c}_{R\beta}^{\dagger}\hat{c}_{R\gamma}\hat{f}_{Rb}\left|\Phi_{R}\right\rangle \times\nonumber \\
 & \times\left[\frac{1}{\left(\mathbb{I}-\Delta_{RR}\right)\Delta_{RR}}\right]_{ba}^{1/2}\,.\label{eq:Fermionic_solutions_examples2}
\end{align}

\begin{widetext}
In summary, at the leading order of the $1/z$ expansion we have that:
\begin{equation}
\frac{\left\langle \Psi\right|H^{Ferm}\left|\Psi\right\rangle }{\left\langle \Psi\mid\Psi\right\rangle }\cong\left\langle \Psi_{0}\right|\tilde{H}_{Eff}^{Ferm}\left|\Psi_{0}\right\rangle \,,\label{eq:Fermionic_equivalence}
\end{equation}
where: 
\begin{equation}
H^{Ferm}=\sum_{R_{1}\neq R_{2}}\sum_{\alpha,\beta=1}^{N}t_{R_{1};R_{2}}^{\alpha;\beta}\big[c_{R_{1}\alpha}^{\dagger}\big]\big[c_{R_{2}\beta}\big]+\left(\sum_{R_{1}\neq R_{2}}\sum_{\alpha,\beta,\gamma,\delta=1}^{N}X_{R_{1};R_{2}}^{\alpha\beta\gamma;\delta}\big[c_{R_{1}\alpha}^{\dagger}c_{R_{1}\beta}^{\dagger}c_{R_{1}\gamma}\big]\big[c_{R_{2}\delta}\big]+h.c.\right)\,,\label{eq:Hamiltonian_fermion}
\end{equation}
\begin{equation}
\tilde{H}_{Eff}^{Ferm}=\sum_{R_{1}\neq R_{2}}\sum_{\alpha,\beta=1}^{N}\sum_{a,b=1}^{N}t_{R_{1};R_{2}}^{\alpha;\beta}[\mathcal{R}_{R\alpha a}f_{R_{1}a}^{\dagger}][\mathcal{R}_{R\alpha b}^{*}f_{R_{2}b}]+\left(\sum_{R_{1}\neq R_{2}}\sum_{\alpha,\beta,\gamma,\delta=1}^{N}\sum_{a,d=1}^{N}X_{R_{1};R_{2}}^{\alpha\beta\gamma;\delta}[\mathcal{S}_{R\alpha\beta\gamma a}f_{R_{1}a}^{\dagger}][\mathcal{R}_{R\delta d}^{*}f_{R_{2}d}]+h.c.\right)\,.\label{eq:Eff_fermion}
\end{equation}
\end{widetext}

Here the coefficients $\mathcal{R}_{R},\,\mathcal{S}_{R}$ are explicitly
expressed in terms of the Gutzwiller variational parameters in Eqs.~(\ref{eq:Fermionic_solutions_examples2})
and (\ref{eq:Fermionic_solutions_examples2}).

In terms of the notation introduced in Eq. (\ref{eq:Generic_Hamiltonian})
all of the above equations can be schematically represented as follows:
\begin{equation}
P_{R}^{\dagger}O_{R\mu}P_{R}\sim\sum_{a}\mathcal{Z}_{R\mu a}f_{Ra}^{\dagger}.\label{eq:fermionic_equivalence-4}
\end{equation}
Here $\mathcal{R}_{R}$ and $\mathcal{S}_{R}$ are specific instances
of the coefficients $\mathcal{Z}_{R\mu a}$. Here the $\mathcal{Z}_{R\mu a}$
are determined by the equation: 
\begin{align}
 & \left\langle \Psi_{0}\right|P_{R}^{\dagger}O_{R\mu}P_{R}f_{Ra}\left|\Psi_{0}\right\rangle =\nonumber \\
 & =\left\langle \Psi_{0}\right|\left[\sum_{b}\mathcal{Z}_{R\mu b}f_{Rb}^{\dagger}\right]f_{Ra}\left|\Psi_{0}\right\rangle .\label{eq:Equations}
\end{align}
The explicit solution of the equation above is the following, see
Appendix \ref{subsec:Fermionic-Equivalences} : 
\begin{equation}
\mathcal{Z}_{R\mu a}=-\sum_{b}\left\langle \Phi_{R}\right|\hat{O}_{R\mu}\hat{f}_{Rb}\left|\Phi_{R}\right\rangle \left[\frac{1}{\left(\mathbb{I}-\Delta_{RR}\right)\Delta_{RR}}\right]_{ba}^{1/2}.\label{eq:Fermionic_R_matrix}
\end{equation}

\subsection{\label{sec:Bosonic-Equivalences-(fermion}Bosonic Equivalences (fermion
number conserving operators)}

Let us derive the equivalence relations for all local operators of
Eq.~(\ref{eq:Equality-1}) that do not change the number of electrons,
i.e., $c_{R\alpha}^{\dagger}c_{R\beta}$. We note that something similar
can be done for the operators $c_{R\alpha}^{\dagger}c_{R\beta}^{\dagger}c_{R\gamma}c_{R\delta}$
though they do not appear as a single site term as a part of two site
terms in the Hamiltonian in Eq. (\ref{eq:Hamiltonian-1}):

As demonstrated in the Appendices \ref{sec:The--scaling}, \ref{subsec:Fermionic-operators}
and \ref{subsec:Main-Result-1}, at the leading order of the $1/z$
expansion the following Gutzwiller equivalences hold: 
\begin{align}
P_{R}^{\dagger}c_{R\alpha}^{\dagger}c_{R\beta}P_{R} & \sim\sum_{ab}\mathcal{T}_{R\alpha\beta ba}f_{Ra}^{\dagger}f_{Rb}+\mathcal{T}_{R\alpha\beta I}I\nonumber \\
P_{R}^{\dagger}c_{R\alpha}^{\dagger}c_{R\beta}^{\dagger}c_{R\gamma}c_{R\delta}P_{R} & \sim\sum_{ab}\mathcal{T}_{R\alpha\beta\gamma\delta ba}f_{Ra}^{\dagger}f_{Rb}+\mathcal{T}_{R\alpha\beta\gamma\delta I}I.\label{eq:Bosonic_number_conserving_example}
\end{align}
Here $\mathcal{T}_{R}$ is an example of the coefficients $\mathcal{Z}_{R\mu i}$.
In this case we show in Appendix \ref{subsec:Main-Result} that $\mathcal{T}_{R\alpha\beta cd},\mathcal{T}_{R\alpha\beta I}$
are determined by the following equations: 
\begin{align}
 & \left\langle \Psi_{0}\right|P_{R}^{\dagger}c_{R\alpha}^{\dagger}c_{R\beta}P_{R}f_{Ra}^{\dagger}f_{Rb}\left|\Psi_{0}\right\rangle =\nonumber \\
 & \left\langle \Psi_{0}\right|\left[\sum_{cd}\mathcal{T}_{R\alpha\beta dc}f_{Rc}^{\dagger}f_{Rd}+\mathcal{T}_{R\alpha\beta I}I\right]f_{Ra}^{\dagger}f_{Rb}\left|\Psi_{0}\right\rangle \nonumber \\
 & \left\langle \Psi_{0}\right|P_{R}^{\dagger}c_{R\alpha}^{\dagger}c_{R\beta}P_{R}\cdot I\left|\Psi_{0}\right\rangle =\nonumber \\
 & \left\langle \Psi_{0}\right|\left[\sum_{cd}\mathcal{T}_{R\alpha\beta dc}f_{Rc}^{\dagger}f_{Rd}+\mathcal{T}_{R\alpha\beta I}I\right]\cdot I\left|\Psi_{0}\right\rangle .\label{eq:bosonic residues}
\end{align}
Furthermore in Appendix \ref{subsec:Mixed-Basis-results-1} we solve
these equations explicitly and show that: 
\begin{align}
\mathcal{T}_{R\alpha\beta dc} & =\sum_{ab}\left[\frac{1}{\left(\mathbb{I}-\Delta_{RR}\right)\Delta_{RR}}\right]_{da}^{1/2}\left\langle \Phi_{R}\right|\hat{c}_{R\alpha}^{\dagger}\hat{c}_{R\beta}\hat{f}_{Rb}\hat{f}_{Ra}^{\dagger}\left|\Phi_{R}\right\rangle \nonumber \\
 & \times\left[\frac{1}{\left(\mathbb{I}-\Delta_{RR}\right)\Delta_{RR}}\right]_{bc}^{1/2}-\nonumber \\
 & -\left\langle \Phi_{R}\right|\hat{c}_{R\alpha}^{\dagger}\hat{c}_{R\beta}\left|\Phi_{R}\right\rangle \times\left[\frac{1}{\left(\mathbb{I}-\Delta_{RR}\right)}\right]_{dc},\label{eq:Bosonic_conserving_solution_example}
\end{align}

\begin{equation}
\mathcal{T}_{\alpha\beta I}=\left\langle \Phi_{R}\right|\hat{c}_{R\alpha}^{\dagger}\hat{c}_{R\beta}\left|\Phi_{R}\right\rangle -\sum_{cd}\mathcal{R}_{R\alpha\beta c}^{d}\left[\Delta_{RR}\right]_{cd}.\label{eq:R_mu_I-1-1}
\end{equation}

\begin{widetext}
In summary, at the leading order of the $1/z$ expansion we have that:

\begin{equation}
\frac{\left\langle \Psi\right|H^{Con}\left|\Psi\right\rangle }{\left\langle \Psi\mid\Psi\right\rangle }\cong\left\langle \Psi_{0}\right|\tilde{H}_{Eff}^{Con}\left|\Psi_{0}\right\rangle \label{eq:Conserving_equivalence}
\end{equation}
where: 
\begin{equation}
H^{Con}=\sum_{R_{1}\neq R_{2}}\sum_{\alpha,\beta,\gamma,\delta=1}^{N}V_{R_{1};R_{2}}^{\alpha\beta;\gamma\delta}\left[c_{R_{1}\alpha}^{\dagger}c_{R_{1}\beta}\right]\left[c_{R_{2}\gamma}^{\dagger}c_{R_{2}\delta}\right],\label{eq:number_conserving}
\end{equation}

\begin{equation}
\tilde{H}_{Eff}^{Con}=\sum_{R_{1}\neq R_{2}}\sum_{\alpha,\beta,\gamma,\delta=1}^{N}V_{R_{1};R_{2}}^{\alpha\beta;\gamma\delta}\left[\sum_{a,b=1}^{N}\mathcal{T}_{R\alpha\beta ab}f_{Ra}^{\dagger}f_{Rb}+\mathcal{T}_{R\alpha\beta I}I\right]\times\left[\sum_{c,d=1}^{N}\mathcal{T}_{R\gamma\delta cd}f_{Rc}^{\dagger}f_{Rd}+\mathcal{T}_{R\gamma\delta I}I\right].\label{eq:Eff_conserving}
\end{equation}
\end{widetext}

Here the coefficients $\mathcal{T}_{R}$ are explicitly expressed
in terms of the Gutzwiller variational parameters in Eqs.~(\ref{eq:Bosonic_conserving_solution_example})
and (\ref{eq:R_mu_I-1-1}).

In terms of the notation introduced in Eq. (\ref{eq:Generic_Hamiltonian})
all of the above equations can be schematically represented as follows:
\begin{equation}
P_{R}^{\dagger}O_{R\mu}P_{R}\sim\sum_{ab}\mathcal{Z}_{R\mu ba}f_{Ra}^{\dagger}f_{Rb}+\mathcal{Z}_{R\mu I}I.\label{eq:Bosonic_equivalence_number conserving}
\end{equation}
Here $\mathcal{T}_{R}$ is a specific instance of the coefficients
$\mathcal{Z}_{R\mu ba}$. Here the $\mathcal{Z}_{R\mu ba}$ are determined
by the equations: 
\begin{align}
 & \left\langle \Psi_{0}\right|P_{R}^{\dagger}O_{R\mu}P_{R}f_{Ra}^{\dagger}f_{Rb}\left|\Psi_{0}\right\rangle =\nonumber \\
 & \left\langle \Psi_{0}\right|\left[\sum_{cd}\mathcal{Z}_{R\mu dc}f_{Rc}^{\dagger}f_{Rd}+\mathcal{Z}_{R\mu I}I\right]f_{Ra}^{\dagger}f_{Rb}\left|\Psi_{0}\right\rangle \nonumber \\
 & \left\langle \Psi_{0}\right|P_{R}^{\dagger}O_{R\mu}P_{R}\cdot I\left|\Psi_{0}\right\rangle =\nonumber \\
 & \left\langle \Psi_{0}\right|\left[\sum_{cd}\mathcal{Z}_{R\mu dc}f_{Rc}^{\dagger}f_{Rd}+\mathcal{Z}_{R\mu I}I\right]\cdot I\left|\Psi_{0}\right\rangle .\label{eq:bosonic residues-4}
\end{align}
The explicit solution of the equation above is the following, see
Appendix \ref{subsec:Mixed-Basis-results-1}:

\begin{align}
\mathcal{Z}_{R\mu dc} & =\sum_{ab}\left[\frac{1}{\left(\mathbb{I}-\Delta_{RR}\right)\Delta_{RR}}\right]_{da}^{1/2}\left\langle \Phi_{R}\right|\hat{O}_{R\mu}\hat{f}_{Rb}\hat{f}_{Ra}^{\dagger}\left|\Phi_{R}\right\rangle \nonumber \\
 & \times\left[\frac{1}{\left(\mathbb{I}-\Delta_{RR}\right)\Delta_{RR}}\right]_{bc}^{1/2}-\nonumber \\
 & -\left\langle \Phi_{R}\right|\hat{O}_{R\mu}\left|\Phi_{R}\right\rangle \times\left[\frac{1}{\left(\mathbb{I}-\Delta_{RR}\right)}\right]_{dc},\label{eq:Indecies-1}
\end{align}
\begin{equation}
\mathcal{Z}_{R\mu I}=\left\langle \Phi_{R}\right|\hat{O}_{R\mu}\left|\Phi_{R}\right\rangle -\sum_{cd}\mathcal{Z}_{R\mu dc}\left[\Delta_{RR}\right]_{cd}.\label{eq:R_mu_I-1}
\end{equation}

\subsection{\label{sec:Bosonic-Equivalences-(fermion-1}Bosonic Equivalences
(fermion number changing operators)}

Let us derive the equivalence relations for all local operators of
Eq.~(\ref{eq:Equality-1}) that increase the number of electrons
by two, i.e., $c_{R\alpha}^{\dagger}c_{R\beta}^{\dagger}$. We note
that something similar can be done for the operators $c_{R\alpha}^{\dagger}c_{R\beta}^{\dagger}c_{R\gamma}^{\dagger}c_{R\delta}$
though they do not appear as any single site terms as a part of two
site terms in the Hamiltonian in Eq. (\ref{eq:Hamiltonian-1}):

As demonstrated in the Appendices \ref{sec:The--scaling}, \ref{subsec:Bosonic-operators-with}
and \ref{subsec:Main-Result-1}, at the leading order of the $1/z$
expansion the following Gutzwiller equivalences hold: 
\begin{align}
P_{R}^{\dagger}c_{R\alpha}^{\dagger}c_{R\beta}^{\dagger}P_{R} & \sim\sum_{ab}\mathcal{U}_{R\alpha\beta ab}f_{Ra}^{\dagger}f_{Rb}^{\dagger}\nonumber \\
P_{R}^{\dagger}c_{R\alpha}^{\dagger}c_{R\beta}^{\dagger}c_{R\gamma}^{\dagger}c_{R\delta}P_{R} & \sim\sum_{ab}\mathcal{U}_{R\alpha\beta\gamma\delta ab}f_{Ra}^{\dagger}f_{Rb}^{\dagger}.\label{eq:Bosonic_pair_hopping_example}
\end{align}
Here $\mathcal{U}_{R}$ is an example of the coefficients $\mathcal{Z}_{R\mu i}$.
The Hermitian conjugate of Eq. (\ref{eq:Bosonic_pair_hopping_example})
also holds. We will not need to consider terms that change the electron
number by more then two (though a similar treatment may be done for
them) as they are not needed in the Hamiltonian in Eq. (\ref{eq:Hamiltonian-1}).
In Appendix \ref{subsec:Main-Result-2} we show that $\mathcal{U}_{R\alpha\beta cd}$
satisfy the following equations: 
\begin{align}
 & \left\langle \Psi_{0}\right|P_{R}^{\dagger}c_{R\alpha}^{\dagger}c_{R\beta}^{\dagger}P_{R}f_{Ra}f_{Rb}\left|\Psi_{0}\right\rangle =\nonumber \\
 & =\left\langle \Psi_{0}\right|\left[\sum_{cd}\mathcal{U}_{R\alpha\beta cd}f_{Rc}^{\dagger}f_{Rd}^{\dagger}\right]f_{Ra}f_{Rb}\left|\Psi_{0}\right\rangle .\label{eq:relation-1}
\end{align}
Furthermore in Appendix \ref{subsec:Mixed-Basis-results} we solve
these equations explicitly and show that: 
\begin{align}
\mathcal{U}_{\alpha\beta ab} & =-\sum_{\gamma\delta}\left[\frac{1}{\left(\mathbb{I}-\Delta_{RR}^{T}\right)\Delta_{RR}^{T}}\right]_{ac}^{1/2}\times\nonumber \\
 & \times\left\langle \Phi_{R}\right|\hat{c}_{R\alpha}^{\dagger}\hat{c}_{R\beta}^{\dagger}\hat{f}_{Rc}\hat{f}_{Rd}\left|\Phi_{R}\right\rangle \times\left[\frac{1}{\left(\mathbb{I}-\Delta\right)\Delta}\right]_{db}^{1/2}.\label{eq:R_matrix_pair_hopping-1-1}
\end{align}

\begin{widetext}
In summary, at the leading order of the $1/z$ expansion we have that:
\begin{equation}
\frac{\left\langle \Psi\right|H^{Chan}\left|\Psi\right\rangle }{\left\langle \Psi\mid\Psi\right\rangle }\cong\left\langle \Psi_{0}\right|\tilde{H}_{Eff}^{Chan}\left|\Psi_{0}\right\rangle ,\label{eq:Changing_equivalence}
\end{equation}
where:

\begin{equation}
H^{Chan}=\sum_{R_{1}\neq R_{2}}\sum_{\alpha,\beta,\gamma,\delta=1}^{N}Y_{R_{1};R_{2}}^{\alpha\beta;\gamma\delta}\big[c_{R_{1}\alpha}^{\dagger}c_{R_{1}\beta}^{\dagger}\big]\big[c_{R_{2}\gamma}c_{R_{2}\delta}\big],\label{eq:Number_changing}
\end{equation}
\begin{equation}
\tilde{H}_{Eff}^{Chan}=\sum_{R_{1}\neq R_{2}}\sum_{\alpha,\beta,\gamma,\delta=1}^{N}Y_{R_{1};R_{2}}^{\alpha\beta;\gamma\delta}\left[\sum_{a,b=1}^{N}\mathcal{U}_{R\alpha\beta ab}f_{Ra}^{\dagger}f_{Rb}^{\dagger}\right]\times\left[\sum_{c,d=1}^{N}\mathcal{U}_{R\gamma\delta dc}^{*}f_{Rc}f_{Rd}\right].\label{eq:Eff_chan}
\end{equation}
\end{widetext}

Here the coefficients $\mathcal{U}_{R}$ are explicitly expressed
in terms of the Gutzwiller variational parameters in Eqs.~(\ref{eq:R_matrix_pair_hopping-1-1}).

In terms of the notation introduced in Eq. (\ref{eq:Generic_Hamiltonian})
all of the above equations can be schematically represented as follows:
\begin{equation}
P_{R}^{\dagger}O_{R\mu}P_{R}\sim\sum_{ab}\mathcal{\bar{Z}}_{R\mu ab}f_{Ra}^{\dagger}f_{Rb}^{\dagger}.\label{eq:Bosonic_equivalence_pair_hopping}
\end{equation}
Here $\mathcal{U}_{R}$ is a specific instance of the coefficients
$\mathcal{\bar{Z}}_{R\mu ab}$. Here the $\bar{\mathcal{Z}}_{R\mu ab}$
are determined by the equations: 
\begin{align}
 & \left\langle \Psi_{0}\right|P_{R}^{\dagger}O_{R\mu}P_{R}f_{Ra}f_{Rb}\left|\Psi_{0}\right\rangle =\nonumber \\
 & =\left\langle \Psi_{0}\right|\left[\sum_{cd}\bar{\mathcal{Z}}_{R\mu cd}f_{Rc}^{\dagger}f_{Rd}^{\dagger}\right]f_{Ra}f_{Rb}\left|\Psi_{0}\right\rangle .\label{eq:relation-1-3}
\end{align}
The explicit solution of the equation above is the following, see
Appendix \ref{subsec:Mixed-Basis-results}:

\begin{align}
\bar{\mathcal{Z}}_{\mu ab} & =-\sum_{\gamma\delta}\left[\frac{1}{\left(\mathbb{I}-\Delta_{RR}^{T}\right)\Delta_{RR}^{T}}\right]_{ac}^{1/2}\times\nonumber \\
 & \times\left\langle \Phi_{R}\right|\hat{O}_{R\mu}\hat{f}_{Rc}\hat{f}_{Rd}\left|\Phi_{R}\right\rangle \times\left[\frac{1}{\left(\mathbb{I}-\Delta\right)\Delta}\right]_{db}^{1/2}.\label{eq:R_matrix_pair_hopping-1}
\end{align}

\section{\label{sec:Gutzwiller-Lagrangian}Gutzwiller Lagrange Function}

In Section \ref{sec:Equivalences} we have expressed explicitly the
total energy as a function of the variational parameters at the leading
order of the $1/z$ and the $P_{R}^{\dagger}P_{R}-I$ expansions.

Here we consider the problem of minimizing the variational energy
with respect to the variational parameters $\left|\Psi_{0}\right\rangle $
and $\left\{ \left|\Phi_{R}\right\rangle \right\} $. To achieve this
goal we need to take into account that: (1) the energy has to be minimized
satisfying the constraints, (2) the renormalization coefficients $\left\{ \mathcal{R}_{R},\mathcal{S}_{R},\mathcal{T}_{R},\mathcal{U}_{R}\right\} $
depend on the variational parameters non-linearly, see Eqs. (\ref{eq:Fermionic_R_matrix}),
(\ref{eq:Bosonic_conserving_solution_example}), (\ref{eq:R_mu_I-1-1})
and (\ref{eq:R_matrix_pair_hopping-1-1})). Following Refs. \citep{Lanata_2015,Lanata_2016,Lanata_2017},
these problems can tackled introducing Lagrange multipliers both for
enforcing the Gutzwiller constraints (Eq. \ref{eq:Conditions-4})
as well as for promoting the coefficients $\left\{ \mathcal{R}_{R},\mathcal{S}_{R},\mathcal{T}_{R},\mathcal{U}_{R}\right\} $
in Eqs. (\ref{eq:Fermionic_R_matrix}), (\ref{eq:Bosonic_conserving_solution_example}),
(\ref{eq:R_mu_I-1-1}) and (\ref{eq:R_matrix_pair_hopping-1-1}))
and $\left[\Delta_{R_{1}R_{2}}\right]_{ab}$ to independent variables.
Furthermore, we promote to independent variable the coefficients $o_{R\alpha\beta}=\left\langle \Phi_{R}\right|\hat{c}_{R\alpha}^{\dagger}\hat{c}_{R\beta}\left|\Phi_{R}\right\rangle $. 
\begin{widetext}
Within this strategy, the energy minimization problem amounts to calculate
the saddle points of the following Lagrange function: 
\begin{align}
 & \mathcal{L}_{N}\left(\left\{ D_{R},E_{R},F_{R},G_{R}\right\} ,\left\{ D_{R}^{\ast},E_{R}^{\ast},F_{R}^{\ast},G_{R}^{\ast}\right\} ,\left\{ \mathcal{R}_{R},\mathcal{S}_{R},\mathcal{T}_{R},\mathcal{U}_{R}\right\} ,\left\{ \mathcal{R}_{R}^{*},\mathcal{S}_{R}^{*},\mathcal{T}_{R}^{\ast},\mathcal{U}_{R}^{\ast}\right\} ,\right.\nonumber \\
 & \left.\left[\lambda_{R_{1}R_{2}}\right]_{ab},\left[\lambda_{R}^{c}\right]_{ab},\lambda_{R\alpha\beta}^{b},o_{R\alpha\beta},E_{R}^{c},E,\left[\Delta_{R_{1}R_{2}}\right]_{ab},\mu,\left|\Phi_{R}\right\rangle ,\left|\Psi_{0}\right\rangle \right)\nonumber \\
 & =\mathcal{L}_{QP}\left(\left\{ \mathcal{R}_{R},\mathcal{S}_{R}\right\} ,\left\{ \mathcal{R}_{R}^{*},\mathcal{S}_{R}^{\ast}\right\} ,\left[\lambda_{R_{1}R_{2}}\right]_{ab},E,\left|\Psi_{0}\right\rangle \right)+\nonumber \\
 & +\mathcal{L}_{Embed}\left(\left\{ D_{R},E_{R},F_{R},G_{R}\right\} ,\left\{ D_{R}^{\ast},E_{R}^{\ast},F_{R}^{\ast},G_{R}^{\ast}\right\} ,\left[\Delta_{R_{1}R_{2}}\right]_{ab},\lambda_{R\alpha\beta}^{b},E_{R}^{c},\left|\Phi_{R}\right\rangle \right)+\nonumber \\
 & +\mathcal{L}_{Mix}\left(\left\{ D_{R},E_{R},F_{R},G_{R}\right\} ,\left\{ D_{R}^{\ast},E_{R}^{\ast},F_{R}^{\ast},G_{R}^{\ast}\right\} ,\left\{ \mathcal{R}_{R},\mathcal{S}_{R},\mathcal{T}_{R},\mathcal{U}_{R}\right\} ,\right.\nonumber \\
 & \left.\left\{ \mathcal{R}_{R}^{*},\mathcal{S}_{R}^{*},\mathcal{T}_{R}^{\ast},\mathcal{U}_{R}^{\ast}\right\} ,\left[\lambda_{R_{1}R_{2}}\right]_{ab},\left[\lambda_{R}^{c}\right]_{ab},\lambda_{R\alpha\beta}^{b},\left[\Delta_{R_{1}R_{2}}\right]_{ab},o_{R\alpha\beta}\right)\nonumber \\
 & +\mathcal{L}_{HF}\left(\left\{ \mathcal{T}_{R},\mathcal{U}_{R}\right\} ,\left\{ \mathcal{T}_{R}^{\ast},\mathcal{U}_{R}^{\ast}\right\} ,\left[\Delta_{R_{1}R_{2}}\right]_{ab},o_{R\alpha\beta}\right),\label{eq:Lagrangian}
\end{align}

where:

\begin{align}
 & \mathcal{L}_{QP}\left(\left\{ \mathcal{R}_{R},\mathcal{S}_{R}\right\} ,\left\{ \mathcal{R}_{R}^{*},\mathcal{S}_{R}^{*}\right\} ,\left[\lambda_{R_{1}R_{2}}\right]_{ab},E,\mu,\left|\Psi_{0}\right\rangle \right)=\nonumber \\
 & =\left\langle \Psi_{0}\right|-\sum_{R_{1}R_{2}\alpha\beta}t_{R_{1};R_{2}}^{\alpha;\beta}\mathcal{R}_{R_{1}\alpha c}\mathcal{R}_{R_{2}\beta d}^{\ast}f_{R_{1}c}^{\dagger}f_{R_{2}d}+\left[\sum_{\alpha\beta\gamma;\delta=1}^{N}\sum_{c,d=1}^{N}X_{R_{1};R_{2}}^{\alpha\beta\gamma;\delta}\mathcal{S}_{R_{1}\alpha\beta;c}^{\gamma}\mathcal{R}_{R_{2}\delta d}^{*}f_{R_{1}c}^{\dagger}f_{R_{2}d}+h.c.\right]+\nonumber \\
 & +\sum_{R_{1}R_{2}}\sum_{a,b=1}^{N}\left[\lambda_{R_{1}R_{2}}\right]_{ab}f_{R_{1}a}^{\dagger}f_{R_{2}b}-\mu\sum_{Ra}f_{Ra}^{\dagger}f_{Ra}\left|\Psi_{0}\right\rangle +E\left(1-\left\langle \Psi_{0}\mid\Psi_{0}\right\rangle \right)+\mu\mathcal{N},\label{eq:Quasiparticle_lagrangian}
\end{align}

\begin{equation}
\mathcal{L}_{Embed}\left(\left\{ D_{R},E_{R},F_{R},G_{R}\right\} ,\left\{ D_{R}^{\ast},E_{R}^{\ast},F_{R}^{\ast},G_{R}^{\ast}\right\} ,E_{R}^{c},\lambda_{R\alpha\beta}^{b},\left[\Delta_{RR}\right]_{ab},\left|\Phi_{R}\right\rangle \right)=\sum_{R}\left\langle \Phi_{R}\right|H_{Embed}^{R}\left|\Phi_{R}\right\rangle +E_{R}^{c}\left(1-\left\langle \Phi_{R}\mid\Phi_{R}\right\rangle \right),\label{eq:Embed}
\end{equation}

\begin{align}
H_{embed}^{R} & =\sum_{\alpha\beta\gamma\delta}U_{R}^{\alpha\beta\gamma\delta}\hat{c}_{R\alpha}^{\dagger}\hat{c}_{R\beta}^{\dagger}\hat{c}_{R\gamma}\hat{c}_{R\delta}+\sum_{\alpha\beta}E_{R}^{\alpha\beta}\hat{c}_{R\alpha}^{\dagger}\hat{c}_{R\beta}+\sum_{ab}\left[\lambda_{R}^{c}\right]_{ab}\hat{f}_{Rb}\hat{f}_{Ra}^{\dagger}+\left[\sum_{\alpha b}\left[D_{R\alpha}\right]_{b}\hat{c}_{R\alpha}^{\dagger}\hat{f}_{Rb}+h.c.\right]\nonumber \\
 & +\left[\sum_{\alpha\beta cd}\left[E_{R\alpha\beta}^{\gamma}\right]_{d}\hat{c}_{R\alpha}^{\dagger}\hat{c}_{R\beta}^{\dagger}\hat{c}_{R\gamma}\hat{f}_{Rd}+h.c.\right]+\sum_{\alpha\beta cd}\left[F_{R\alpha\beta}\right]_{cd}\left(\hat{c}_{R\alpha}^{\dagger}\hat{c}_{R\beta}\hat{f}_{Rd}\hat{f}_{Rc}^{\dagger}-\hat{c}_{R\alpha}^{\dagger}\hat{c}_{R\beta}\left[\Delta\right]_{cd}\right)\nonumber \\
 & -\left[\sum_{\alpha>\beta;c>d}\left[G_{R\alpha\beta}\right]_{cd}\hat{c}_{R\alpha}^{\dagger}\hat{c}_{R\beta}^{\dagger}\hat{f}_{Rd}\hat{f}_{Rc}+h.c.\right]+\sum_{\alpha\beta}\lambda_{R\alpha\beta}^{b}\hat{c}_{R\alpha}^{\dagger}\hat{c}_{R\beta},\label{eq:Embedding_Hamiltonian}
\end{align}

\begin{align}
 & \mathcal{L}_{Mix}\left(\left\{ D_{R},E_{R},F_{R},G_{R}\right\} ,\left\{ D_{R}^{\ast},E_{R}^{\ast},F_{R}^{\ast},G_{R}^{\ast}\right\} ,\left\{ \mathcal{R}_{R},\mathcal{S}_{R},\mathcal{T}_{R},\mathcal{U}_{R}\right\} ,\left\{ \mathcal{R}_{R}^{*},\mathcal{S}_{R}^{*},\mathcal{T}_{R}^{\ast},\mathcal{U}_{R}^{\ast}\right\} ,\left[\lambda_{R_{1}R_{2}}\right]_{ab},\left[\lambda_{R}^{c}\right]_{ab},\lambda_{R\alpha\beta}^{b},\left[\Delta_{R_{1}R_{2}}\right]_{ab},o_{R\alpha\beta}\right)\nonumber \\
 & =-\sum_{Rab}\left[\lambda_{RR}+\lambda_{R}^{c}\right]_{ab}\left[\Delta_{RR}\right]_{ab}-\sum_{R_{1}\neq R_{2}ab}\left[\lambda_{R_{1}R_{2}}\right]_{ab}\left[\Delta{}_{R_{1}R_{2}}\right]_{ab}-\sum_{R\alpha\beta}\lambda_{R\alpha\beta}^{b}o_{R\alpha\beta}-\nonumber \\
 & -\sum_{R\beta a}\left[\left[D_{R\beta}\right]_{a}\left[\mathcal{R}_{R\beta}\sqrt{\left(\mathbb{I}-\Delta_{RR}\right)\Delta_{RR}}\right]_{a}+c.c.\right]-\sum_{R\alpha\beta\gamma d}\left[\left[E_{R\alpha\beta}^{\gamma}\right]_{d}\left[\mathcal{S}_{R\alpha\beta}^{\gamma}\sqrt{\left(\mathbb{I}-\Delta_{RR}\right)\Delta_{RR}}\right]_{d}+c.c.\right]\nonumber \\
 & -\sum_{R\alpha\beta ab}\left[F_{R\alpha\beta}\right]_{ab}\left(\sqrt{\left(\mathbb{I}-\Delta_{RR}\right)\Delta_{RR}}\mathcal{T}_{R\beta}^{\alpha}\sqrt{\left(\mathbb{I}-\Delta_{RR}\right)\Delta_{RR}}\right)_{ab}\nonumber \\
 & -\sum_{R\alpha>\beta ab}\left[\left[G_{R\alpha\beta}\right]_{ab}\left[\sqrt{\left(\mathbb{I}-\Delta_{RR}\right)\Delta_{RR}}\mathcal{U}_{R\alpha\beta}\sqrt{\left(\mathbb{I}-\Delta_{RR}^{T}\right)\Delta_{RR}^{T}}\right]_{ab}+c.c.\right],\label{eq:Mixing_Lagrangian}
\end{align}
\begin{align}
 & \mathcal{L}_{HF}\left(\left\{ \mathcal{T}_{R},\mathcal{U}_{R}\right\} ,\left\{ \mathcal{T}_{R}^{\ast},\mathcal{U}_{R}^{\ast}\right\} ,\left[\Delta_{R_{1}R_{2}}\right]_{ab},o_{R\alpha\beta}\right)=\nonumber \\
 & =\sum_{R_{1}R_{2}\alpha\beta\gamma\delta}V_{R_{1};R_{2}}^{\alpha\beta;\gamma\delta}o_{R_{1}\alpha\beta}o_{R_{2}\gamma\delta}-\sum_{R_{1}\neq R_{2}}\sum_{\alpha\beta\gamma\delta}\sum_{abcd}V_{R_{1};R_{2}}^{\alpha\beta;\gamma\delta}\left[\Delta_{R_{1}R_{2}}\right]_{ac}\left[\Delta_{R_{2}R_{1}}\right]_{db}\left[\mathcal{T}_{R_{1}\alpha}^{\beta}\right]_{ba}\left[\mathcal{T}_{R_{2}\gamma}^{\delta}\right]_{dc}+\nonumber \\
 & +\sum_{R_{1}\neq R_{2}}\sum_{a>b;c>d}\left[Y_{R_{1};R_{2}}^{\alpha\beta;\gamma\delta}\sum_{a>b}\sum_{c>d}\left(\left[\Delta_{R_{1}R_{2}}\right]_{ad}\left[\Delta_{R_{1}R_{2}}\right]_{bc}-\left[\Delta_{R_{1}R_{2}}\right]_{ac}\left[\Delta_{R_{1}R_{2}}\right]_{bd}\right)\left[\mathcal{U}_{R_{1}\alpha\beta}\right]_{ab}\left[\mathcal{U}_{R_{2}\gamma\delta}^{*}\right]_{cd}\right]\,.\label{eq:L_HF}
\end{align}
\end{widetext}

Here $[E_{R\alpha}^{\beta}]_{cd}$, $[F_{R\alpha\beta}]_{cd}$, $[G_{R\alpha\beta}^{\gamma}]_{d}$
and $[\mathcal{T}_{R\alpha}^{\beta}]_{cd}$, $[\mathcal{U}_{R\alpha\beta}]_{cd}$,
$[\mathcal{S}_{R\alpha\beta}^{\gamma}]_{d}$ represent tensors of
the size $N\times N\times N\times N$. Furthermore $\left[D_{R\alpha}\right]_{b}$
and $\left[\mathcal{R}_{R\alpha}\right]_{b}$ and $o_{R\alpha\beta},\left[\lambda_{R}^{c}\right]_{ab},\lambda_{R\alpha\beta}^{b}$
represent matrices of size $N\times N$ for each site $R$. Further
$\left[\Delta_{R_{1}R_{2}}\right]_{ab},\left[\lambda_{R_{1}R_{2}}\right]_{ab}$
represent matrices of size $N\times N$ indexed by all the pairs of
sites $R_{1},R_{2}$. Here $E$ and $E_{R}^{c}$ are single numbers
indexed by nothing or the site $R$ respectively. We point out that
these tensors satisfy the relations:

\begin{align}
\left[F_{R\alpha\beta}\right]_{cd} & =\left[F_{R\beta\alpha}^{*}\right]_{cd}\label{eq:Equivalnce}\\
\left[\mathcal{U}_{R\alpha\beta}\right]_{cd} & =-\left[\mathcal{U}_{R\alpha\beta}\right]_{dc}\\
\left[\lambda_{R_{1}R_{2}}\right]_{ab} & =\left[\lambda_{R_{1}R_{2}}^{*}\right]_{ba}\\
\left[\Delta_{R_{1}R_{2}}\right]_{ab} & =\left[\Delta_{R_{2}R_{1}}^{*}\right]_{ba}\\
\lambda_{R\alpha\beta}^{b} & =\lambda_{R\beta\alpha}^{b*}\\
o_{R\alpha\beta} & =o_{R\beta\alpha}^{*}.
\end{align}
Note that, if we delete the terms $V,X,Y$ and the Lagrange multipliers
relevant to them (in particular delete the Hartree-Fock part of the
Lagrangian), we reduce our problem to known results \citep{Lanata2009,Lanata2012,Lanata_2015,Lanata_2016}.
Extensions to Ghost Gutzwiller construction is straightforward though
cumbersome \citep{Lanata_2017(2)}.

\section{\label{sec:Conclusions}Conclusions}

In this work we have introduced a new method to study the GA for a
broad class of multi-band extended Hubbard Hamiltonians with two site
interactions by combing the large $z$ and the $P_{R}^{\dagger}P_{R}-I$
expansions. We have presented the final result in terms of a Gutzwiller
Lagrange function valid for multi-band Hubbard models. Using this
formalism, we have studied the single band extended Hubbard model,
showing that this method is highly practical and leads to new qualitative
results. In particular, we have recovered a Brinkman-Rice transition
\citep{Brinkman_1970} for the extended Hubbard model and observed
that a valence skipping phase emerges for large intersite interactions.
Our work can enable a more refined treatment of intersite interactions
in the ab-initio calculations. In the future this may lead to parameter
free theories of realistic solids and molecules.

Acknowledgements: This work was supported by the Computational Materials
Sciences Program funded by the US Department of Energy, Office of
Science, Basic Energy Sciences, Materials Sciences and Engineering
Division. N. L. was supported by the VILLUM FONDEN via the Centre
of Excellence for Dirac Materials (Grant No. 11744).

\appendix

\section{\label{sec:Weak-coupling-expansion} $P_{R}^{\dagger}P_{R}-I$ expansion}

In this Appendix we will describe the $P_{R}^{\dagger}P_{R}-I$ expansion
used in Section \ref{sec:Gutzwiller-Approximation-and} and use it
to derive Eq. (\ref{eq:Equality_ish-1}) at leading order in the $P_{R}^{\dagger}P_{R}-I$
expansion. We will focus on the case of the two point operator, though
higher point operators may be handled similarly. Corrections to Eq.
(\ref{eq:Equality_ish-1}) appear as higher order terms in the $P_{R}^{\dagger}P_{R}-I$
expansion parameter $x$, which we now introduce through the following
relation: 
\begin{equation}
P_{R}^{\dagger}P_{R}\equiv I+x\left(P_{R}^{\dagger}P_{R}-I\right)\equiv I+x\Theta_{R}.\label{eq:Operator-1}
\end{equation}
The correct value of $x$ is given by $x=1$, but we will treat $x$
as a small parameter; which stems from the physical assumption that
all the eigenvalues of $\Theta_{R}\ll1$. As such we will drop all
terms proportional to any positive power of $x$.

\subsection{\label{subsec:Warm-Up-I:}The case with two sites}

Consider the simplest case where the system is composed of just two
sites $R$ and $R'$. In this case we have that: 
\begin{align}
 & \frac{\left\langle \Psi\right|O_{R}O_{R'}\left|\Psi\right\rangle }{\left\langle \Psi\mid\Psi\right\rangle }\nonumber \\
 & =\frac{\left\langle \Psi_{0}\right|P_{R}^{\dagger}O_{R}P_{R}P_{R'}^{\dagger}O_{R'}P_{R'}\left|\Psi_{0}\right\rangle }{\left\langle \Psi_{0}\right|P_{R}^{\dagger}P_{R}P_{R'}^{\dagger}P_{R'}\left|\Psi_{0}\right\rangle }\nonumber \\
 & =\frac{\left\langle \Psi_{0}\right|P_{R}^{\dagger}O_{R}P_{R}P_{R'}^{\dagger}O_{R'}P_{R'}\left|\Psi_{0}\right\rangle }{\mathscr{D}},\label{eq:Two_site}
\end{align}
where 
\begin{align}
\mathscr{D} & =1+x\left\langle \Psi_{0}\right|\Theta_{R}\left|\Psi_{0}\right\rangle \nonumber \\
 & +x\left\langle \Psi_{0}\right|\Theta_{R'}\left|\Psi_{0}\right\rangle +x^{2}\left\langle \Psi_{0}\right|\Theta_{R}\Theta_{R'}\left|\Psi_{0}\right\rangle \label{eq:Denominator}
\end{align}
We see that 
\begin{equation}
\frac{\left\langle \Psi\right|O_{R}O_{R'}\left|\Psi\right\rangle }{\left\langle \Psi\mid\Psi\right\rangle }\cong\left\langle \Psi_{0}\right|P_{R}^{\dagger}O_{R}P_{R}P_{R'}^{\dagger}O_{R'}P_{R'}\left|\Psi_{0}\right\rangle \label{eq:Approximate}
\end{equation}
with corrections being order $x$ or higher. It is not too hard to
see that this is still true in the general case of many sites. Indeed
any terms not in the form of the right hand side of Eq. (\ref{eq:Approximate})
in the expansion of $\frac{\left\langle \Psi\right|O_{R}O_{R'}\left|\Psi\right\rangle }{\left\langle \Psi\mid\Psi\right\rangle }$,
both in the denominator and the numerator, come with positive powers
of $x$, and therefore are neglected in the $P_{R}^{\dagger}P_{R}-I$
approximation.

\section{\label{sec:The--scaling} $1/z$ scaling }

\subsection{\label{sec:The-single-particle}$1/z$ Scaling for a single particle}

As a first step to understand the various terms that enter the Gutzwiller
energy function in Eq. (\ref{eq:Generic_Hamiltonian}), we will consider
the single particle Hamiltonian in the limit of large dimensions.
We will follow closely \citep{Metzner_1989}. We will show that: 
\begin{align}
t_{n}^{\alpha;\beta} & \sim\frac{1}{z^{n/2}}\nonumber \\
\left\langle \Psi_{0}\right|f_{Ra}^{\dagger}f_{R'b}\left|\Psi_{0}\right\rangle  & \sim\frac{1}{z^{n/2}}.\label{eq:Correlations_a_b}
\end{align}
For simplicity we assume a hypercubic lattice. Here $n$ is the Manhattan
distance between $R$ and $R'$. Where the Manhattan distance, $D\left(R,R'\right)$,
is the shortest distance between two points $R$ and $R'$ that can
be travelled by a particle that can only move on the edges of the
hypercubic lattice.

For simplicity we will consider a spinless, single band, tight binding
Hamiltonian on the hypercubic lattice (orbital and spin degrees may
be added straightforwardly): 
\begin{equation}
H=-\sum_{RR'}t_{RR'}c_{R}^{\dagger}c_{R'}+h.c.\label{eq:Hamiltonian}
\end{equation}
Lets decompose the Hamiltonian into pieces with equal Manhattan distance:
\begin{equation}
H=-\sum_{n=1}^{\infty}\sum_{D\left(R,R'\right)=n}t_{n}c_{R}^{\dagger}c_{R'}+h.c.\equiv H_{n}\label{eq:Manhattan}
\end{equation}
The eigenvalues of the Hamiltonian $H_{n}$ are given by: 
\begin{align}
 & E_{n}\left(k_{1},k_{2},...k_{d}\right)\nonumber \\
 & =-t_{n}\sum_{P_{n}}\left[\exp\left(i\vec{P}_{n}\cdot\vec{k}\right)+\exp\left(-i\vec{P}_{n}\cdot\vec{k}\right)\right]\nonumber \\
 & =-t_{n}\sum_{P_{n}}E_{P_{n}}\left(\vec{k}\right)\label{eq:Energy}
\end{align}
Here the $P_{n}$ are all the Manhattan paths of total length $n$
with distinct endpoints (we choose only one path for each endpoint)
and $\vec{P}_{n}$ is the vector displacement of the Manhattan path
(note we are grouping path $\vec{P}_{n}$ and $-\vec{P}_{n}$ together
to obtain a real value for the energy). It is not too hard to see
that there are: 
\begin{equation}
2^{n-1}\left(\begin{array}{c}
d+n-1\\
n
\end{array}\right)\equiv\mathcal{N}\left(n\right)\sim z^{n}\label{eq:scaling}
\end{equation}
such paths. We now have that: 
\begin{align}
\int\frac{d^{d}k}{\left(2\pi\right)^{d}}E_{P_{n}}\left(\vec{k}\right) & =0\nonumber \\
\int\frac{d^{d}k}{\left(2\pi\right)^{d}}E_{P_{n}}\left(\vec{k}\right)E_{P'_{n}}\left(\vec{k}\right) & =2\delta_{P_{n}P'_{n}}\label{eq:Energy_variance}
\end{align}
Therefore by the central limit theorem we have that the energy has
a distribution given by: 
\begin{align}
\mathscr{P}_{n}\left(E\right) & \equiv\int\frac{d^{d}k}{\left(2\pi\right)^{d}}\delta\left(-t_{n}\sum_{P_{n}}E_{P_{n}}\left(\vec{k}\right)-E\right)\nonumber \\
 & \cong\frac{1}{\mathscr{Z}}\exp\left(-t_{n}^{2}\cdot\mathcal{N}\left(n\right)\cdot E^{2}\right)\label{eq:Energy_width}
\end{align}
Here $\mathscr{Z}$ is a normalization constant. We now demand that
$\mathscr{P}_{n}\left(E\right)$ be independent of $z$ in which case
we must have that: 
\begin{equation}
t_{n}\sim\frac{1}{\sqrt{\mathcal{N}\left(n\right)}}\sim\frac{1}{z^{n/2}}\label{eq:Hopping}
\end{equation}
We now have that 
\begin{equation}
\int_{-\infty}^{0}E\cdot\mathscr{P}_{n}\left(E\right)=\left\langle H_{n}\right\rangle \sim1\label{eq:Energy-1}
\end{equation}
We further have that 
\begin{equation}
\left\langle H_{n}\right\rangle =-t_{n}\cdot2\cdot\mathcal{N}\left(n\right)\cdot\left\langle \Psi_{0}\right|c_{R}^{\dagger}c_{R'}\left|\Psi_{0}\right\rangle \sim1\label{eq:Energy_correlation}
\end{equation}
This means that: 
\begin{equation}
\left\langle \Psi_{0}\right|c_{R}^{\dagger}c_{R'}\left|\Psi_{0}\right\rangle \sim\frac{1}{z^{n/2}}\label{eq:Correlation}
\end{equation}
from which Eq. (\ref{eq:Correlations_a_b}) follows. Below we use
these results to show how the various terms in the main Hamiltonian
in Eq. (\ref{eq:Generic_Hamiltonian}) scale.

\subsection{\label{sec:Scaling-of-operator}Scaling of operator expectation values}

We will consider the large co-ordination number, large $z$, approximation.
As a first step towards towards obtaining the results in Appendices
\ref{sec:Leading-order-in} and \ref{sec:Main-construction} (which
present key results needed in the main text in Section \ref{sec:Equivalences})
we calculate the scaling of various operators in the large $z$limit.
We will also only take the leading order in the $P_{R}^{\dagger}P_{R}-I$
expansion.

\subsubsection{\label{sec:Scaling-fermionic-operators}Scaling fermionic operators}

We now used Wick's theorem to obtain the scaling for the expectation
values of various terms in Eq. (\ref{eq:Generic_Hamiltonian}). Lets
assume that the Hamiltonian contains $J_{R;R'}^{\mu;\nu}O_{R\mu}O_{R'\nu}$,
with $O_{R\mu}$ and $O_{R'\nu}$ fermionic, then we know that the
lowest order contribution to $J_{RR'}^{\mu;\nu}\left\langle \Psi\right|O_{R\mu}O_{R'\nu}\left|\Psi\right\rangle /\left\langle \Psi\mid\Psi\right\rangle $
may be written as: 
\begin{align}
 & J_{R;R'}^{\mu;\nu}\left\langle \Psi\right|O_{R\mu}O_{R'\nu}\left|\Psi\right\rangle /\left\langle \Psi\mid\Psi\right\rangle \nonumber \\
 & \sim J_{R;R'}^{\mu;\nu}\sum_{ab=1}^{N}\mathcal{Z}_{\mu a}^{*}\mathcal{Z}_{\nu b}\left\langle \Psi_{0}\right|f_{Ra}^{\dagger}f_{R'b}\left|\Psi_{0}\right\rangle \nonumber \\
 & \sim J_{R;R'}^{\mu;\nu}\cdot\frac{1}{z^{n/2}}.\label{eq:Energy_fermion}
\end{align}
Here we have used the operator equivalences in Section \ref{sec:Equivalences}.
Now there $\mathcal{N}\left(n\right)\sim z^{n}$ such terms (see Appendix
\ref{sec:The-single-particle}) so we have the total contribution
for such terms scales as: 
\begin{equation}
\sim J_{R;R'}^{\mu;\nu}\cdot\frac{1}{z^{n/2}}\cdot z^{n}\sim1.\label{eq:Scaling}
\end{equation}
In which case we have that: 
\begin{equation}
J_{R;R'}^{\mu;\nu}\sim\frac{1}{z^{n/2}}.\label{eq:Scaling-1}
\end{equation}

\subsubsection{\label{sec:Scaling-Bosonic-operator}Scaling Bosonic operator that
changes fermion number by two}

Lets assume that the Hamiltonian contains $J_{R;R'}^{\mu;\nu}O_{R\mu}O_{R'\nu}$,
with $O_{R\mu}$ and $O_{R'\nu}$ bosonic and changing fermion number
by two, then we know that the lowest order contribution to $J_{R;R'}^{\mu;\nu}\left\langle \Psi\right|O_{R\mu}O_{R'\nu}\left|\Psi\right\rangle /\left\langle \Psi\mid\Psi\right\rangle $
may be written as: 
\begin{align}
 & J_{R;R'}^{\mu;\nu}\left\langle \Psi\right|O_{R\mu}O_{R'\nu}\left|\Psi\right\rangle /\left\langle \Psi\mid\Psi\right\rangle \nonumber \\
 & \sim J_{R;R'}^{\mu;\nu}\sum_{a,b,c,d=1}^{N}\bar{\mathcal{Z}}_{\mu ac}^{*}\bar{\mathcal{Z}}_{\nu bc}\left\langle \Psi_{0}\right|f_{Ra}^{\dagger}f_{R'b}\left|\Psi_{0}\right\rangle \times\nonumber \\
 & \times\left\langle \Psi_{0}\right|f_{Rc}^{\dagger}f_{R'd}\left|\Psi_{0}\right\rangle \nonumber \\
 & \sim J_{R;R'}^{\mu;\nu}\cdot\frac{1}{z^{n}}.\label{eq:Energy_num,ber_changing}
\end{align}
Now there $N\left(n\right)\sim z^{n}$ such terms, so we have the
total contribution for such terms scales as: 
\begin{equation}
\sim J_{R;R'}^{\mu;\nu}\cdot\frac{1}{z^{n}}\cdot z^{n}\sim1.\label{eq:Scaling-2}
\end{equation}
In which case we have that: 
\begin{equation}
J_{R;R'}^{\mu;\nu}\sim1.\label{eq:Scaling-1-1}
\end{equation}

\subsubsection{\label{sec:Scaling-Bosonic-operator-1}Scaling Bosonic operator that
does not change fermion number}

\paragraph{\label{subsec:Hartree-Term}Hartree Term}

Lets assume that the Hamiltonian contains $J_{R;R'}^{\mu;\nu}O_{R\mu}O_{R'\nu}$,
, with $O_{R\mu}$ and $O_{R'\nu}$ bosonic and fermion number conserving,
then we know that the lowest order, Hartree, term for $J_{R;R'}^{\mu;\nu}\left\langle \Psi\right|O_{R\mu}O_{R'\nu}\left|\Psi\right\rangle /\left\langle \Psi\mid\Psi\right\rangle $
may be written as: 
\begin{align}
 & J_{R;R'}^{\mu;\nu}\left\langle \Psi\right|O_{R\mu}O_{R'\nu}\left|\Psi\right\rangle /\left\langle \Psi\mid\Psi\right\rangle \nonumber \\
 & \sim J_{R;R'}^{\mu;\nu}\left\langle \Psi_{0}\right|P_{R}^{\dagger}O_{R\mu}P_{R}\left|\Psi_{0}\right\rangle \left\langle \Psi_{0}\right|P_{R'}^{\dagger}O_{R'\nu}P_{R'}\left|\Psi_{0}\right\rangle \nonumber \\
 & \sim J_{R;R'}^{\mu;\nu}\label{eq:energy_scaling_fermion}
\end{align}
Now there $\mathcal{N}\left(n\right)\sim z^{n}$ such terms so we
have the total contribution for such terms is: 
\begin{equation}
\sim J_{R;R'}^{\mu;\nu}\cdot z^{n}\sim1\label{eq:Scaling-2-1}
\end{equation}
In which case we have that: 
\begin{equation}
J_{R;R'}^{\mu;\nu}\sim\frac{1}{z^{n}}\label{eq:Scaling-1-1-1}
\end{equation}

\paragraph{\label{subsec:Fock-term}Fock term}

The scaling of the Fock term may be given by: 
\begin{align}
 & J_{R;R'}^{\mu;\nu}\left\langle \Psi\right|O_{R\mu}O_{R'\nu}\left|\Psi\right\rangle /\left\langle \Psi\mid\Psi\right\rangle \nonumber \\
 & \sim J_{R;R'}^{\mu;\nu}\sum_{a,b,c,d=1}^{N}\mathcal{Z}_{\mu ac}^{*}\mathcal{Z}_{\nu bd}\left\langle \Psi_{0}\right|f_{Ra}^{\dagger}f_{R'd}\left|\Psi_{0}\right\rangle \times\nonumber \\
 & \times\left\langle \Psi_{0}\right|f_{R'c}^{\dagger}f_{Ra}\left|\Psi_{0}\right\rangle \nonumber \\
 & \sim J_{R;R'}^{\mu;\nu}\cdot\frac{1}{z^{n}}\sim\frac{1}{z^{2n}}.\label{eq:Energy_fock}
\end{align}
Now there $\mathcal{N}\left(n\right)\sim z^{n}$ such terms so we
have the total contribution for such terms is: 
\begin{equation}
\sim J_{R;R'}^{\mu;\nu}\cdot\frac{1}{z^{n}}\cdot z^{n}\sim\frac{1}{z^{n}}\rightarrow0\label{eq:Scaling_fock}
\end{equation}
This means that the Fock term is highly suppressed, however we still
are able to keep its effects in the Gutzwiller Lagrange function in
Eq. (\ref{eq:Lagrangian}).

\subsubsection{\label{subsec:Discussion}Discussion}

We would like to note that these scaling arguments are very interesting
from a formal and theoretical point of view, but need to be significantly
modified for the case of realistic extended Hubbard model Hamiltonians.
For example, the coefficients in $V_{R_{1};R_{2}}^{\alpha\beta;\gamma\delta}$
which are proportional to a Coulomb mediated density density interaction
scale as $\sim1/\sqrt{n}$ which is not $z$ dependent, which is markedly
different then Eq. (\ref{eq:Scaling-1-1-1}). We note that the reason
for the square root in the previous formula is that for $z\gg n$
the typical Manhattan path makes $n-1$ turns in different directions.
Furthermore our scaling analysis shows that the order of magnitude
of the density density terms is the same as those for spin spin interactions,
all $V_{R_{1};R_{2}}^{\alpha\beta;\gamma\delta}$ scale the same way
with $z$. However because one term involves the overlap between electron
wavefunctions on different sites while the other doesn't; as such
the spin spin interaction is exponentially suppressed with distance
while the density density is not. This is true even though the exponent
has nothing to do with the co-ordination number $z$ but with the
exponentially decaying tails of the Wannier functions used to generate
the Hubbard model \citep{Haule_2010,Airchorn_2009,Amadon_2008,Anisimov_2005,Zhao_2012,Kunes_2011}.
Furthermore it is almost impossible to have the pair hopping matrix
elements $Y_{R_{1};R_{2}}^{\alpha\beta;\gamma\delta}$ to be distance
independent (see Eq. (\ref{eq:Scaling-1-1})) as these are exponentially
decaying with the decay length being given by the overlap of various
Wannier functions. As such, realistic Hamiltonians do not scale exactly
as in the limit that $z\rightarrow\infty$. More precisely, as the
various terms in $J_{R_{1};R_{2}}^{\mu;\nu}$, $J_{R_{1};R_{2};R_{3}}^{\mu;\nu;\eta}$
and $J_{R_{1};R_{2};R_{3};R_{4}}^{\mu;\nu;\eta;\rho}$ do not follow
the scaling needed for a rigorous $1/z$ expansion in realistic materials.
As such we will scale each different type of term separately and maintain
only leading order term or terms in the $1/z$ expansion for each
type of interaction. More explicitly the terms we keep in the equivalences
in Eqs. (\ref{eq:fermionic_equivalence-4}), (\ref{eq:Bosonic_conserving_solution_example}),
(\ref{eq:R_mu_I-1-1}) and (\ref{eq:R_matrix_pair_hopping-1-1}) are
leading order for each individual term in the Hamiltonian in Eq. (\ref{eq:Hamiltonian-1}).
We note that in Eqs. (\ref{eq:Bosonic_conserving_solution_example}),
(\ref{eq:R_mu_I-1-1}) we keep both the Hartree and the Fock terms,
despite the fact that the Fock terms are subleading. This is done
because the bosonic fermion number conserving terms in the Hamiltonian
in Eq. (\ref{eq:Hamiltonian-1}) are usually the biggest terms as
in many cases they do not involve overlaps of Wannier functions on
different sites and as such must be handled as carefully as possible.
Furthermore the Hartree term is handled exactly by our formalism (see
Appendix \ref{subsec:Bosonic-Operator-swith}) so no corrections to
the Hartree term effect the precision to which the Fock terms may
be handled.

\subsection{\label{sec:Leading-order-in}Leading order in $1/z$ terms for Gutzwiller
(failure of Eq. (\ref{eq:Equality_ish-1}) without additional assumptions
besides large co-ordination number)}

\subsubsection{\label{subsec:Manhattan-path-example}Manhattan path example}

\begin{figure}
\begin{centering}
\includegraphics[width=0.75\columnwidth]{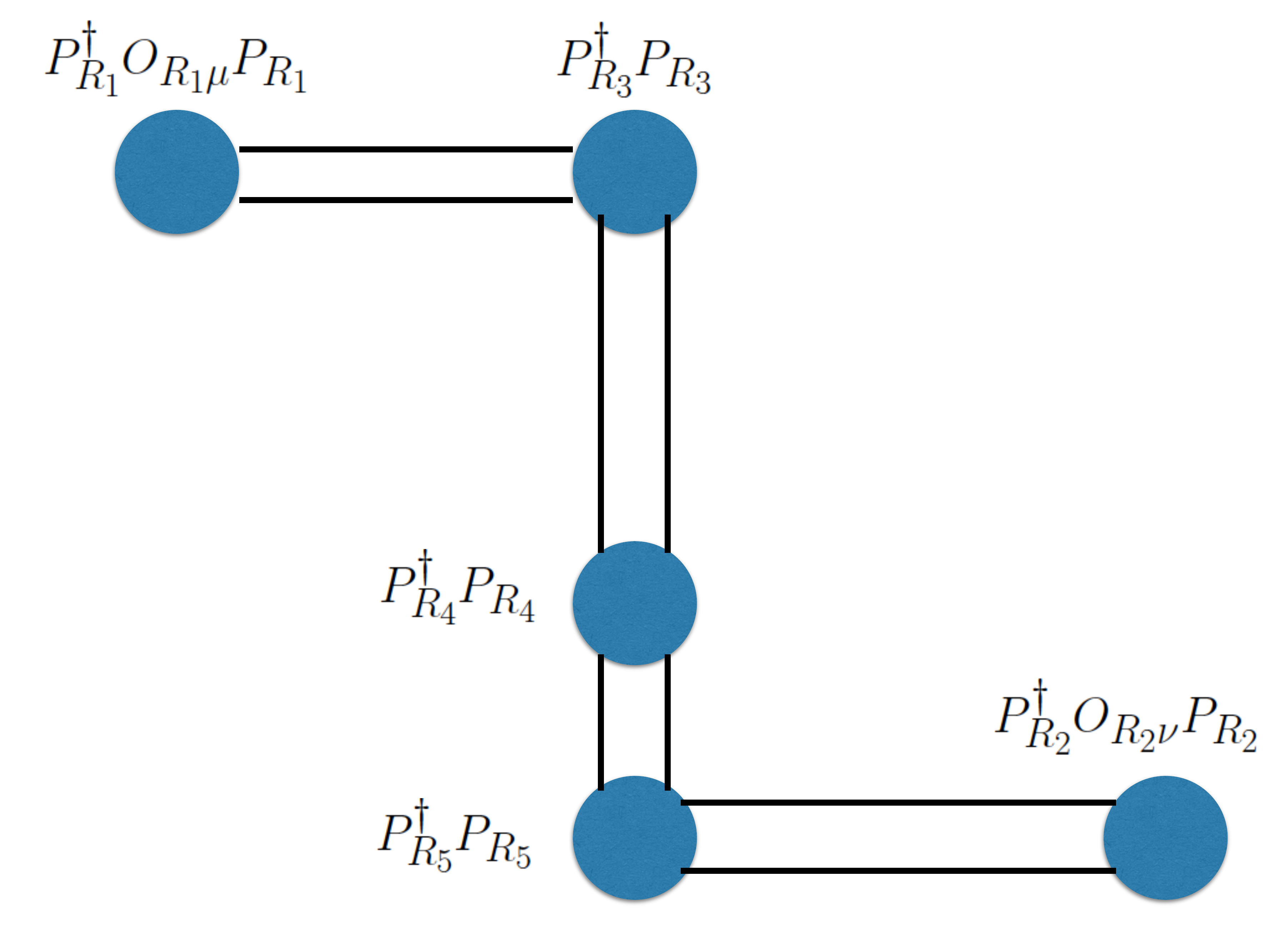} 
\par\end{centering}
\caption{\label{fig-Manhattan} Manhattan path showing order one term, not
considered on the right hand side of Eq. (\ref{eq:Equality_ish-1}).
Lines indicate Wick's theorem contractions.}
\end{figure}

In this section we will motivate the need for the $P_{R}^{\dagger}P_{R}-I$
expansion used in Section \ref{sec:Gutzwiller-Approximation-and}.
We show that already at leading order Eq. (\ref{eq:Equality_ish-1})
fails without additional assumptions, besides the Gutzwiller constraints
and the $1/z$ expansion, for the extended GA presented in the main
text. In particular, the usual arguments \citep{Sandri_2014,Sandri2014,Lanata_2016,Lanata_2015,Lanata2012,Lanata2009}
about the validity Eq. (\ref{eq:Equality_ish-1}) based on the Gutzwiller
constraints in Eq. (\ref{eq:Conditions-4-1}) and the large co-ordination
limit, which work for the regular Hubbard model, fails without additional
assumptions, such as the $P_{R}^{\dagger}P_{R}-I$ expansion. We now
we consider the pair hopping terms $-$ those that change the particle
number by two. We show that already at leading order for those terms
Eq. (\ref{eq:Equality_ish-1}) fails without additional assumptions,
as we need terms where we insert projectors of the form $P_{R_{i}}^{\dagger}P_{R_{i}}$
at various sites to obtain the correct expectation value for the left
hand side of Eq. (\ref{eq:Equality_ish-1}) to make the equality true
even at leading order. Indeed, if we take the pair hopping operator
$J_{R_{1};R_{2}}^{\mu;\nu}O_{R_{1}\mu}O_{R_{2}\nu}$ and consider
any Manhattan path between $R_{1}$ and $R_{2}$: $P_{R_{1}R_{2}}$.
Then for every Manhattan path $P_{R_{1}R_{2}}$ we may insert an arbitrary
number of operators $P_{R_{i}}^{\dagger}P_{R_{i}}$ at any set of
sites (with at most one insertion per site) along this path with and
obtain an operator: 
\begin{equation}
J_{R_{1}R_{2}}^{\mu;\nu}[P_{R_{1}}^{\dagger}O_{R_{1}\mu}P_{R_{1}}][\prod_{i}P_{R_{i}}^{\dagger}P_{R_{i}}][P_{R_{2}}^{\dagger}O_{R_{2}\nu}P_{R_{2}}].\label{eq:Operator}
\end{equation}
We now contract the operators along the Manhattan path with nearest
neighbors being contracted with each other (two contractions per neighbor)
and obtain a term on the left hand side of Eq. (\ref{eq:Equality_ish-1})
which is of the same order of magnitude as the terms on the right
hand side in Eq. (\ref{eq:Equality_ish-1}), but not included on the
right hand side (as such the equation fails). This is pictured in
Fig. (\ref{fig-Manhattan}). We see that Eq. (\ref{eq:Equality_ish-1})
fails even in the large coordination limit without additional assumptions.
In Section \ref{sec:Gutzwiller-Approximation-and} we make such an
assumption: the $P_{R}^{\dagger}P_{R}-I$ expansion.

\subsubsection{\label{subsec:Tadpole-example}Tadpole example}

\begin{figure}
\begin{centering}
\includegraphics[width=0.75\columnwidth]{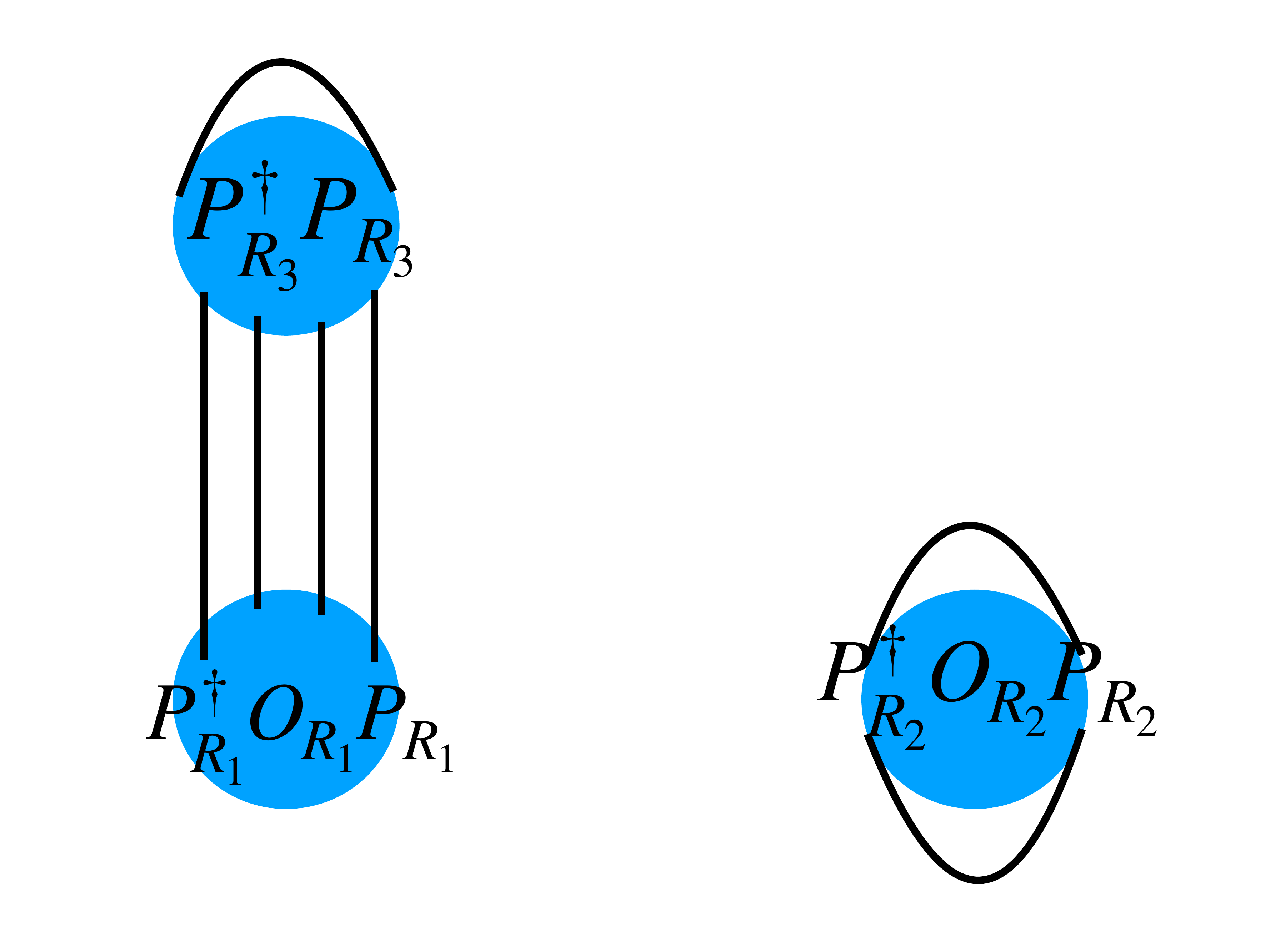} 
\par\end{centering}
\caption{\label{fig:tadpole} One site tadpole contraction (involving $P_{R}^{\dagger}P_{R}-I$
terms) that has the same scaling with $1/z$ as the Fock terms.}
\end{figure}

We notice that the example presented in Section \ref{subsec:Manhattan-path-example}
does not apply for nearest neighbor sites. Here we will show that,
using just the $1/z$ expansion, the Fock terms in Eq. ((\ref{eq:Eff_conserving}))
have the same scaling in $1/z$ as additional multisite terms not
considered on the right hand side in Eq. (\ref{eq:Equality_ish-1})
but needed for the left hand side, making an additional assumption,
such as the $P_{R}^{\dagger}P_{R}-I$ expansion, necessary for Eq.
(\ref{eq:Equality_ish-1}) to be true for them. We now assume that
$P_{R_{1}}^{\dagger}O_{R_{1}}P_{R_{1}}$ and $P_{R_{2}}^{\dagger}O_{R_{2}}P_{R_{2}}$
are operators on nearest neighbor sites which conserve fermion number.
We again consider Eq. (\ref{eq:Equality_ish-1}) for these operators.
We now consider site $R_{3}$ which is a nearest neighbor of site
$R_{1}$ where there are four contractions between $P_{R_{1}}^{\dagger}O_{R_{1}}P_{R_{1}}$
and $P_{R_{3}}^{\dagger}P_{R_{3}}$ on the right hand side of Eq.
(\ref{eq:Equality_ish-1}). This term scales as $\sim1/z^{2}$ and
there are $\sim z$ such terms which means that that the total scaling
of such terms is $\sim z\times1/z^{2}\sim1/z$ which means that it
has the same scaling as any Fock term on the right hand side of Eq.
(\ref{eq:Equality_ish-1}). This is true for both for bosonic fermion
number conserving operators and bosonic fermion number changing operators
(in the case of superconductivity, not considered in this work). As
such an additional assumption such as the $P_{R}^{\dagger}P_{R}-I$
expansion is mandatory for those terms even for nearest neighbors.
This is illustrated in Fig. (\ref{fig:tadpole}).

\section{\label{sec:Towards-ab-intio}Towards ab initio research}

In order to perform true ab initio research one needs to deal with
all of the terms in Eq. (\ref{eq:Hamiltonian-1}), not just the one
and two site terms as in Eq. (\ref{eq:Equality-1}) studied in the
main text. In this Appendix we outline a method to do just that.

\begin{figure}
\begin{centering}
\includegraphics[width=0.75\columnwidth]{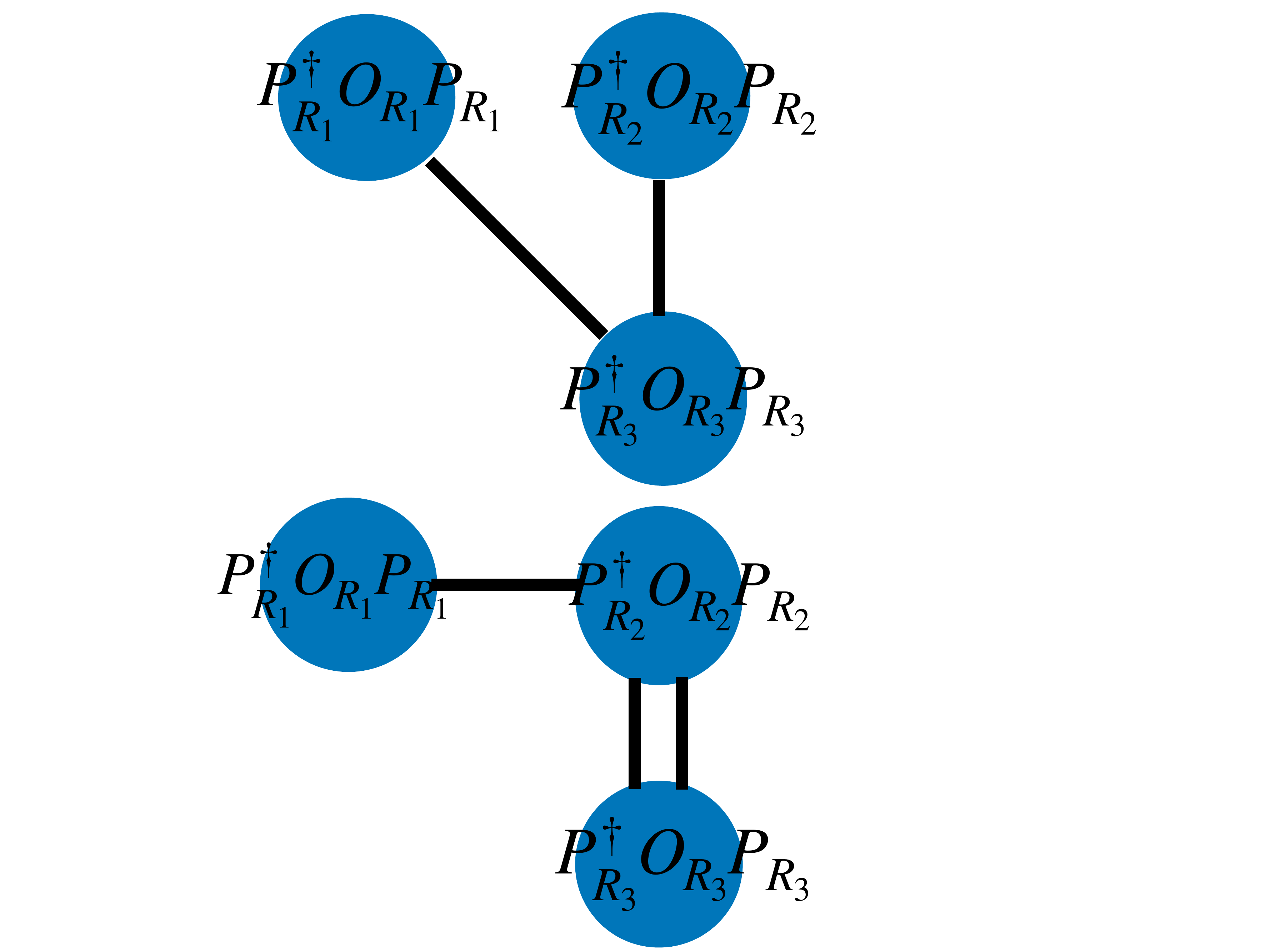} 
\par\end{centering}
\caption{\label{fig-Y-shape} Two diagrams for three sites as described in
the main text with a different number of contractions but the same
$1/z$ scaling. Lines indicate Wick's theorem contractions.}
\end{figure}

\subsection{\label{sec:Main-construction}The need for three Wick's contractions
for calculating expectation values with three and four point operators}

As a first step towards handling three and four site terms, we will
describe why the treatment presented in the main text fails for them.

\subsubsection{\label{sec:Other-geometries}Three site terms}

\paragraph{\label{par:Fock-contribution}Fock contribution}

We would like to show an example of the need for three fermion contractions
to evaluate the expectation value of the Hamiltonian in Eq. (\ref{eq:Hamiltonian-1})
for three point operators, even in leading order in the $1/z$ expansion
and the $P_{R}^{\dagger}P_{R}-I$ expansion. Consider two single fermion
operators and one bosonic operator on three neighbor sites such that
$R_{2}$ (fermionic site) is the nearest neighbor of both $R_{1}$
(fermionic site) and $R_{3}$ (bosonic site). We will consider the
terms $\sum_{R_{1}\neq R_{2}\neq R_{3}}\mathcal{V}_{R_{3};R_{2};R_{1}}^{\alpha\beta;\gamma;\delta}[c_{R_{3}\alpha}^{\dagger}c_{R_{3}\beta}][c_{R_{2}\gamma}^{\dagger}][c_{R_{1}\delta}]$
in Eq. (\ref{eq:Hamiltonian-1}) (one can check that the terms of
the form $\left[\sum_{R_{1}\neq R_{2}\neq R_{3}}\mathcal{Y}_{R_{3};R_{2};R_{1}}^{\alpha\beta;\gamma;\delta}[c_{R_{3}\alpha}^{\dagger}c_{R_{3}\beta}^{\dagger}][c_{R_{2}\gamma}][c_{R_{1}\delta}]+h.c.\right]$
have a similar scaling problem). We now consider the Fock terms (those
with atleast one contraction between each site). We now consider possible
contractions relevant to this scenario. Both diagrams shown in figure
\ref{fig-Y-shape} contribute to order $\sim\frac{1}{z^{3/2}}$ (which
is leading order for the Fock term). Here we are ignoring the scaling
for the prefactor for this diagram as well as the number of similar
diagrams (see Appendix \ref{sec:Scaling-of-operator}) which is the
same for both sets of contractions. However the first diagram has
three contractions on the central site and therefore is higher order
then we considered until now (as such we need the terms described
in Section \ref{subsec:Alternate-approach} below).

\paragraph{\label{par:Hartree-terms}Hartree terms }

One can check that the Hartree contribution to the terms $\sum_{R_{1}\neq R_{2}\neq R_{3}}\mathcal{V}_{R_{3};R_{2};R_{1}}^{\alpha\beta;\gamma;\delta}[c_{R_{3}\alpha}^{\dagger}c_{R_{3}\beta}][c_{R_{2}\gamma}^{\dagger}][c_{R_{1}\delta}]$
in Eq. (\ref{eq:Hamiltonian-1}) (the one with no contractions between
the bosonic site and the two fermionic sites) does not have an inconsistency
in scaling (there are no terms of the same magnitude). As such it
is possible to consistently keep only the Hartree contribution for
these terms. Again, one cannot keep the Fock contribution for the
terms $\left[\sum_{R_{1}\neq R_{2}\neq R_{3}}\mathcal{Y}_{R_{3};R_{2};R_{1}}^{\alpha\beta;\gamma;\delta}[c_{R_{3}\alpha}^{\dagger}c_{R_{3}\beta}^{\dagger}][c_{R_{2}\gamma}][c_{R_{3}\delta}]+h.c.\right]$
and there is no Hartree piece. To see this clearly, note that the
Hartree contribution for the example in Fig. \ref{fig-Y-shape} scales
as $\sim\frac{1}{\sqrt{z}}$ while the Fock and the three contractions
diagram scale as $\frac{1}{z^{3/2}}$ meaning that the Hartree is
dominant. To keep this term, one needs only modify the Lagrange function
in Section \ref{sec:Gutzwiller-Lagrangian} by changing $\mathcal{L}_{HF}$
to: 
\begin{widetext}
\begin{align}
 & \mathcal{L}_{HF}\left(\left\{ \mathcal{R}_{R},\mathcal{T}_{R},\mathcal{U}_{R}\right\} ,\left\{ \mathcal{R}_{R}^{*},\mathcal{T}_{R}^{\ast},\mathcal{U}_{R}^{\ast}\right\} ,\left[\Delta_{R_{1}R_{2}}\right]_{ab},o_{R\alpha\beta}\right)=\nonumber \\
 & =\sum_{R_{1}R_{2}\alpha\beta\gamma\delta}V_{R_{1};R_{2}}^{\alpha\beta;\gamma\delta}o_{R\alpha\beta}o_{R\gamma\delta}-\sum_{R_{1}\neq R_{2}}\sum_{\alpha\beta\gamma\delta}\sum_{abcd}V_{R_{1};R_{2}}^{\alpha\beta;\gamma\delta}\left[\Delta_{R_{1}R_{2}}\right]_{ac}\left[\Delta_{R_{2}R_{1}}\right]_{db}\left[\mathcal{T}_{R_{1}\alpha}^{\beta}\right]_{ba}\left[\mathcal{T}_{R_{2}\gamma}^{\delta}\right]_{dc}+\nonumber \\
 & +\sum_{R_{1}\neq R_{2}}\sum_{\alpha>\beta\gamma>\delta}\left[Y_{R_{1};R_{2}}^{\alpha\beta;\gamma\delta}\sum_{a>b}\sum_{c>d}\left(\left[\Delta_{R_{1}R_{2}}\right]_{ad}\left[\Delta_{R_{1}R_{2}}\right]_{bc}-\left[\Delta_{R_{1}R_{2}}\right]_{ac}\left[\Delta_{R_{1}R_{2}}\right]_{bd}\right)\left[\mathcal{U}_{R\alpha\beta}\right]_{ab}\left[\mathcal{U}_{R\gamma\delta}^{*}\right]_{cd}\right]+\nonumber \\
 & +\sum_{R_{1}\neq R_{2}\neq R_{3}}\sum_{\alpha\beta\gamma\delta}\sum_{ab}\mathcal{V}_{R_{1};R_{2},R_{3}}^{\alpha\beta;\gamma;\delta}\left[\Delta_{R_{2}R_{3}}\right]_{ab}\left[\mathcal{R}_{R_{2}\gamma}\right]_{a}\left[\mathcal{R}_{R_{3}\delta}^{*}\right]_{b}o_{R_{1}\alpha\beta}.\label{eq:Hartree-Fock_three_particle}
\end{align}
As compared to Eq. (\ref{eq:L_HF}). Below we will see the problem
gets worse for four site terms as there are no Hartree terms and there
is no way to keep the Fock piece consistently. 
\end{widetext}

\subsubsection{\label{subsec:Four-site-terms}Four site terms}

\begin{figure}
\begin{centering}
\includegraphics[width=0.75\columnwidth]{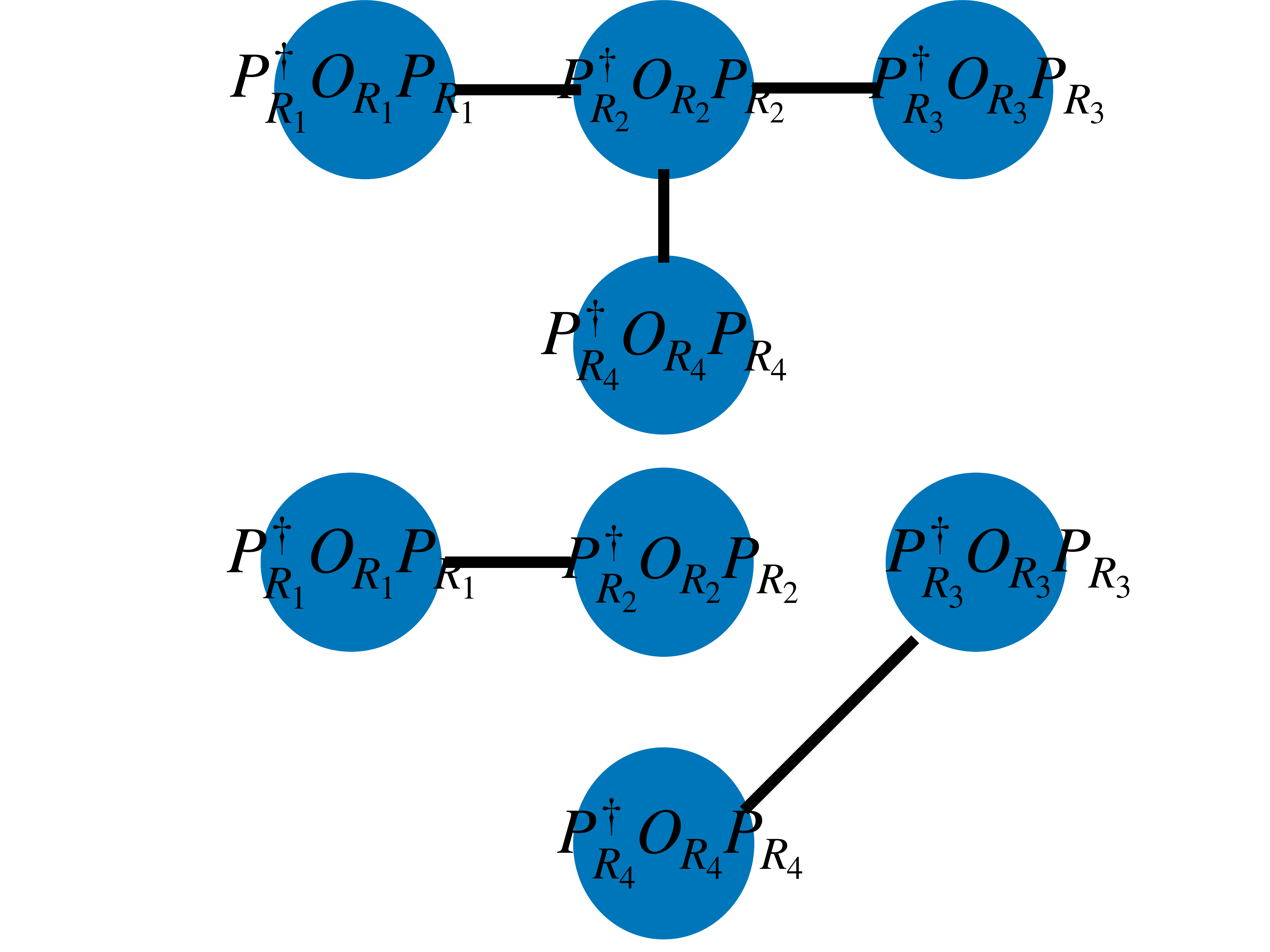} 
\par\end{centering}
\caption{\label{fig-T-shape} Two diagrams for four sites as described in the
main text with a different number of contractions but the same $1/z$
scaling. Lines indicate Wick's theorem contractions.}
\end{figure}

We would like to present an example where, despite making both the
large coordination number assumption and the leading order in $P_{R}^{\dagger}P_{R}-I$
expansion assumption, one needs three fermion contractions to obtain
the leading order term for four site operators. Consider four single
fermion operators on four neighbor sites such that $R_{2}$ is nearest
neighbor to $R_{1},\,R_{3}$ and $R_{4}$ (which are all next nearest
neighbors to each other). We now consider the two possible sets of
contractions shown in Figure \ref{fig-T-shape}, both sets of contractions
contribute to order $\sim\frac{1}{z^{3/2}}$ (which is leading order
for this diagram). Here we are ignoring the scaling for the prefactor
for this diagram as well as the number of similar diagrams, so that
the total scaling for both terms can be made order $\sim1$ (see Appendix
\ref{sec:Scaling-of-operator}). However the first diagram has three
contractions on the central site and therefore is higher order then
we considered so far. As such we need to add three contraction terms
to the site $R_{2}$ as is described in Section \ref{subsec:Alternate-approach}.

\subsection{\label{subsec:Ad-Hoc-approach}Approximate approach}

The procedure, to properly account for the terms considered in Appendix
\ref{sec:Main-construction}, described in Appendix \ref{subsec:Alternate-approach}
below, is rather cumbersome. In this Appendix in order to perform
crude ab-initio research we present a simplified method to treat multisite
interactions to deal with all terms in Eq. (\ref{eq:Hamiltonian-1}).
To do this very quickly but crudely one needs an effective Hamiltonian
much like Eq. (\ref{eq:Generic_hamiltonian-1}) but for the generalized
model given by the Hamiltonian in Eq. (\ref{eq:Hamiltonian-1}). One
natural guess for such an effective Hamiltonian is that: 
\begin{equation}
\left\langle \Psi_{0}\right|H_{P}\left|\Psi_{0}\right\rangle \cong\left\langle \Psi_{0}\right|H_{Eff}\left|\Psi_{0}\right\rangle .\label{eq:Effective_generalized}
\end{equation}

\begin{widetext}
Where $H_{Eff}$ is given by: 
\begin{align}
H_{Eff} & =\sum_{R}P_{R}^{\dagger}H_{R}^{loc}P_{R}+\sum_{R_{1}\neq R_{2}}\sum_{\mu\nu}J_{R_{1};R_{2}}^{\mu;\nu}\left[\sum_{i}\mathcal{Z}_{R_{1}\mu i}O_{Ri}\right]\left[\sum_{j}\mathcal{Z}_{R_{2}\nu j}O_{R'j}\right]+\nonumber \\
 & +\sum_{R_{1}\neq R_{2}\neq R_{3}}\sum_{\mu\nu\eta}J_{R_{1};R_{2};R_{3}}^{\mu;\nu;\eta}\left[\sum_{i}\mathcal{Z}_{R_{1}\mu i}O_{Ri}\right]\left[\sum_{j}\mathcal{Z}_{R_{2}\nu j}O_{R'j}\right]\left[\sum_{k}\mathcal{Z}_{R_{3}\mu k}O_{R_{3}k}\right]+\nonumber \\
 & +\sum_{R_{1}\neq R_{2}\neq R_{3}\neq R_{4}}\sum_{\mu\nu\eta\rho}J_{R_{1};R_{2};R_{3};R_{4}}^{\mu;\nu;\eta;\rho}\left[\sum_{i}\mathcal{Z}_{R_{1}\mu i}O_{R_{1}i}\right]\left[\sum_{j}\mathcal{Z}_{R_{2}\nu j}O_{R_{2}j}\right]\left[\sum_{k}\mathcal{Z}_{R_{3}\mu k}O_{R_{3}k}\right]\left[\sum_{l}\mathcal{Z}_{R_{4}\rho l}O_{R_{4}l}\right]\label{eq:Generic_hamiltonian-1-1}
\end{align}
We again note that as explained in Appendices \ref{sec:Other-geometries}
and \ref{subsec:Four-site-terms} this is not completely satisfactory,
even for the leading order in the $P_{R}^{\dagger}P_{R}-I$ expansion
and the leading order in the $1/z$ expansion. There are terms of
the same order of magnitude in the $1/z$ expansion as the terms in
the Hamiltonian in Eq. (\ref{eq:Generic_hamiltonian-1-1}) which are
missed by the Hamiltonian in Eq. (\ref{eq:Generic_hamiltonian-1-1}).
However using the effective Hamiltonian in Eq. (\ref{eq:Generic_hamiltonian-1-1})
is a very simple approach and may be good enough for preliminary calculations.
If this approximate approach is chosen, then the Lagrange function
in Section \ref{sec:Gutzwiller-Lagrangian} need only be modified
by changing $\mathcal{L}_{HF}$, which now should be given by: 
\begin{align}
 & \mathcal{L}_{HF}\left(\left\{ \mathcal{R}_{R},\mathcal{T}_{R},\mathcal{U}_{R}\right\} ,\left\{ \mathcal{R}_{R}^{*},\mathcal{T}_{R}^{\ast},\mathcal{U}_{R}^{\ast}\right\} ,\left[\Delta_{R_{1}R_{2}}\right]_{ab},o_{R\alpha\beta}\right)=\nonumber \\
 & =\sum_{R_{1}R_{2}\alpha\beta\gamma\delta}V_{R_{1};R_{2}}^{\alpha\beta;\gamma\delta}o_{R\alpha\beta}o_{R\gamma\delta}-\sum_{R_{1}\neq R_{2}}\sum_{\alpha\beta\gamma\delta}\sum_{abcd}V_{R_{1};R_{2}}^{\alpha\beta;\gamma\delta}\left[\Delta_{R_{1}R_{2}}\right]_{ac}\left[\Delta_{R_{2}R_{1}}\right]_{db}\left[\mathcal{T}_{R_{1}\alpha}^{\beta}\right]_{ba}\left[\mathcal{T}_{R_{2}\gamma}^{\delta}\right]_{dc}+\nonumber \\
 & +\sum_{R_{1}\neq R_{2}}\sum_{\alpha>\beta\gamma>\delta}\left[Y_{R_{1};R_{2}}^{\alpha\beta;\gamma\delta}\sum_{a>b}\sum_{c>d}\left(\left[\Delta_{R_{1}R_{2}}\right]_{ad}\left[\Delta_{R_{1}R_{2}}\right]_{bc}-\left[\Delta_{R_{1}R_{2}}\right]_{ac}\left[\Delta_{R_{1}R_{2}}\right]_{bd}\right)\left[\mathcal{U}_{R\alpha\beta}\right]_{ab}\left[\mathcal{U}_{R\gamma\delta}^{*}\right]_{cd}\right]+\nonumber \\
 & +\sum_{R_{1}\neq R_{2}\neq R_{3}}\sum_{\alpha\beta\gamma\delta}\sum_{ab}\mathcal{V}_{R_{1};R_{2},R_{3}}^{\alpha\beta;\gamma;\delta}\left[\Delta_{R_{2}R_{3}}\right]_{ab}\left[\mathcal{R}_{R_{2}\gamma}\right]_{a}\left[\mathcal{R}_{R_{3}\delta}^{*}\right]_{b}o_{R_{1}\alpha\beta}-\nonumber \\
 & -\sum_{R_{1}\neq R_{2}\neq R_{3}}\sum_{\alpha\beta\gamma\delta}\sum_{abcd}\mathcal{V}_{R;R';R"}^{\alpha\beta;\gamma;\delta}\left[\Delta_{R_{2}R_{1}}\right]_{cb}\left[\Delta_{R_{1}R_{3}}\right]_{ad}\left[\mathcal{R}_{R\gamma}\right]_{c}\left[\mathcal{R}_{R\delta}^{*}\right]_{d}\left[\mathcal{T}_{R\alpha}^{\beta}\right]_{ba}+\nonumber \\
 & +\sum_{R_{1}\neq R_{2}\neq R_{3}}\sum_{\alpha>\beta\gamma\delta}\sum_{abcd}\left[\mathcal{Y}_{R;R';R"}^{\alpha\beta;\gamma;\delta}\left(\left[\Delta_{R_{1}R_{3}}\right]_{ad}\left[\Delta_{R_{1}R_{2}}\right]_{bc}-\left[\Delta_{R_{1}R_{2}}\right]_{ac}\left[\Delta_{R_{1}R_{3}}\right]_{bd}\right)\left[\mathcal{R}_{R_{2}\gamma}\right]_{c}\left[\mathcal{R}_{R_{3}\delta}^{*}\right]_{d}\left[\mathcal{U}_{R\alpha\beta}\right]_{ab}+c.c.\right]+\nonumber \\
 & +\sum_{R_{1}\neq R_{2}\neq R_{3}\neq R_{4}}\sum_{\alpha\beta\gamma\delta}\sum_{abcd}S{}_{R_{1};R_{2};R_{3};R_{4}}^{\alpha;\beta;\gamma;\delta}\left(\left[\Delta_{R_{1}R_{4}}\right]_{ad}\left[\Delta_{R_{2}R_{3}}\right]_{bc}-\left[\Delta_{R_{1}R_{3}}\right]_{ac}\left[\Delta_{R_{2}R_{4}}\right]_{bd}\right)\left[\mathcal{R}_{R_{3}\gamma}^{\ast}\right]_{c}\left[\mathcal{R}_{R_{4}\delta}^{*}\right]_{d}\left[\mathcal{R}_{R_{1}\alpha}\right]_{a}\left[\mathcal{R}_{R_{2}\beta}\right]_{b}.\label{eq:Hartree_Fock_Lagrangian}
\end{align}
As compared to Eq. (\ref{eq:L_HF}). With this modification the Lagrange
function can be minimized directly. 
\end{widetext}

\subsection{\label{subsec:Alternate-approach}Extended approach}

A rigorous approach to the problem of three contractions outlined
in Section \ref{sec:Main-construction} is to replace some of the
terms of the form $P_{R}^{\dagger}O_{R}P_{R}$ on various sites sites
in our Hamiltonian (the precise sites to be chosen are explained in
Section \ref{sec:Main-construction}) with three and four fermion
terms instead of the one and two fermion terms found in the equivalences
in Eqs. (\ref{eq:fermionic_equivalence-4}), (\ref{eq:Indecies-1}),
(\ref{eq:R_mu_I-1}) and (\ref{eq:R_matrix_pair_hopping-1}). This
allows one to reproduce the three and four contraction terms shown
in Section \ref{sec:Main-construction}. This approach is needed only
for some of the three and four site terms in Eq. (\ref{eq:Hamiltonian-1})
and the two site analysis done previously is unaffected. One then
uses the generalized equivalences which will be given below for these
sites and proceeds much like in the main text. We now describe the
more general equivalences needed for these three and four fermion
equivalences.

\subsubsection{\label{sec:Fermionic-Equivalences-1}Fermionic Equivalences}

We consider fermionic single site operators $O_{R\mu}$. In Appendix
\ref{subsec:Fermionic-equivalences} we will show that there is an
equivalence : 
\begin{equation}
P_{R}^{\dagger}O_{R\mu}P_{R}\sim\sum_{a}\tilde{\mathcal{Z}}_{R\mu a}f_{Ra}^{\dagger}+\sum_{a}\tilde{\mathcal{Z}}_{R\mu ab}^{c}f_{Ra}^{\dagger}f_{Rb}^{\dagger}f_{Rc}\label{eq:fermionic_equivalence-4-1}
\end{equation}
and similarly for its Hermitian conjugate. In Appendix \ref{subsec:Fermionic-equivalences}
we show that the coefficients $\tilde{\mathcal{Z}}_{R\mu ab}^{c},\tilde{\mathcal{Z}}_{R\mu a}$
are determined by the following equations: 
\begin{align}
 & \left\langle \Psi_{0}\right|P_{R}^{\dagger}O_{R\mu}P_{R}f_{Rd}\left|\Psi_{0}\right\rangle \nonumber \\
 & =\left\langle \Psi_{0}\right|\left[\sum_{b}\tilde{\mathcal{Z}}_{R\mu b}f_{Rb}^{\dagger}+\sum_{a}\tilde{\mathcal{Z}}_{R\mu ab}^{c}f_{Ra}^{\dagger}f_{Rb}^{\dagger}f_{Rc}\right]f_{Rd}\left|\Psi_{0}\right\rangle ;\nonumber \\
 & \left\langle \Psi_{0}\right|P_{R}^{\dagger}O_{R\mu}P_{R}f_{Rd}^{\dagger}f_{Re}f_{Rf}\left|\Psi_{0}\right\rangle =\nonumber \\
 & \left\langle \Psi_{0}\right|\left[\sum_{b}\tilde{\mathcal{Z}}_{R\mu b}f_{Rb}^{\dagger}+\sum_{a}\tilde{\mathcal{Z}}_{R\mu ab}^{c}f_{Ra}^{\dagger}f_{Rb}^{\dagger}f_{Rc}\right]\cdot\nonumber \\
 & \cdot f_{Rd}^{\dagger}f_{Re}f_{Rf}\left|\Psi_{0}\right\rangle \label{eq:Equations_fermionic}
\end{align}

\subsubsection{\label{sec:Bosonic-Equivalences-(fermion-2}Bosonic Equivalences
(fermion number conserving operators)}

We consider bosonic operators that conserve fermion number. In Appendix
\ref{subsec:Bosonic_equivalences} we show the following equivalence:
\begin{align}
P_{R}^{\dagger}O_{R\mu}P_{R} & \sim\sum_{abcd}\tilde{\mathcal{Z}}_{R\mu ab}^{cd}f_{Ra}^{\dagger}f_{Rb}^{\dagger}f_{Rc}f_{Rd}+\nonumber \\
 & +\sum_{ab}\tilde{\mathcal{Z}}_{R\mu a}^{b}f_{Ra}^{\dagger}f_{Rb}+\tilde{\mathcal{Z}}_{R\mu I}I\label{eq:Equivalence_bosonic_number_conserving}
\end{align}
We show that $\tilde{\mathcal{Z}}_{R\mu ab}^{cd},\tilde{\mathcal{Z}}_{R\mu c}^{d},\tilde{\mathcal{Z}}_{R\mu I}$
are determined by the following equations (see Appendix \ref{subsec:Bosonic_equivalences}):
\begin{align}
 & \left\langle \Psi_{0}\right|P_{R}^{\dagger}O_{R\mu}P_{R}f_{Re}^{\dagger}f_{Rf}^{\dagger}f_{Rg}f_{Rh}\left|\Psi_{0}\right\rangle =\nonumber \\
 & =\left\langle \Psi_{0}\right|\left[\sum_{abcd}\tilde{\mathcal{Z}}_{R\mu ab}^{cd}f_{Ra}^{\dagger}f_{Rb}^{\dagger}f_{Rc}f_{Rd}+\right.\nonumber \\
 & \left.\sum_{ab}\tilde{\mathcal{Z}}_{R\mu a}^{b}f_{Ra}^{\dagger}f_{Rb}+\tilde{\mathcal{Z}}_{R\mu I}I\right]f_{Re}^{\dagger}f_{Rf}^{\dagger}f_{Rg}f_{Rh}\left|\Psi_{0}\right\rangle ;\nonumber \\
 & =\left\langle \Psi_{0}\right|P_{R}^{\dagger}O_{R\mu}P_{R}f_{Re}^{\dagger}f_{Rf}\left|\Psi_{0}\right\rangle =\nonumber \\
 & \left\langle \Psi_{0}\right|\left[\sum_{abcd}\tilde{\mathcal{Z}}_{R\mu ab}^{cd}f_{Ra}^{\dagger}f_{Rb}^{\dagger}f_{Rc}f_{Rd}\right.\nonumber \\
 & \left.+\sum_{ab}\tilde{\mathcal{Z}}_{R\mu a}^{b}f_{Ra}^{\dagger}f_{Rb}+\tilde{\mathcal{Z}}_{R\mu I}I\right]\cdot f_{Re}^{\dagger}f_{Rf}\left|\Psi_{0}\right\rangle ;\nonumber \\
 & \left\langle \Psi_{0}\right|P_{R}^{\dagger}O_{R\mu}P_{R}\cdot I\left|\Psi_{0}\right\rangle =\nonumber \\
 & =\left\langle \Psi_{0}\right|\left[\sum_{abcd}\tilde{\mathcal{Z}}_{R\mu ab}^{cd}f_{Ra}^{\dagger}f_{Rb}^{\dagger}f_{Rc}f_{Rd}+\right.\nonumber \\
 & \left.\sum_{ab}\tilde{\mathcal{Z}}_{R\mu a}^{b}f_{Ra}^{\dagger}f_{Rb}+\tilde{\mathcal{Z}}_{R\mu I}I\right]\cdot I\left|\Psi_{0}\right\rangle \label{eq:Bosonic_equations_number_conserving}
\end{align}

\subsubsection{\label{sec:Bosonic-Equivalences-(fermion-1-1}Bosonic Equivalences
(fermion number changing operators)}

We now consider bosonic operators that change the total number of
electrons by two, such as pair hopping terms. In Appendix \ref{subsec:Bosonic_equivalences-1}
we show that: 
\begin{equation}
P_{R}^{\dagger}O_{R\mu}P_{R}\sim\sum_{ab}\tilde{\mathcal{Z}}_{R\mu ab}f_{Ra}^{\dagger}f_{Rb}^{\dagger}+\sum_{ab}\tilde{\mathcal{Z}}_{R\mu abc}^{d}f_{Ra}^{\dagger}f_{Rb}^{\dagger}f_{Rc}^{\dagger}f_{Rd}\label{eq:Bosonic_equivalence_pair_hopping-2}
\end{equation}
and similarly for its complex conjugate. We show that $\tilde{\mathcal{Z}}_{R\mu abc}^{d}$
and $\tilde{\mathcal{Z}}_{R\mu ab}$ are determined by the following
equations (see Appendix \ref{subsec:Bosonic_equivalences}): 
\begin{align}
 & \left\langle \Psi_{0}\right|P_{R}^{\dagger}O_{R\mu}P_{R}f_{Re}^{\dagger}f_{Rf}f_{Rg}f_{Rh}\left|\Psi_{0}\right\rangle =\nonumber \\
 & =\left\langle \Psi_{0}\right|\left[\sum_{ab}\tilde{\mathcal{Z}}_{R\mu ab}f_{Ra}^{\dagger}f_{Rb}^{\dagger}+\sum_{ab}\tilde{\mathcal{Z}}_{R\mu ab}f_{Ra}^{\dagger}f_{Rb}^{\dagger}f_{Rc}^{\dagger}f_{Rd}\right]\cdot\nonumber \\
 & \cdot f_{Re}^{\dagger}f_{Rf}f_{Rg}f_{Rh}\left|\Psi_{0}\right\rangle ;\nonumber \\
 & \left\langle \Psi_{0}\right|P_{R}^{\dagger}O_{R\mu}P_{R}f_{Re}f_{Rf}\left|\Psi_{0}\right\rangle =\nonumber \\
 & =\left\langle \Psi_{0}\right|\left[\sum_{ab}\tilde{\mathcal{Z}}_{R\mu ab}f_{Ra}^{\dagger}f_{Rb}^{\dagger}+\sum_{ab}\tilde{\mathcal{Z}}_{R\mu ab}f_{Ra}^{\dagger}f_{Rb}^{\dagger}f_{Rc}^{\dagger}f_{Rd}\right]\cdot\nonumber \\
 & \cdot f_{Re}f_{Rf}\left|\Psi_{0}\right\rangle \label{eq:Bosonic_equations_number_changing}
\end{align}

\section{\label{sec:Derivations-of-Eqs.-1}Derivations of equivalence relations}

\subsection{\label{sec:Some-useful-functions}Useful identities}

In this section we drop the index $R$ as only one site is considered
and $\Delta=\Delta_{R,R}$. Below to derive the Lagrangian in Section
\ref{sec:Derivations-of-Eqs.} we will need the functions:

\begin{align}
Tr\left\{ \rho_{0}O_{i}O_{j}\right\}  & =F_{i;j}\left(\Delta\right)\nonumber \\
\frac{1}{\sqrt{\rho_{0}}}O_{i}\sqrt{\rho_{0}} & =\sum_{k}\mathcal{G}_{i;k}\left(\Delta\right)O_{k}\label{eq:Functions}
\end{align}
For $O_{i}\subset I,f_{a},f_{a}^{\dagger},f_{a}^{\dagger}f_{b},f_{a}f_{b},f_{a}^{\dagger}f_{b}^{\dagger}$.
Indeed we have by Wick's theorem: 
\begin{align}
Tr\left\{ \rho_{0}f_{a}^{\dagger}f_{b}\right\} = & \Delta_{ab}\nonumber \\
Tr\left\{ \rho_{0}f_{a}f_{b}^{\dagger}\right\} = & \left(\mathbb{I}-\Delta\right)_{ba}\nonumber \\
Tr\left\{ \rho_{0}f_{a}^{\dagger}f_{b}f_{c}^{\dagger}f_{d}\right\} = & \Delta_{ab}\Delta_{cd}+\Delta_{ac}\left(\mathbb{I}-\Delta\right)_{db}\nonumber \\
Tr\left\{ \rho_{0}f_{a}^{\dagger}f_{b}^{\dagger}f_{c}f_{d}\right\} = & -\Delta_{ac}\Delta_{bd}+\Delta_{ad}\Delta_{bc}\nonumber \\
Tr\left\{ \rho_{0}f_{a}f_{b}f_{c}^{\dagger}f_{d}^{\dagger}\right\} = & -\left(\mathbb{I}-\Delta\right)_{ca}\left(\mathbb{I}-\Delta\right)_{db}\nonumber \\
 & +\left(\mathbb{I}-\Delta\right)_{da}\left(\mathbb{I}-\Delta\right)_{cb}\label{eq:Wick's_theorem}
\end{align}
We further have that: 
\begin{align}
\frac{1}{\sqrt{\rho_{0}}}f_{a}\sqrt{\rho_{0}} & =\sum_{b}\left[\frac{\Delta^{T}}{\mathbb{I}-\Delta^{T}}\right]_{ab}^{1/2}f_{b}\nonumber \\
\frac{1}{\sqrt{\rho_{0}}}f_{a}^{\dagger}\sqrt{\rho_{0}} & =\sum_{b}\left[\frac{\mathbb{I}-\Delta^{T}}{\Delta^{T}}\right]_{ba}^{1/2}f_{b}^{\dagger}\nonumber \\
\frac{1}{\sqrt{\rho_{0}}}f_{a}^{\dagger}f_{b}\sqrt{\rho_{0}} & =\sum_{cd}\left[\frac{\Delta^{T}}{\mathbb{I}-\Delta^{T}}\right]_{bd}^{1/2}\left[\frac{\mathbb{I}-\Delta^{T}}{\Delta^{T}}\right]_{ca}^{1/2}f_{c}^{\dagger}f_{d}\nonumber \\
\frac{1}{\sqrt{\rho_{0}}}f_{a}^{\dagger}f_{b}^{\dagger}\sqrt{\rho_{0}} & =\sum_{cd}\left[\frac{\mathbb{I}-\Delta^{T}}{\Delta^{T}}\right]_{ca}^{1/2}\left[\frac{\mathbb{I}-\Delta^{T}}{\Delta^{T}}\right]_{db}^{1/2}f_{c}^{\dagger}f_{d}^{\dagger}\nonumber \\
\frac{1}{\sqrt{\rho_{0}}}f_{a}f_{b}\sqrt{\rho_{0}} & =\sum_{cd}\left[\frac{\Delta^{T}}{\mathbb{I}-\Delta^{T}}\right]_{ac}^{1/2}\left[\frac{\Delta^{T}}{\mathbb{I}-\Delta^{T}}\right]_{bd}^{1/2}f_{c}f_{d}\label{eq:Rotations-1}
\end{align}
Where repetitive use of Eq. (\ref{eq:Density_matrix-1}) has been
made. Explicitly this means that:

\begin{align}
\mathcal{G}_{f_{a};f_{b}}\left(\Delta\right) & =\left[\frac{\Delta^{T}}{\mathbb{I}-\Delta^{T}}\right]_{ab}^{1/2}\nonumber \\
\mathcal{G}_{f_{a}^{\dagger};f_{b}^{\dagger}}\left(\Delta\right) & =\left[\frac{\mathbb{I}-\Delta^{T}}{\Delta^{T}}\right]_{ba}^{1/2}\nonumber \\
\mathcal{G}_{f_{a}^{\dagger}f_{b};f_{c}^{\dagger}f_{d}}\left(\Delta\right) & =\left[\frac{\Delta^{T}}{\mathbb{I}-\Delta^{T}}\right]_{bd}^{1/2}\left[\frac{\mathbb{I}-\Delta^{T}}{\Delta^{T}}\right]_{ca}^{1/2}\nonumber \\
\mathcal{G}_{f_{a}^{\dagger}f_{b}^{\dagger};f_{c}^{\dagger}f_{d}^{\dagger}}\left(\Delta\right) & =\left[\frac{\mathbb{I}-\Delta^{T}}{\Delta^{T}}\right]_{ca}^{1/2}\left[\frac{\mathbb{I}-\Delta^{T}}{\Delta^{T}}\right]_{db}^{1/2}\nonumber \\
\mathcal{G}_{f_{a}f_{b};f_{c}f_{d}}\left(\Delta\right) & =\left[\frac{\Delta^{T}}{\mathbb{I}-\Delta^{T}}\right]_{ac}^{1/2}\left[\frac{\Delta^{T}}{\mathbb{I}-\Delta^{T}}\right]_{bd}^{1/2}\label{eq:Rotations}
\end{align}

\subsection{\label{sec:Operators-in-the-1}Operators in the embedding mapping}

In this section we have suppressed the site index $R$ in our notation
as we will be dealing with a single site only. We would like to show
that: 
\begin{widetext}
\begin{equation}
Tr\left[\phi^{\dagger}O_{\mu}\phi O_{i}\right]=\left\langle \Phi\right|\hat{O}_{\mu}\hat{O}_{i}^{R}\left|\Phi\right\rangle \exp\left[i\frac{\pi}{2}\left[\left[N\left(n'\right)-N\left(n\right)\right]^{2}-\left[N\left(n'\right)-N\left(n\right)\right]\right]\right]\label{eq:Main_relation_embedding}
\end{equation}

Here we define: 
\begin{equation}
O_{i}=\prod_{i=0}^{m}f_{a_{i}}^{\left(\dagger\right)},\,\hat{O}_{i}^{R}=\prod_{i=0}^{m}\hat{f}_{a_{m-i}}^{\left(\dagger\right)}\label{eq:Reverse-1}
\end{equation}

and 
\[
O_{\mu}=\prod_{i=0}^{K}c_{\alpha_{i}}^{\left(\dagger\right)},\,\hat{O}_{\mu}=\prod_{i=0}^{K}\hat{c}_{\alpha_{i}}^{\left(\dagger\right)}
\]
Here $R$ stands for reverse order. Here $\left(\dagger\right)$ means
that there may or may not be Hermitian conjugation (so that Eq. (\ref{eq:Main_relation_embedding})
can be used to handle strings of both creation and annihilation operators).
We have that: 
\begin{align}
 & \left\langle \Phi\right|\hat{O}_{\mu}\hat{O}_{i}\left|\Phi\right\rangle \nonumber \\
 & =\sum_{n',\Gamma'}\left\langle \hat{n}'\right|\left\langle \hat{\Gamma}'\right|\phi_{n'\Gamma'}^{\ast}\hat{O}_{\mu}\times\nonumber \\
 & \times\sum_{\Gamma,n}\phi_{\Gamma n}U_{PH}^{\dagger}\hat{O}_{i}U_{PH}\left|\hat{\Gamma}\right\rangle \left|\hat{n}\right\rangle \exp\left(i\frac{\pi}{2}N\left(n\right)\left(N\left(n\right)-1\right)\right)\nonumber \\
 & \times\exp\left(-i\frac{\pi}{2}N\left(n'\right)\left(N\left(n'\right)-1\right)\right)\nonumber \\
 & =\sum_{n',\Gamma'}\sum_{\Gamma,n}\phi_{\Gamma n}\left\langle \hat{n}'\right|\left\langle \hat{\Gamma}'\right|\phi_{n'\Gamma'}^{\ast}\hat{O}_{\mu}\hat{O}_{i}^{R\dagger}\left|\hat{\Gamma}\right\rangle \left|\hat{n}\right\rangle \times\exp\left(i\frac{\pi}{2}N\left(n\right)\left(N\left(n\right)-1\right)\right)\nonumber \\
 & \times\exp\left(-i\frac{\pi}{2}N\left(n'\right)\left(N\left(n'\right)-1\right)\right)\nonumber \\
 & =\sum_{n',\Gamma'}\sum_{\Gamma,n}\exp\left(i\frac{\pi}{2}N\left(n\right)\left(N\left(n\right)-1\right)-i\frac{\pi}{2}N\left(n'\right)\left(N\left(n'\right)-1\right)\right)\left(-1\right)^{\left[N\left(\Gamma\right)\right]\left[m+1\right]}\nonumber \\
 & \times\phi_{n'\Gamma'}^{\ast}\phi_{\Gamma n}\times\left\langle \hat{\Gamma}'\right|\hat{O}_{\mu}\left|\hat{\Gamma}\right\rangle \left\langle \hat{n}'\right|\hat{O}_{i}^{R\dagger}\left|\hat{n}\right\rangle \label{eq:Main_computation}
\end{align}
Now we have that: 
\begin{equation}
N\left(\Gamma\right)=N\left(n\right)\label{eq:Mod}
\end{equation}
This means that $\left(-1\right)^{\left[N\left(\Gamma\right)\right]\left[m+1\right]}=\left(-1\right)^{\left[N\left(n\right)\right]\left[m+1\right]}$.
Furthermore we have that: 
\begin{equation}
m+1=N\left(n'\right)-N\left(n\right)\left[mod\,2\right]\label{eq:Mod-1}
\end{equation}
This means that 
\begin{align}
 & =\exp\left(i\frac{\pi}{2}N\left(n\right)\left(N\left(n\right)-1\right)-i\frac{\pi}{2}N\left(n'\right)\left(N\left(n'\right)-1\right)\right)\left(-1\right)^{\left[N\left(\Gamma\right)\right]\left[m+1\right]}\nonumber \\
 & =\exp\left[i\frac{\pi}{2}\left[N\left(n\right)\left(N\left(n\right)-1\right)-N\left(n'\right)\left(N\left(n'\right)-1\right)-2\left[N\left(n'\right)-N\left(n\right)\right]\left[N\left(n\right)\right]\right]\right]\label{eq:Phases}
\end{align}
\end{widetext}

We now write: 
\begin{equation}
N\left(n'\right)=N\left(n\right)+\left[N\left(n'\right)-N\left(n\right)\right]\label{eq:relation}
\end{equation}
This means that: 
\begin{align}
 & \exp\left[i\frac{\pi}{2}\left[N\left(n\right)\left(N\left(n\right)-1\right)-N\left(n'\right)\left(N\left(n'\right)-1\right)\right.\right.\nonumber \\
 & \left.\left.-2\left[N\left(n'\right)-N\left(n\right)\right]\left[N\left(n\right)\right]\right]\right]\nonumber \\
 & =\exp\left[i\frac{\pi}{2}\left[\left[N\left(n'\right)-N\left(n\right)\right]^{2}-\left[N\left(n'\right)-N\left(n\right)\right]\right]\right]\nonumber \\
 & =\left\{ \begin{array}{cc}
1, & N\left(n'\right)-N\left(n\right)=0\left[mod\,4\right]\\
1, & N\left(n'\right)-N\left(n\right)=1\left[mod\,4\right]\\
-1, & N\left(n'\right)-N\left(n\right)=2\left[mod\,4\right]\\
-1, & N\left(n'\right)-N\left(n\right)=3\left[mod\,4\right]
\end{array}\right.\label{eq:Mod_4}
\end{align}
We now write: 
\begin{align}
 & \left\langle \Phi\right|\hat{O}_{\mu}\hat{O}_{i}\left|\Phi\right\rangle \nonumber \\
 & =\sum_{n',\Gamma'}\sum_{\Gamma,n}\exp\left[i\frac{\pi}{2}\left[\left[N\left(n'\right)-N\left(n\right)\right]^{2}-\left[N\left(n'\right)-N\left(n\right)\right]\right]\right]\nonumber \\
 & \times\phi_{n'\Gamma'}^{\ast}\phi_{\Gamma n}\times\left\langle \hat{\Gamma}'\right|\hat{O}_{\mu}\left|\hat{\Gamma}\right\rangle \left\langle \hat{n}'\right|\hat{O}_{i}^{R\dagger}\left|\hat{n}\right\rangle \nonumber \\
 & =Tr\left[\phi^{\dagger}O_{\mu}\phi O_{i}^{R}\right]\times\nonumber \\
 & \times\exp\left[i\frac{\pi}{2}\left[\left[N\left(n'\right)-N\left(n\right)\right]^{2}-\left[N\left(n'\right)-N\left(n\right)\right]\right]\right]\label{eq:Final}
\end{align}
and relation (\ref{eq:Main_relation_embedding}) follows. We have
also assumed that $\left\langle n'\right|\hat{O}_{i}^{R\dagger}\left|n\right\rangle =\left\langle n\right|\hat{O}^{R}\left|n'\right\rangle $,
e.g. the operator $\hat{O}_{i}^{R}$ is real, however all operators
of the form in Eq. (\ref{eq:Reverse-1}) are of this form.

\subsection{\label{sec:Derivations-of-Eqs.}Derivations of equivalence relations}

We will focus on the case of two point operators. The general case
in Eq. (\ref{eq:Generic_hamiltonian-1}) can not be done similarly,
as was shown in Appendix \ref{sec:Main-construction}.

\subsubsection{\label{subsec:Fermionic-operators}Fermionic operators}

\paragraph{\label{subsec:Main-Result-1}Equivalence results}

By Wicks theorem in the limit of large dimensions we may write that:
\begin{align}
 & \left\langle \Psi_{0}\right|P_{R}^{\dagger}O_{R\mu}P_{R}P_{R'}^{\dagger}O_{R'\nu}P_{R'}\left|\Psi_{0}\right\rangle =\nonumber \\
 & =[P_{R}^{\dagger}\overbracket{O_{R\mu}P_{R}][P_{R'}^{\dagger}O_{R'\nu}}P_{R'}]\nonumber \\
 & \times\left(1+O\left(\frac{1}{z}\right)\right)\label{eq:Single_contractions}
\end{align}
Where $[P_{R}^{\dagger}\overbracket{O_{R\mu}P_{R}][P_{R'}^{\dagger}O_{R'\nu}}P_{R'}]$
means perform all intrasite contractions and then a single intersite
contractions between $R$ and $R'$. Terms with more then one contraction
between $R$ and $R'$ are higher order in $1/z$. We may write that
\citep{Sandri_2014,Lanata2012,Lanata2009,Lanata_2015,Lanata_2016,Lanata_2017}:
\begin{align}
 & [P_{R}^{\dagger}\overbracket{O_{R\mu}P_{R}][P_{R'}^{\dagger}O_{R'\nu}}P_{R'}]\nonumber \\
 & =\sum_{ab}T_{R\mu a}T_{R'\nu b}\left\langle f_{Ra}^{\dagger}f_{R'b}\right\rangle \label{eq:Fock_piece-1}
\end{align}
For some coefficients $T_{R\mu a}$. Indeed to obtain the expectation
in the first line of Eq. (\ref{eq:Single_contractions}) we can contract
the operators $P_{R}^{\dagger}O_{R\mu}P_{R}$ and $P_{R'}^{\dagger}O_{R'\nu}P_{R'}$
in all possible ways while leaving only one operator uncontracted.
We can then write contractions between the two remaining single fermion
terms. Since this is a well defined procedure there is unique $T_{R\mu a},\,T_{R'\nu b}$
that represent the final states after all but one of the contractions
have been made. As such to derive that Eq. (\ref{eq:Generic_hamiltonian-1})
is reproduced to order $1/z$ we need only show that: 
\begin{equation}
\mathcal{Z}_{R\mu a}=T_{R\mu a}\label{eq:Main_relation-1}
\end{equation}
However we have that: 
\begin{align}
 & \left\langle \Psi_{0}\right|\left[\sum_{a}\mathcal{Z}_{R\mu a}f_{Ra}^{\dagger}\right]f_{Rb}\left|\Psi_{0}\right\rangle \nonumber \\
 & =\left\langle \Psi_{0}\right|P_{R}^{\dagger}O_{R\mu}P_{R}f_{Rb}\left|\Psi_{0}\right\rangle =\nonumber \\
 & =\left\langle \Psi_{0}\right|\left[\sum_{a}T_{R\mu a}f_{Ra}^{\dagger}\right]f_{Rb}\left|\Psi_{0}\right\rangle .\label{eq:Derivation}
\end{align}
From which Eq. (\ref{eq:Main_relation-1}) follows.

\paragraph{\label{subsec:Fermionic-Equivalences}solutions to equivalence relations}

In this Appendix we have suppressed the site index $R$ in our notation
as we will be dealing with a single site only and $\Delta=\Delta_{R,R}$.
We will also not distinguish between linear operators and their matrices
in the natural basis. We know that the equivalence relationship may
be written as: 
\begin{equation}
Tr\left[\sqrt{\rho_{0}}P_{R}^{\dagger}O_{\mu}P_{R}f_{a}\sqrt{\rho_{0}}\right]=\sum_{b}\mathcal{Z}_{\mu b}\left\langle f_{b}^{\dagger}f_{a}\right\rangle \label{eq:Eq}
\end{equation}
We have that 
\begin{equation}
\sum_{b}\mathcal{Z}_{\mu b}\left\langle f_{b}^{\dagger}f_{a}\right\rangle =\sum_{b}\mathcal{Z}_{\mu b}\Delta_{ba}\label{eq:Density-1}
\end{equation}
Furthermore using the results in Section \ref{sec:Some-useful-functions}:
\begin{align}
 & Tr\left[\sqrt{\rho_{0}}P_{R}^{\dagger}O_{\mu}P_{R}f_{a}\sqrt{\rho_{0}}\right]\nonumber \\
 & =\sum_{b}\left[\frac{\Delta}{\mathbb{I}-\Delta}\right]_{ba}^{1/2}Tr\left[\phi^{\dagger}O_{\mu}\phi f_{b}\right]\label{eq:Transformation}\\
 & \equiv\Gamma\sqrt{\frac{\Delta}{\mathbb{I}-\Delta}}
\end{align}
Here $\Gamma_{b}$ is the vector $\left[\phi^{\dagger}O_{\mu}\phi f_{b}\right]$
and we are using matrix notation. We now have that in matrix notation
\begin{align}
\mathcal{Z}_{\mu}\Delta & =\Gamma\sqrt{\frac{\Delta}{\mathbb{I}-\Delta}}\nonumber \\
\mathcal{Z}_{\mu a} & =\sum_{b}Tr\left[\phi^{\dagger}O_{\mu}\phi f_{b}\right]\left[\frac{1}{\left(\mathbb{I}-\Delta\right)\Delta}\right]_{ba}^{1/2}\label{eq:R_fermion}
\end{align}
Now using Eq. (\ref{eq:Main_relation_embedding}) we have Eq. (\ref{eq:Fermionic_R_matrix})
follows. Or alternatively: 
\begin{equation}
\left[\mathcal{Z}\left[\left(\mathbb{I}-\Delta\right)\Delta\right]^{1/2}\right]_{b}=Tr\left[\phi^{\dagger}O_{\mu}\phi f_{b}\right]\label{eq:Final_expression}
\end{equation}
We note that by the complex conjugation property in Eq. (\ref{eq:Conjugate})
we have that: 
\begin{align}
\mathcal{\hat{Z}}_{\mu^{\dagger}a} & =\mathcal{Z}_{\mu a}^{\ast}=\Gamma_{\mu}^{\dagger}\sqrt{\frac{\Delta^{*}}{\mathbb{I}-\Delta^{*}}}\nonumber \\
 & =\sum_{b}Tr\left[\phi^{\dagger}O_{\mu}^{\dagger}\phi f_{b}^{\dagger}\right]\left[\frac{1}{\left(\mathbb{I}-\Delta^{T}\right)\Delta^{T}}\right]_{ab}^{1/2}\label{eq:conjugate}
\end{align}
Here we have used the following notation: 
\[
P_{R}^{\dagger}O_{R\mu}^{\dagger}P_{R}\sim\sum_{a}\mathcal{Z}_{R\mu^{\dagger}a}f_{Ra}.
\]

\subsubsection{\label{subsec:Bosonic-Operator-swith}Bosonic operators with same
number of creation operators as annihilation operators}

\paragraph{\label{subsec:Main-Result}Equivalence results}

We would like to make Eq. (\ref{eq:Generic_hamiltonian-1}) more precise.
Explicitly below we show that the Gutzwiller equivalences are set
up so that if $P_{R}^{\dagger}O_{R\mu}P_{R}\sim\sum_{i}\mathcal{Z}_{R\mu i}O_{Ri}$
then: 
\begin{align}
 & \left\langle \Psi_{0}\right|P_{R}^{\dagger}O_{R\mu}P_{R}P_{R'}^{\dagger}O_{R'\nu}P_{R'}\left|\Psi_{0}\right\rangle _{Hartree}=\nonumber \\
 & =\left\langle \Psi_{0}\right|\left[\sum_{i}\mathcal{Z}_{R\mu i}O_{Ri}\right]\left[\sum_{j}\mathcal{Z}_{R'\nu j}O_{R'j}\right]\left|\Psi_{0}\right\rangle _{Hartree}\label{eq:Hartree-1}
\end{align}
with no corrections of any form. Here $Hartree$ means that only intersite
Wick contractions are done \citep{Wysokinski_2015}. Furthermore we
will show that the Fock term only renormalizes weakly: 
\begin{align}
 & \left\langle \Psi_{0}\right|P_{R}^{\dagger}O_{R\mu}P_{R}P_{R'}^{\dagger}O_{R'\nu}P_{R'}\left|\Psi_{0}\right\rangle _{Fock}=\nonumber \\
 & =\left\langle \Psi_{0}\right|\left[\sum_{i}\mathcal{Z}_{R\mu i}O_{Ri}\right]\left[\sum_{j}\mathcal{Z}_{R'\nu j}O_{R'j}\right]\left|\Psi_{0}\right\rangle _{Fock}\times\nonumber \\
 & \times\left(1+O\left(\frac{1}{z}\right)\right)\label{eq:Renormalization-1}
\end{align}
Here $Fock$ means that there are two or more intersite Wick contraction\citep{Wysokinski_2015}
(or equivalently all the terms not included in the $Hartree$ piece).
Here $z$ is the number of nearest neighbors. In some sense the Gutzwiller
equivalence procedure is designed to change as little of the expectation
values as possible and use the simplest operators possible to do so
\citep{Lanata2012,Lanata2009,Lanata_2015,Lanata_2016,Sandri_2014}.

\paragraph{\label{sec:Hartree-and-Fock}Hartree contributions}

We would like to show that the Hartree contribution to correlation
functions does not renormalize under the equivalences in Eq. (\ref{eq:bosonic residues}).
For this we first consider the equation (see Eq. (\ref{eq:bosonic residues}):
\begin{align}
 & Tr\left[\rho_{R0}P_{R}^{\dagger}O_{R\mu}P_{R}\cdot I\right]=\nonumber \\
 & Tr\left[\rho_{R0}\left[\sum_{cd}\mathcal{Z}_{R\mu dc}f_{Rc}^{\dagger}f_{Rd}+\mathcal{Z}_{R\mu I}I\right]\cdot I\right]\label{eq:Traces}
\end{align}
That says that trace of an operator with respect to the non-interacting
density matrix is preserved under equivalences. This means that the
Hartree term does not renormalize under these equivalences, indeed
the Hartree term may be written as: 
\begin{align*}
 & \left\langle \Psi_{0}\right|P_{R}^{\dagger}O_{R\mu}P_{R}P_{R'}^{\dagger}O_{R'\nu}P_{R'}\left|\Psi_{0}\right\rangle _{Hartree}\\
 & =Tr\left\{ \rho_{R,0}P_{R}^{\dagger}O_{R\mu}P_{R}\right\} \cdot Tr\left\{ \rho_{R',0}P_{R'}^{\dagger}O_{R'\nu}P_{R'}\right\} \\
 & =Tr\left[\rho_{R,0}\left[\sum_{cd}\mathcal{Z}_{R\mu dc}f_{Rc}^{\dagger}f_{Rd}+\mathcal{Z}_{R\mu I}I\right]\right]\times\\
 & \times Tr\left[\rho_{R'0}\left[\sum_{cd}\mathcal{Z}_{R'\nu dc}f_{R'c}^{\dagger}f_{R'd}+\mathcal{Z}_{R'\nu I}I\right]\right]\\
 & =\left\langle \Psi_{0}\right|\left[\sum_{cd}\mathcal{Z}_{R\mu dc}f_{Rc}^{\dagger}f_{Rc}+\mathcal{Z}_{R\mu I}I\right]\times\\
 & \times\left[\sum_{cd}\mathcal{Z}_{R'\nu dc}f_{R'c}^{\dagger}f_{R'd}+\mathcal{Z}_{R'\nu I}I\right]\left|\Psi_{0}\right\rangle _{Hartree}
\end{align*}
Which shows explicitly that in all cases the Hartree term does not
renormalize. We will see below that the Fock does. However due to
operator mixing and Eqs. \ref{eq:bosonic residues} the Fock term
only renormalizes by factors of order $1/z$ see Section \ref{sec:Derivation}.

\paragraph{\label{sec:Derivation}Fock contributions}

Consider bosonic operators with an equal number of creation and annihilation
operators: $O_{R\mu}$ and $O_{R'\nu}$. By Wicks theorem in the limit
of large dimensions we may write that: 
\begin{align}
 & \left\langle \Psi_{0}\right|P_{R}^{\dagger}O_{R\mu}P_{R}P_{R'}^{\dagger}O_{R'\nu}P_{R'}\left|\Psi_{0}\right\rangle =\nonumber \\
 & =\left\langle \Psi_{0}\right|P_{R}^{\dagger}O_{R\mu}P_{R}P_{R'}^{\dagger}O_{R'\nu}P_{R'}\left|\Psi_{0}\right\rangle _{Hartree}+\nonumber \\
 & +[\overbracket{P_{R}^{\dagger}O_{R\mu}\overbracket{P_{R}^{\dagger}][P_{R'}^{\dagger}}O_{R'\nu}}P_{R'}]\times\nonumber \\
 & \times\left(1+O\left(\frac{1}{z}\right)\right)\label{eq:Contractions}
\end{align}
Where {[}$\overbracket{P_{R}^{\dagger}O_{R\mu}\overbracket{P_{R}^{\dagger}][P_{R'}^{\dagger}}O_{R'\nu}}P_{R'}]$
means perform all intrasite contractions and then two intersite contractions
between $R$ and $R'$. Terms with more then two contractions between
$R$ and $R'$ are higher order in $1/z$. We may write that \citep{Sandri2014,Lanata2012,Lanata2009,Lanata_2015,Lanata_2016,Lanata_2017}:
\begin{align}
 & [\overbracket{P_{R}^{\dagger}O_{R\mu}\overbracket{P_{R}^{\dagger}][P_{R'}^{\dagger}}O_{R'\nu}}P_{R'}]\nonumber \\
 & =\sum_{abcd}T_{R\mu ba}T_{R'\nu dc}\left\langle f_{Ra}^{\dagger}f_{Rb}f_{R'c}^{\dagger}f_{R'd}\right\rangle _{Fock}\label{eq:Fock_piece}
\end{align}
For some coefficients $T_{R\mu ba}$. Indeed to obtain the expectation
in the first line of Eq. (\ref{eq:Single_contractions}) we can contract
the operators $P_{R}^{\dagger}O_{R\mu}P_{R}$ and $P_{R'}^{\dagger}O_{R'\nu}P_{R'}$
in all possible ways while leaving only two operator uncontracted.
We can then write contractions between the four remaining single fermion
terms. Since this is a well defined procedure there is unique $T_{R\mu ba},\,T_{R'\nu dc}$
that represent the final states after all but two of the contractions
have been made. Now the Hartree terms in Eq. (\ref{eq:Contractions})
are exactly reproduced by Eq. (\ref{eq:Hartree-1}) as derived in
Section \ref{sec:Hartree-and-Fock}, as such we need only derive that
the Fock terms are reproduced to order $1/z$ or: 
\begin{equation}
\mathcal{Z}_{R\mu ba}=T_{R\mu ba}\label{eq:Main_relation}
\end{equation}
However we have that 
\begin{align}
 & =\left\langle \Psi_{0}\right|\left[\sum_{ab}\mathcal{Z}_{R\mu ba}f_{a}^{\dagger}f_{b}+\mathcal{Z}_{R\mu I}I\right]\left|\Psi_{0}\right\rangle \left\langle \Psi_{0}\right|f_{Rc}^{\dagger}f_{Rd}\left|\Psi_{0}\right\rangle \nonumber \\
 & +\left\langle \Psi_{0}\right|\left[\sum_{ab}\mathcal{Z}_{R\mu ba}f_{a}^{\dagger}f_{b}+\mathcal{Z}_{R\mu I}I\right]f_{Rc}^{\dagger}f_{Rd}\left|\Psi_{0}\right\rangle _{Fock}\nonumber \\
 & =\left\langle \Psi_{0}\right|P_{R}^{\dagger}O_{R\mu}P_{R}f_{Rc}^{\dagger}f_{Rc}\left|\Psi_{0}\right\rangle =\nonumber \\
 & =\left\langle \Psi_{0}\right|P_{R}^{\dagger}O_{R\mu}P_{R}\left|\Psi_{0}\right\rangle \left\langle \Psi_{0}\right|f_{Rc}^{\dagger}f_{Rc}\left|\Psi_{0}\right\rangle \nonumber \\
 & +\left\langle \Psi_{0}\right|\left[\sum_{ab}T_{R\mu ba}f_{Ra}^{\dagger}f_{Rb}\right]f_{Rc}^{\dagger}f_{Rd}\left|\Psi_{0}\right\rangle _{Fock}\nonumber \\
 & =\left\langle \Psi_{0}\right|\left[\sum_{ab}\mathcal{Z}_{R\mu ba}f_{Ra}^{\dagger}f_{Rb}+\mathcal{Z}_{R\mu I}I\right]\left|\Psi_{0}\right\rangle \left\langle \Psi_{0}\right|f_{Rc}^{\dagger}f_{Rd}\left|\Psi_{0}\right\rangle \nonumber \\
 & +\left\langle \Psi_{0}\right|\left[\sum_{ab}T_{R\mu ba}f_{Ra}^{\dagger}f_{Rb}+\mathcal{Z}_{R\mu I}I\right]f_{Rc}^{\dagger}f_{Rd}\left|\Psi_{0}\right\rangle _{Fock}\label{eq:Main_derivation}
\end{align}
From which Eq. (\ref{eq:Main_relation}) follows.

\paragraph{\label{subsec:Mixed-Basis-results-1}Solutions to equivalence relations}

In this Appendix we have suppressed the site index $R$ in our notation
as we will be dealing with a single site only and $\Delta=\Delta_{R,R}$.
We will also not distinguish between linear operators and their matrices
in the natural basis. With this definition due to the relations in
Section \ref{sec:Hartree-and-Fock} we know that the equivalence relationship
may be written as: 
\begin{align}
 & Tr\left[\sqrt{\rho_{0}}P_{R}^{\dagger}O_{\mu}P_{R}f_{a}^{\dagger}f_{b}\sqrt{\rho_{0}}\right]\nonumber \\
 & =Tr\left[\sqrt{\rho_{0}}P_{R}^{\dagger}O_{\mu}P_{R}\sqrt{\rho_{0}}\right]\cdot Tr\left[\rho_{0}f_{a}^{\dagger}f_{b}\right]\\
 & +\sum_{\gamma\delta}\mathcal{Z}_{\mu dc}\left\langle f_{c}^{\dagger}f_{d}f_{a}^{\dagger}f_{b}\right\rangle _{Fock}\label{eq:Main_equivalence}
\end{align}
We now have that 
\begin{align}
 & Tr\left[\sqrt{\rho_{0}}P_{R}^{\dagger}O_{\mu}P_{R}\sqrt{\rho_{0}}\right]\cdot Tr\left[\rho_{0}f_{a}^{\dagger}f_{b}\right]\nonumber \\
 & =Tr\left[\phi^{\dagger}O_{\mu}\phi\right]\cdot\Delta_{ab}\label{eq:Trace_density}
\end{align}
Furthermore we have that: 
\begin{align}
\sum_{\gamma\delta}\mathcal{Z}_{\mu dc}\left\langle f_{c}^{\dagger}f_{d}f_{a}^{\dagger}f_{b}\right\rangle _{Fock} & =\sum_{\gamma\delta}\mathcal{Z}_{\mu dc}\Delta_{cb}\left(\mathbb{I}-\Delta_{ad}\right)\nonumber \\
 & =\left[\left(\mathbb{I}-\Delta\right)\mathcal{Z}_{\mu}\Delta\right]_{ab}\label{eq:Fock_trace}
\end{align}
Furthermore by the results of Section \ref{sec:Some-useful-functions}
we have that: 
\begin{align}
 & Tr\left[\sqrt{\rho_{0}}P_{R}^{\dagger}O_{\mu}P_{R}f_{a}^{\dagger}f_{b}\sqrt{\rho_{0}}\right]\nonumber \\
 & =\sum_{\delta}\left[\frac{\Delta}{\mathbb{I}-\Delta}\right]_{db}^{1/2}\sum_{\gamma}\left[\frac{\mathbb{I}-\Delta}{\Delta}\right]_{ac}^{1/2}Tr\left[\phi^{\dagger}O_{\mu}\phi f_{c}^{\dagger}f_{d}\right]\nonumber \\
 & =\sum_{\delta}\left[\frac{\Delta}{\mathbb{I}-\Delta}\right]_{db}^{1/2}\sum_{\gamma}\left[\frac{\mathbb{I}-\Delta}{\Delta}\right]_{ac}^{1/2}\Upsilon_{cd}\nonumber \\
 & =\left[\sqrt{\frac{\mathbb{I}-\Delta}{\Delta}}\Upsilon\sqrt{\frac{\Delta}{\mathbb{I}-\Delta}}\right]_{ab}\label{eq:Transform_computation}
\end{align}
Here we have defined the matrix $\Upsilon_{cd}=Tr\left[\phi^{\dagger}O_{\mu}\phi f_{c}^{\dagger}f_{d}\right]$.
This means that: 
\begin{align}
\left(\mathbb{I}-\Delta\right)\mathcal{Z}_{\mu}\Delta & =\sqrt{\frac{\mathbb{I}-\Delta}{\Delta}}\Upsilon\sqrt{\frac{\Delta}{\mathbb{I}-\Delta}}\nonumber \\
 & -Tr\left[\phi^{\dagger}O_{\mu}\phi\right]\cdot\Delta\label{eq:matrix_equality}
\end{align}
Or equivalently 
\begin{align}
\mathcal{Z}_{\mu} & =\sqrt{\frac{1}{\left(\mathbb{I}-\Delta\right)\Delta}}\Upsilon\sqrt{\frac{1}{\left(\mathbb{I}-\Delta\right)\Delta}}\nonumber \\
 & -Tr\left[\phi^{\dagger}O_{\mu}\phi\right]\frac{\mathbb{I}}{\left(\mathbb{I}-\Delta\right)}\label{eq:Final_equation}
\end{align}
Furthermore we have that in index notation: 
\begin{align}
\mathcal{Z}_{\mu dc} & =\sum_{ab}\left[\frac{1}{\left(\mathbb{I}-\Delta\right)\Delta}\right]_{da}^{1/2}\times Tr\left[\phi^{\dagger}O_{\mu}\phi f_{a}^{\dagger}f_{b}\right]\times\nonumber \\
 & \times\left[\frac{1}{\left(\mathbb{I}-\Delta\right)\Delta}\right]_{bc}^{1/2}-Tr\left[\phi^{\dagger}O_{\mu}\phi\right]\cdot\left[\frac{\mathbb{I}}{\left(\mathbb{I}-\Delta\right)}\right]_{dc}\label{eq:Indecies}
\end{align}
Now using Eq. (\ref{eq:Main_relation_embedding}) we have Eqs. (\ref{eq:Indecies-1})
and (\ref{eq:R_mu_I-1}) follow. Or alternatively: 
\begin{align}
 & \left[\left[\left(\mathbb{I}-\Delta\right)\Delta\right]^{1/2}\mathcal{Z}\left[\left(\mathbb{I}-\Delta\right)\Delta\right]^{1/2}\right]_{ab}\nonumber \\
 & =Tr\left[\phi^{\dagger}O_{\mu}\phi f_{a}^{\dagger}f_{b}\right]\nonumber \\
 & -Tr\left[\phi^{\dagger}O_{\mu}\phi\right]\cdot\left[\Delta\right]_{ab}\label{eq:Traces_mapping}
\end{align}
Furthermore we have that 
\begin{equation}
\mathcal{Z}_{\mu I}=Tr\left[\phi^{\dagger}O_{\mu}\phi\right]-\sum_{\gamma\delta}\mathcal{Z}_{\mu dc}\Delta_{cd}\label{eq:R_mu_I}
\end{equation}

\subsubsection{\label{subsec:Bosonic-operators-with}Bosonic operators with two
more creation operators then annihilation operators}

\paragraph{\label{subsec:Main-Result-2}Equivalence results}

By Wicks theorem in the limit of large dimensions we may write that:
\begin{align}
 & \left\langle \Psi_{0}\right|P_{R}^{\dagger}O_{R\mu}P_{R}P_{R'}^{\dagger}O_{R'\nu}P_{R'}\left|\Psi_{0}\right\rangle =\nonumber \\
 & =[\overbracket{P_{R}^{\dagger}O_{R\mu}\overbracket{P_{R}^{\dagger}][P_{R'}^{\dagger}}O_{R'\nu}}P_{R'}]\nonumber \\
 & \times\left(1+O\left(\frac{1}{z}\right)\right)\label{eq:Double_contractions}
\end{align}
Where $[\overbracket{P_{R}^{\dagger}O_{R\mu}\overbracket{P_{R}^{\dagger}][P_{R'}^{\dagger}}O_{R'\nu}}P_{R'}]$
means perform all intrasite contractions and then two intersite contractions
between $R$ and $R'$. Terms with more then two contraction between
$R$ and $R'$ are higher order in $1/z$. We may write that \citep{Sandri_2014,Lanata2012,Lanata2009,Lanata_2015,Lanata_2016,Lanata_2017}:
\begin{align}
 & [\overbracket{P_{R}^{\dagger}O_{R\mu}\overbracket{P_{R}^{\dagger}][P_{R'}^{\dagger}}O_{R'\nu}}P_{R'}]\nonumber \\
 & =\sum_{abcd}T_{R\mu ab}T_{R'\nu cd}\left\langle f_{Ra}^{\dagger}f_{Rb}^{\dagger}f_{R'c}f_{R'd}\right\rangle \label{eq:Fock_piece-1-1}
\end{align}
For some coefficients $T_{R\mu ab}$. Indeed to obtain the expectation
in the first line of Eq. (\ref{eq:Single_contractions}) we can contract
the operators $P_{R}^{\dagger}O_{R\mu}P_{R}$ and $P_{R'}^{\dagger}O_{R'\nu}P_{R'}$
in all possible ways while leaving only two operator uncontracted.
We can then write contractions between the four remaining single fermion
terms. Since this is a well defined procedure there is unique $T_{R\mu ab},\,T_{R'\nu cd}$
that represent the final states after all but two of the contractions
have been made. As such to derive that Eq. (\ref{eq:Generic_hamiltonian-1})
is reproduced to order $1/z$ we need to show that: 
\begin{equation}
\bar{\mathcal{Z}}_{R\mu ab}=T_{R\mu ab}\label{eq:Main_relation-1-1}
\end{equation}
However we have that: 
\begin{align}
 & \left\langle \Psi_{0}\right|\left[\sum_{ab}\mathcal{\bar{Z}}_{R\mu ab}c_{Ra}^{\dagger}c_{Rb}^{\dagger}\right]c_{Rc}c_{Rd}\left|\Psi_{0}\right\rangle \nonumber \\
 & =\left\langle \Psi_{0}\right|P_{R}^{\dagger}O_{R\mu}P_{R}c_{Rc}c_{Rd}\left|\Psi_{0}\right\rangle =\nonumber \\
 & =\left\langle \Psi_{0}\right|\left[\sum_{ab}T_{R\mu ab}c_{Ra}^{\dagger}c_{Rb}^{\dagger}\right]c_{Rc}c_{Rd}\left|\Psi_{0}\right\rangle \label{eq:Derivation-1}
\end{align}
From which Eq. (\ref{eq:Main_relation-1-1}) follows.

\paragraph{\label{subsec:Mixed-Basis-results}Solutions to equivalence relations}

In this Appendix we have suppressed the site index $R$ in our notation
as we will be dealing with a single site only and $\Delta=\Delta_{R,R}$.
We will also not distinguish between linear operators and their matrices
in the natural basis. We know that the equivalence relationship may
be written as: 
\begin{equation}
Tr\left[\sqrt{\rho_{0}}P_{R}^{\dagger}O_{\mu}P_{R}f_{a}f_{b}\sqrt{\rho_{0}}\right]=\sum_{c>d}\mathcal{\bar{Z}}_{\mu cd}\left\langle f_{c}^{\dagger}f_{d}^{\dagger}f_{a}f_{b}\right\rangle _{Fock}\label{eq:Eq-1}
\end{equation}
Now we have that: 
\begin{align}
\sum_{c>d}\mathcal{\bar{Z}}_{\mu cd}\left\langle f_{c}^{\dagger}f_{d}^{\dagger}f_{a}f_{b}\right\rangle _{Fock} & =\sum_{c>c}\mathcal{\bar{Z}}_{\mu cc}\times\nonumber \\
 & \times\left(-\Delta_{ca}\Delta_{db}+\Delta_{cb}\Delta_{da}\right)\label{eq:Fock_boson_pait_hopping}
\end{align}
Now define: 
\begin{equation}
\mathcal{\bar{Z}}_{\mu cd}=-\mathcal{\bar{Z}}_{\mu dc}\label{eq:Convention}
\end{equation}
This means that: 
\begin{align*}
\sum_{c>d}\mathcal{\bar{Z}}_{\mu cd}\left\langle f_{c}^{\dagger}f_{d}^{\dagger}f_{a}f_{b}\right\rangle _{Fock} & =\sum_{cd}\mathcal{\bar{Z}}_{\mu cd}\Delta_{cb}\Delta_{da}\\
 & =\left[\Delta^{T}\mathcal{Z}_{\mu}\Delta\right]_{ab}
\end{align*}
Furthermore by the results of Section \ref{sec:Some-useful-functions}
we have that: 
\begin{align}
 & Tr\left[\sqrt{\rho_{0}}P_{R}^{\dagger}O_{\mu}P_{R}f_{a}f_{b}\sqrt{\rho_{0}}\right]\nonumber \\
 & =\sum_{\gamma}\left[\frac{\Delta^{T}}{\mathbb{I}-\Delta^{T}}\right]_{ac}^{1/2}\sum_{\delta}\left[\frac{\Delta^{T}}{\mathbb{I}-\Delta^{T}}\right]_{bd}^{1/2}Tr\left[\phi^{\dagger}O_{\mu}\phi f_{c}f_{d}\right]\nonumber \\
 & =\sum_{\gamma}\left[\frac{\Delta^{T}}{\mathbb{I}-\Delta^{T}}\right]_{ac}^{1/2}\sum_{\delta}\left[\frac{\Delta^{T}}{\mathbb{I}-\Delta^{T}}\right]_{bd}^{1/2}\Omega_{cd}\nonumber \\
 & =\left[\sqrt{\frac{\Delta^{T}}{\mathbb{I}-\Delta^{T}}}\cdot\Omega\cdot\sqrt{\frac{\Delta}{\mathbb{I}-\Delta}}\right]_{ab}\label{eq:Trace_equality}
\end{align}
Here we have defined the matrix $\Omega_{cd}=Tr\left[\phi^{\dagger}O_{\mu}\phi f_{c}f_{d}\right]$.
This means that 
\begin{equation}
\mathcal{\bar{Z}}_{\mu}=\sqrt{\frac{1}{\left(\mathbb{I}-\Delta^{T}\right)\Delta^{T}}}\cdot\Omega\cdot\sqrt{\frac{1}{\left(\mathbb{I}-\Delta\right)\Delta}}\label{eq:R_matrix_two_extra}
\end{equation}
Or equivalently 
\begin{align}
\mathcal{\bar{Z}}_{\mu ab} & =\sum_{cd}\left[\frac{1}{\left(\mathbb{I}-\Delta^{T}\right)\Delta^{T}}\right]_{ac}^{1/2}\cdot\nonumber \\
 & Tr\left[\phi^{\dagger}O_{\mu}\phi f_{c}f_{d}\right]\cdot\left[\frac{1}{\left(\mathbb{I}-\Delta\right)\Delta}\right]_{db}^{1/2}\label{eq:R_matrix_pair_hopping}
\end{align}
Now using Eq. (\ref{eq:Main_relation_embedding}) we have Eq. (\ref{eq:R_matrix_pair_hopping-1})
follows. Or alternatively:

\begin{align}
 & \left[\left[\left(\mathbb{I}-\Delta^{T}\right)\Delta^{T}\right]^{1/2}\bar{\mathcal{Z}}\left[\left(\mathbb{I}-\Delta\right)\Delta\right]^{1/2}\right]_{cd}\nonumber \\
 & =Tr\left[\phi^{\dagger}O_{\mu}\phi f_{c}f_{d}\right]\label{eq:Simplified_relation}
\end{align}
Now we know that: 
\begin{align}
\mathcal{\hat{Z}}_{\mu^{\dagger}ba} & =\mathcal{\bar{Z}}_{\mu ab}^{*}=\sum_{cd}\left[\frac{1}{\left(\mathbb{I}-\Delta\right)\Delta}\right]_{ac}^{1/2}\cdot\nonumber \\
 & Tr\left[\phi^{\dagger}O_{\mu}^{\dagger}\phi f_{d}^{\dagger}f_{c}^{\dagger}\right]\cdot\left[\frac{1}{\left(\mathbb{I}-\Delta^{T}\right)\Delta^{T}}\right]_{db}^{1/2}\label{eq:R_matrix_dagger}
\end{align}

\subsection{\label{sec:Proofs}Proofs for Section \ref{subsec:Alternate-approach}}

We would like to show that the equivalences presented in Section \ref{subsec:Alternate-approach}
represent all three contraction terms. We will show this on the example
of just two operators, where these terms may be used to generate higher
order terms in $1/z$, but they can directly be generalized to three
and four site terms where they can be used as part of the leading
order contribution.

\subsubsection{\label{subsec:Fermionic-equivalences}Fermionic equivalences}

We have that: 
\begin{align}
 & \left\langle \Psi_{0}\right|P_{R}^{\dagger}O_{R\mu}P_{R}P_{R'}^{\dagger}O_{R'\nu}P_{R'}\left|\Psi_{0}\right\rangle \nonumber \\
 & =[P_{R}^{\dagger}\overbracket{O_{R\mu}P_{R}][P_{R'}^{\dagger}O_{R'\nu}}P_{R'}]\nonumber \\
 & +[\overbracket{P_{R}^{\dagger}\overbracket{O_{R\mu}\overbracket{P_{R}][P_{R'}^{\dagger}}O_{R'\nu}}P_{R'}}]\times\nonumber \\
 & \times\left(1+O\left(\frac{1}{z}\right)\right)\label{eq:Fermionic_contractions}
\end{align}
From this it is clear that all we have to show for Eq. (\ref{eq:Equations_fermionic})
is to show that all one and three contraction terms are reproduced.
However the Equations in Eq. (\ref{eq:Equations_fermionic}) are linear
combinations of these equations with invertible coefficients (for
generic contractions $\left\langle f_{a}^{\dagger}f_{b}\right\rangle $)
from which Eq. (\ref{eq:Equations_fermionic}) follows.

\subsubsection{\label{subsec:Bosonic_equivalences}Bosonic equivalences (fermion
number preserving)}

We have that: 
\begin{align}
 & \left\langle \Psi_{0}\right|P_{R}^{\dagger}O_{R\mu}P_{R}P_{R'}^{\dagger}O_{R'\nu}P_{R'}\left|\Psi_{0}\right\rangle \nonumber \\
 & =\left\langle \Psi_{0}\right|P_{R}^{\dagger}O_{R\mu}P_{R}P_{R'}^{\dagger}O_{R'\nu}P_{R'}\left|\Psi_{0}\right\rangle _{Hartree}\nonumber \\
 & +[\overbracket{P_{R}^{\dagger}O_{R\mu}\overbracket{P_{R}^{\dagger}][P_{R'}^{\dagger}}O_{R'\nu}}P_{R'}]+\nonumber \\
 & +[\overbracket{P_{R}^{\dagger}\overbracket{O_{R\mu}\overbracket{P_{R}\overbracket{][}P_{R'}^{\dagger}}O_{R'\nu}}P_{R'}}]\times\nonumber \\
 & \times\left(1+O\left(\frac{1}{z}\right)\right)\label{eq:zer_two_four_contractions}
\end{align}

From this it is clear that all we have to show for Eq. (\ref{eq:Equivalence_bosonic_number_conserving})
is to show that all zero, two and four contraction terms are reproduced.
However the Equations in Eq. (\ref{eq:Equivalence_bosonic_number_conserving})
are linear combinations of these equations with invertible coefficients
(for generic contractions $\left\langle f_{a}^{\dagger}f_{b}\right\rangle $)
from which Eq. (\ref{eq:Equivalence_bosonic_number_conserving}) follows.

\subsubsection{\label{subsec:Bosonic_equivalences-1}Bosonic equivalences (fermion
number changing) }

We have that: 
\begin{align}
 & \left\langle \Psi_{0}\right|P_{R}^{\dagger}O_{R\mu}P_{R}P_{R'}^{\dagger}O_{R'\nu}P_{R'}\left|\Psi_{0}\right\rangle =\nonumber \\
 & =[\overbracket{P_{R}^{\dagger}O_{R\mu}\overbracket{P_{R}][P_{R'}^{\dagger}}O_{R'\nu}}P_{R'}]+\nonumber \\
 & +[\overbracket{P_{R}^{\dagger}\overbracket{O_{R\mu}\overbracket{P_{R}\overbracket{][}P_{R'}}O_{R'\nu}}P_{R'}}]\times\nonumber \\
 & \times\left(1+O\left(\frac{1}{z}\right)\right)\label{eq:two_four_contractions}
\end{align}
\textbackslash From this it is clear that all we have to show for
Eq. (\ref{eq:Bosonic_equations_number_changing}) is to show that
all two and four contraction terms are reproduced. However the Equations
in Eq. (\ref{eq:Bosonic_equations_number_changing}) are linear combinations
of these equations with invertible coefficients (for generic contractions
$\left\langle f_{a}^{\dagger}f_{b}\right\rangle $) from which Eq.
(\ref{eq:Bosonic_equations_number_changing}) follows.

\section{\label{sec:Sanity-Check}Sanity checks}

\subsection{\label{subsec:The-identity-operator}Gutzwiller constraints: the
identity operator}

So far we have not used the Gutzwiller constrains in Eq. (\ref{eq:Conditions-4-1})
in our derivations. We would like to use these constraints to show
that it is possible to insert an identity operators in Eq. (\ref{eq:Generic_hamiltonian-1})
without changing the ground state energy within our calculations and
as such provide a consistency check. More precisely we would like
to show that for: 
\begin{equation}
O_{R\mu}=\kappa I,\label{eq:Identity}
\end{equation}
we have that 
\begin{equation}
\mathcal{Z}_{R\mu I}=\kappa,\,\mathcal{Z}_{R\mu ba}=0,\label{eq:Sanity}
\end{equation}
and as such 
\begin{equation}
P_{R}^{\dagger}IP_{R}\sim I.\label{eq:Sanity_equivalence}
\end{equation}
To do so lets rewrite the Gutzwiller constraints in Eq. (\ref{eq:Conditions-4-1})
in the following suggestive way:

\begin{align}
 & \left\langle \Psi_{0}\right|P_{R}^{\dagger}\kappa IP_{R}f_{Ra}^{\dagger}f_{Rb}\left|\Psi_{0}\right\rangle =\nonumber \\
 & \left\langle \Psi_{0}\right|\left[\sum_{cd}0\cdot f_{Rc}^{\dagger}f_{Rd}+\kappa I\right]f_{Ra}^{\dagger}f_{Rb}\left|\Psi_{0}\right\rangle \nonumber \\
 & \left\langle \Psi_{0}\right|P_{R}^{\dagger}\kappa IP_{R}\cdot I\left|\Psi_{0}\right\rangle =\nonumber \\
 & \left\langle \Psi_{0}\right|\left[\sum_{cd}0\cdot f_{Rc}^{\dagger}f_{Rd}+\kappa I\right]\cdot I\left|\Psi_{0}\right\rangle .\label{eq:bosonic residues-1}
\end{align}
From which Eqs. (\ref{eq:Sanity}) and (\ref{eq:Sanity_equivalence})
follow.

\subsection{\label{subsec:Hermicity-check}Hermicity check}

We would like to show that under the equivalence in Eqs. (\ref{eq:fermionic_equivalence-4}),
(\ref{eq:Indecies-1}) , (\ref{eq:R_mu_I-1}) and (\ref{eq:R_matrix_pair_hopping-1})
Hermitian Hamiltonians are mapped onto Hermitian Hamiltonians. The
main idea will to show that the Hermitian conjugate of an operator
is equivalent to the Hermitian conjugate of the equivalent operator.
Below in Appendix \ref{subsec:Explicit-hermicity-proof} we will show
that this is sufficient for the Hamiltonian in Eq. (\ref{eq:Generic_hamiltonian-1})
to be Hermitian.

\subsubsection{\label{subsec:Fermionic-operators-1}Fermionic operators}

We start with the case where the operator $O_{R\mu}$ is fermionic.
For this case we would like to check that if: 
\begin{equation}
P_{R}^{\dagger}O_{R\mu}P_{R}\sim\sum_{a}\mathcal{Z}_{R\mu a}f_{Ra}\label{eq:fermionic_equivalence-1}
\end{equation}
then 
\begin{equation}
P_{R}^{\dagger}O_{R\mu}^{\dagger}P_{R}\sim\sum_{a}\mathcal{Z}_{R\mu a}^{*}f_{Ra}^{\dagger}\label{eq:Hermitian_conjugate}
\end{equation}
More precisely we would like to show that: 
\begin{equation}
\mathcal{Z}_{R\mu^{\dagger}a}=\mathcal{\hat{Z}}_{R\mu a}^{*}.\label{eq:Conjugate}
\end{equation}
To do so, we notice that Eq. (\ref{eq:Equations}) may be rewritten
as: 
\begin{equation}
\left\langle \Psi_{0}\right|f_{Ra}P_{R}^{\dagger}O_{R\mu}P_{R}\left|\Psi_{0}\right\rangle =\left\langle \Psi_{0}\right|f_{Ra}\left[\sum_{b}\mathcal{Z}_{\mu b}f_{Rb}^{\dagger}\right]\left|\Psi_{0}\right\rangle .\label{eq:Equation+other order}
\end{equation}
Indeed after we perform all contractions and reduced $P_{R}^{\dagger}O_{R\mu}P_{R}$
to a single annihilation operator in every possible way using Wick's
theorem we get that 
\begin{equation}
P_{R}^{\dagger}O_{R\mu}P_{R}\sim\sum_{a}\mathcal{Z}_{R\mu a}f_{Ra}\label{eq:fermionic_equivalence-1-1}
\end{equation}
and it does not matter which order we put the operators $f_{Ra}$
in order to determine the coefficients $\mathcal{Z}_{R\mu a}$, e.g.
$\mathcal{Z}_{R\mu a}$ are unique and it does not matter which of
the two types of equations we use to determine them. Taking the Hermitian
conjugate of this equation (Eq. (\ref{eq:Equation+other order}))
we get that: 
\begin{equation}
\left\langle \Psi_{0}\right|P_{R}^{\dagger}O_{R\mu}^{\dagger}P_{R}f_{Ra}^{\dagger}\left|\Psi_{0}\right\rangle =\left\langle \Psi_{0}\right|\left[\sum_{b}\mathcal{Z}_{R\mu b}^{*}f_{Rb}\right]f_{Ra}^{\dagger}\left|\Psi_{0}\right\rangle .\label{eq:Conjugate-1}
\end{equation}
As such Eq. (\ref{eq:Conjugate}) follows. The bosonic Hermicity checks
are highly similar and done below.

\subsubsection{\label{subsec:Bosonic-operator-with}Bosonic operator with the same
number of creation and annihilation operators}

We continue with the case where the operator $O_{R\mu}$ is bosonic
number conserving operator. For this case we would like to check that
if: 
\begin{equation}
P_{R}^{\dagger}O_{R\mu}P_{R}\sim\sum_{\alpha\beta}\mathcal{Z}_{R\mu ba}f_{Ra}^{\dagger}f_{Rb}+\mathcal{Z}_{R\mu I}I\label{eq:Bosonic_equivalence_number conserving-1}
\end{equation}
then 
\begin{equation}
P_{R}^{\dagger}O_{R\mu}^{\dagger}P_{R}\sim\sum_{\alpha\beta}\mathcal{Z}_{R\mu ba}^{*}f_{Rb}^{\dagger}f_{Ra}+\mathcal{Z}_{R\mu I}^{*}I\label{eq:Bosonic_equivalence_number conserving-1-1}
\end{equation}
e.g. 
\begin{equation}
\mathcal{Z}_{R\mu ba}^{*}=\mathcal{Z}_{R\mu^{\dagger}ab},\,\mathcal{Z}_{R\mu I}^{*}=\mathcal{Z}_{R\mu^{\dagger}I}.\label{eq:Hermicirty_bosonic_conserving}
\end{equation}
However in a manner similar to Appendix \ref{subsec:Fermionic-operators-1}
we can show that $\mathcal{Z}_{R\mu cd},\mathcal{Z}_{R\mu I}$ satisfy
the following equations: 
\begin{align}
 & \left\langle \Psi_{0}\right|f_{Rb}^{\dagger}f_{Ra}P_{R}^{\dagger}O_{R\mu}P_{R}\left|\Psi_{0}\right\rangle =\nonumber \\
 & \left\langle \Psi_{0}\right|f_{Rb}^{\dagger}f_{Ra}\left[\sum_{cd}\mathcal{Z}_{R\mu cd}f_{Rd}^{\dagger}f_{Rc}+\mathcal{Z}_{R\mu I}I\right]\left|\Psi_{0}\right\rangle \nonumber \\
 & \left\langle \Psi_{0}\right|I\cdot P_{R}^{\dagger}O_{R\mu}P_{R}\left|\Psi_{0}\right\rangle =\nonumber \\
 & \left\langle \Psi_{0}\right|I\cdot\left[\sum_{cd}\mathcal{Z}_{R\mu cd}f_{Rd}^{\dagger}f_{Rc}+\mathcal{Z}_{R\mu I}I\right]\left|\Psi_{0}\right\rangle .\label{eq:bosonic residues-2}
\end{align}
Taking the Hermitian conjugate of these equations we get that: 
\begin{align}
 & \left\langle \Psi_{0}\right|P_{R}^{\dagger}O_{R\mu}P_{R}f_{Ra}^{\dagger}f_{Rb}\left|\Psi_{0}\right\rangle =\nonumber \\
 & \left\langle \Psi_{0}\right|\left[\sum_{cd}\mathcal{Z}_{R\mu cd}^{*}f_{Rc}^{\dagger}f_{Rd}+\mathcal{Z}_{R\mu I}^{*}I\right]f_{Ra}^{\dagger}f_{Rb}\left|\Psi_{0}\right\rangle \nonumber \\
 & \left\langle \Psi_{0}\right|P_{R}^{\dagger}O_{R\mu}P_{R}\cdot I\left|\Psi_{0}\right\rangle =\nonumber \\
 & \left\langle \Psi_{0}\right|\left[\sum_{cd}\mathcal{Z}_{R\mu cd}^{*}f_{Rc}^{\dagger}f_{Rd}+\mathcal{Z}_{R\mu I}^{*}I\right]\cdot I\left|\Psi_{0}\right\rangle \label{eq:bosonic residues-3}
\end{align}
and Eqs. (\ref{eq:Bosonic_equivalence_number conserving-1-1}) and
(\ref{eq:Hermicirty_bosonic_conserving}) follow.

\subsubsection{\label{subsec:Bosonic-operators-that}Bosonic operators that change
fermion number by two}

We continue with the case where the operator $O_{R\mu}$ is bosonic
operator that changes the fermion number by two. For this case we
would like to check that if:

\begin{equation}
P_{R}^{\dagger}O_{R\mu}P_{R}\sim\sum_{\alpha\beta}\mathcal{\bar{\mathcal{Z}}}_{R\mu ab}f_{Ra}^{\dagger}f_{Rb}^{\dagger}\label{eq:Bosonic_equivalence_pair_hopping-1}
\end{equation}
then: 
\begin{equation}
P_{R}^{\dagger}O_{R\mu}^{\dagger}P_{R}\sim\sum_{\alpha\beta}\mathcal{\hat{Z}}_{R\mu ab}^{*}f_{Rb}f_{Ra}\label{eq:Bosonic_equivalence_pair_hopping-1-1}
\end{equation}
e.g. 
\begin{equation}
\mathcal{\bar{Z}}_{R\mu ab}^{*}=\mathcal{\hat{Z}}_{R\mu^{\dagger}ba}.\label{eq:Hermicirty_bosonic_changing}
\end{equation}
However we have that $\bar{\mathcal{Z}}_{R\mu\alpha\beta}$ satisfy
the following equations: 
\begin{align}
 & \left\langle \Psi_{0}\right|f_{Rb}f_{Ra}P_{R}^{\dagger}O_{R\mu}P_{R}\left|\Psi_{0}\right\rangle \nonumber \\
 & =\left\langle \Psi_{0}\right|f_{Rb}f_{Ra}\left[\sum_{cd}\mathcal{\bar{Z}}_{R\mu dc}f_{Rd}^{\dagger}f_{Rc}^{\dagger}\right]\left|\Psi_{0}\right\rangle .\label{eq:Backwords}
\end{align}
Taking the hermitian conjugate of this equation we get that: 
\begin{align}
 & \left\langle \Psi_{0}\right|P_{R}^{\dagger}O_{R\mu}^{\dagger}P_{R}f_{Ra}^{\dagger}f_{Rb}^{\dagger}\left|\Psi_{0}\right\rangle =\nonumber \\
 & \left\langle \Psi_{0}\right|\left[\sum_{cd}\mathcal{\bar{Z}}_{R\mu dc}^{*}f_{Rc}f_{Rd}\right]f_{Ra}^{\dagger}f_{Rb}^{\dagger}\left|\Psi_{0}\right\rangle \label{eq:relation-1-1}
\end{align}
and Eqs. (\ref{eq:Bosonic_equivalence_pair_hopping-1-1}) and (\ref{eq:Hermicirty_bosonic_changing})
follow.

\subsubsection{\label{subsec:Explicit-hermicity-proof}Explicit hermicity proof}

We note that for a hermitian Hamiltonian we may write 
\begin{align}
H & =\frac{1}{2}\left(H+H^{\dagger}\right)\nonumber \\
 & =\frac{1}{2}\left[\sum_{n}\sum_{R_{1}...R_{n}}\sum_{\mu_{1}...\mu_{n}}J_{R_{1}...R_{n}}^{\mu_{1};....\mu_{n}}O_{R_{1}\mu_{1}}....O_{R_{n}\mu_{n}}\right]\nonumber \\
 & +\frac{1}{2}\left[\sum_{n}\sum_{R_{1}...R_{n}}\sum_{\mu_{1}...\mu_{n}}J_{R_{1}...R_{n}}^{\mu_{1};....\mu_{n}*}O_{R_{n}\mu_{n}}^{\dagger}....O_{R_{1}\mu_{1}}^{\dagger}\right].\label{eq:Hamiltonian-2}
\end{align}
In which case 
\begin{widetext}
\begin{align}
H\sim H_{Eff} & =\frac{1}{2}\left[\sum_{n}\sum_{R_{1}...R_{n}}\sum_{\mu_{1}...\mu_{n}}J_{R_{1}...R_{n}}^{\mu_{1};....\mu_{n}}\left[\sum_{i_{1}}\mathcal{Z}_{R_{1}\mu_{1}i_{1}}O_{R_{1}\mu_{1}i_{1}}.\right]...\left[\sum_{i_{n}}\mathcal{Z}_{R_{1}\alpha_{n}i_{n}}O_{R_{n}\mu_{n}i_{n}}\right]\right]\nonumber \\
 & +\frac{1}{2}\left[\sum_{n}\sum_{R_{1}...R_{n}}\sum_{\mu_{1}...\mu_{n}}J_{R_{1}...R_{n}}^{\mu_{1};....\mu_{n}*}\left[\sum_{i_{n}}\mathcal{Z}_{R_{1}\alpha_{n}i_{n}}^{*}O_{R_{n}\mu_{n}i_{n}}^{\dagger}\right]...\left[\sum_{i_{1}}\mathcal{Z}_{R_{1}\alpha_{1}i_{1}}^{*}O_{R_{1}\mu_{1}i_{1}}^{\dagger}\right]\right].\label{eq:Effective}
\end{align}
Which is clearly Hermitian. 
\end{widetext}

\section{\label{sec:Short-Lagrangian}Short Lagrangian}
\begin{widetext}
We can now obtain and analog of Eq. (\ref{eq:Lagrangian}) by replacing
the various constraints by Lagrange multipliers and the main Hamiltonian
by the embedding Hamiltonian. As such we have the following action
(which is for simplicity specialized to the case where we only consider
the first two lines of Eq. (\ref{eq:Hamiltonian-1})): 
\begin{align}
 & \mathcal{L}_{N}\left(D_{R\mu i},E_{R}^{c},\mathcal{Z}_{R\mu i},\phi_{R},\left[\lambda_{R_{1}R_{2}}^{s}\right]_{ab},\left[\lambda_{R}^{cs}\right]_{ab},\left[\Delta_{R_{1}R_{2}}\right]_{ab},\left|\Psi_{0}\right\rangle \right)=\left\langle \Psi_{0}\right|\varepsilon_{HF}+\left[\lambda_{R_{1}R_{2}}^{s}\right]_{ab}-\mu\left|\Psi_{0}\right\rangle \nonumber \\
 & +\sum_{R}E_{R}^{c}-\sum_{R,ab}\left(\left[\lambda_{RR}^{s}\right]_{ab}+\left[\lambda_{R}^{cs}\right]_{ab}\right)\left[\Delta_{RR}\right]_{ab}-\mathcal{F}_{D}\left(\left\{ \left[\Delta_{R_{1}R_{2}}\right]_{ab}\right\} \right)-\sum_{R\mu i,j}D_{R\mu j}\mathcal{Z}_{R\mu i}F_{ij}\left(\left[\Delta_{RR}\right]_{ab}\right)-\nonumber \\
 & -\sum_{R_{1}\neq R_{2},}\sum_{ab}\left[\lambda_{R_{1}R_{2}}^{s}\right]_{ab}\left[\Delta_{R_{1}R_{2}}\right]_{ab}+\mathcal{L}_{embed}+\mu N\label{eq:Main_lagrangian}
\end{align}
Where we define define $\mathcal{L}_{embed}=\sum_{R}\left\langle \Phi_{R}\right|H_{embed}^{R}\left|\Phi_{R}\right\rangle +\sum_{R}E_{R}^{cs}$
, where: 
\begin{align}
H_{embed} & =\sum_{\mu i}D_{R\mu i}\left(\sum_{k}\hat{O}_{R\mu}\hat{O}_{Rk}^{R}\mathcal{G}_{i;k}\left(\Delta_{R}\right)\right)\exp\left[i\frac{\pi}{2}\left[\left[N\left(n'\right)-N\left(n\right)\right]^{2}-\left[N\left(n'\right)-N\left(n\right)\right]\right]\right]\nonumber \\
 & +H_{R}^{loc}-\sum_{R}E_{R}^{cs}+\sum_{ab}\left[\lambda_{R}^{cs}\right]_{ab}\hat{f}_{Rb}\hat{f}_{Ra}^{\dagger}\label{eq:Embedding}
\end{align}
Where $\varepsilon$ is the Hartree Fock energy, e.g. 
\begin{equation}
\varepsilon_{HF}=\sum_{\mu\nu}\sum_{R_{1},R_{2}}\sum_{ij}J_{R_{1};R_{2}}^{\mu;\nu}\mathcal{Z}_{R_{1}\mu i}\mathcal{Z}_{R_{2}\nu j}\left[O_{R_{1}i}O_{R_{2}j}\right]_{HF}\label{eq:hartree_fock}
\end{equation}
Where for example 
\begin{equation}
\left[f_{R_{1}a}^{\dagger}f_{R_{1}b}f_{R_{2}c}^{\dagger}f_{R_{2}d}\right]_{HF}=\left[\Delta_{R_{1}R_{1}}\right]_{ab}f_{R_{2}c}^{\dagger}f_{R_{2}d}+\left[\Delta_{R_{2}R_{2}}\right]_{cd}f_{R_{1}a}^{\dagger}f_{R_{1}b}+\left[\Delta_{R_{2}R_{1}}\right]_{ad}f_{R_{1}b}f_{R_{2}c}^{\dagger}-\left[\Delta_{R_{1}R_{2}}\right]_{bc}f_{R_{1}a}^{\dagger}f_{R_{2}d}\label{eq:Hartree_Fock-1}
\end{equation}
And 
\begin{equation}
\mathcal{F}_{D}\left(\left\{ \left[\Delta_{R_{1}R_{2}}\right]_{ab}\right\} \right)\supset J_{R_{1};R_{2}}^{\mu;\nu}\left[\left[\Delta_{R_{1}R_{1}}\right]_{ab}\left[\Delta_{R_{2}R_{2}}\right]_{cd}-\left[\Delta_{R_{2}R_{1}}\right]_{ad}\left[\Delta_{R_{1}R_{2}}\right]_{bc}\right]\label{eq:Double_counting}
\end{equation}
The extermination is carried out over all variables including Lagrange
multipliers \citep{Sandri_2014,Lanata2009,Lanata2012,Lanata_2015,Lanata_2016,Lanata_2017}.
We have that at saddle point $\left[\lambda_{R_{1}\neq R_{2}}^{s}\right]_{ab}=0$,
see Appendix \ref{sec:Quasiparticle-energies}. We note that the Lagrangian
in Eq. (\ref{eq:Main_lagrangian}) is not real so it cannot be minimized
only extremized (the saddle point energy is real though). For a more
complex but completely real Lagrangian that is specialized to the
Hamiltonian in Eq. (\ref{sec:Main-Hamiltonian}) see Section \ref{sec:Gutzwiller-Lagrangian}
which is our main result. 
\end{widetext}

\subsection{\label{sec:Quasiparticle-energies}Quasiparticle energies}

\subsubsection{\label{sec:Argument-why}Argument why $\left[\lambda_{R_{1}\protect\neq R_{2}}^{s}\right]_{ab}=0$
at saddle point }

In the formulation in Appendix \ref{sec:Short-Lagrangian} we have
that the energy functional is given by: 
\begin{align}
\mathcal{L}_{N} & =\left\langle \Psi_{0}\right|\varepsilon_{HF}+\left[\lambda_{R_{1}R_{2}}^{s}\right]_{ab}-\mu\left|\Psi_{0}\right\rangle \nonumber \\
 & -\sum_{R_{1}\neq R_{2},ab}\left[\lambda_{R_{1}R_{2}}^{s}\right]_{ab}\left[\Delta_{R_{1}R_{2}}\right]_{ab}-\nonumber \\
 & -\sum_{R_{1}\neq R_{2},\mu\nu}J_{R_{1};R_{2}}^{\mu;\nu}\sum_{i,j}\mathcal{Z}_{R_{1}\mu i}\mathcal{Z}_{R_{2}\nu i}\left[\Delta_{R_{1}R_{2}}\right]_{ab}\cdot\nonumber \\
 & \cdot\left[\Delta_{R_{2}R_{1}}\right]_{cd}+...\label{eq:Gutzwiller_functional}
\end{align}
Where $+....$ will not effect our derivations. We now have that extremizing
with respect to the variables in the Lagrange function: 
\begin{align}
0 & =\frac{\partial\mathcal{L}_{N}}{\partial\left[\Delta_{R_{1}\neq R_{2}}\right]_{ab}}=\sum_{\mu\nu}J_{R_{1};R_{2}}^{\mu;\nu}\sum_{i,j}\mathcal{Z}_{R_{1}\mu i}\mathcal{Z}_{R_{2}\nu i}\left[\Delta_{R_{2}R_{1}}\right]_{cd}\nonumber \\
 & +\sum_{\mu\nu}J_{R_{1};R_{2}}^{\mu;\nu}\sum_{i,j}\mathcal{Z}_{R_{1}\mu i}\mathcal{Z}_{R_{2}\nu i}\left\langle f_{R_{2}c}^{\dagger}f_{R_{1}d}\right\rangle -\nonumber \\
 & -\left[\lambda_{R_{1}R_{2}}^{s}\right]_{ab}=\left[\lambda_{R_{1}R_{2}}^{s}\right]_{ab}\label{eq:Lambda_equation}
\end{align}
This greatly simplifies the saddle point, and in particular the quasiparticle
Hamiltonian $H_{QP}$ see below.

\subsubsection{\label{sec:Introduction-2}Excitations}

We know that the proper way to study the quasiparticle energy is to
take the thermodynamic limit and to study the following wave function:
\begin{equation}
P_{G}\sum_{\alpha}\psi_{R\alpha}f_{\alpha}^{\dagger}\left(R\right)\left|\Psi_{Slater}\right\rangle =P_{G}\left|\Psi_{Slater}^{\psi}\right\rangle \label{eq:Wave_function}
\end{equation}
Now the single particle density matrix $\Delta_{RR;\alpha\beta}$
for $\left|\Psi_{Slater}^{\psi}\right\rangle $ is almost the same
as for $\left|\Psi_{Slater}\right\rangle $ plus $\frac{1}{N}$ corrections
(here $N$ is the number of atoms which blows up in the thermodynamic
limit). This means that the equivalence relations 
\begin{align}
O_{R\mu} & \sim\sum_{i}\mathcal{Z}_{R\mu i}^{\psi}O_{Ri}=\sum_{i}\mathcal{Z}_{R\mu i}O_{i}+O\left(\frac{1}{N}\right)\label{eq:correction}\\
\mathcal{Z}_{R\mu i}^{\psi} & =\mathcal{R}_{R\mu i}+O\left(\frac{1}{N}\right)
\end{align}
However we know that the energy functional is stationary with respect
to variations of $\mathcal{Z}_{R\mu i}$ which means that: 
\begin{equation}
Z\mathcal{L}_{N}\left(\mathcal{Z}_{R\mu i}^{\psi}\right)=\mathcal{L}_{N}\left(\mathcal{Z}_{R\mu i}\right)+O\left(N\cdot\frac{1}{N^{2}}\right)\label{eq:Variation}
\end{equation}
This means that we may as well ignore the changes in $\mathcal{Z}_{R\mu i}$
and set 
\begin{equation}
\mathcal{Z}_{R\mu i}^{\psi}\cong\mathcal{Z}_{R\mu i}\label{eq:No_change}
\end{equation}
As such the Hamiltonian for the original problem 
\begin{equation}
H=\sum_{R,R',\mu\nu}J_{R_{1};R_{2}}^{\mu;\nu}O_{R_{1}\mu}O_{R_{2}\nu}+H^{loc}+\sum_{R}\left[\lambda_{RR}^{s}\right]_{ab}f_{Ra}^{\dagger}f_{Rb}\label{eq:Hamitlonian}
\end{equation}
While $H^{loc}$ and the other terms in $\mathcal{L}_{N}$ depend
on $\left[\Delta_{R_{1}R_{2}}\right]_{ab}$ because of the stationarity
conditions their effect may be neglected. As such the problem of excitations
maps onto the problem of excitations of the following Hamiltonian:
\begin{align}
H_{eff} & =\sum_{R,R',\mu\nu}J_{R_{1};R_{2}}^{\mu;\nu}\sum_{i,j}\mathcal{Z}_{R\mu i}\mathcal{Z}_{R'\nu i}O_{R_{1}i}O_{R_{2}j}\nonumber \\
 & +\sum_{Rab}\left[\lambda_{RR}^{s}\right]_{ab}f_{Ra}^{\dagger}f_{Rb}\label{eq:Effective_Hamiltonian}
\end{align}
Which has no dependence on $\psi$ of any form. This means that the
excitation energy for the Gutzwiller Lagrange function is the same
as the excitation energy of $H_{eff}$ which is the Hartree Fock excitation
energy which is given by the eigenenergies of the Hamiltonian: 
\begin{align}
H_{QP} & =\sum_{R,R',\mu\nu}J_{R_{1};R_{2}}^{\mu;\nu}\sum_{i,j}\mathcal{Z}_{R_{1}\mu i}\mathcal{Z}_{R_{2}\nu i}\left[O_{R_{1}i}O_{R_{2}j}\right]_{HF}\nonumber \\
 & +\sum_{Rab}\left[\lambda_{RR}^{s}\right]_{ab}f_{Ra}^{\dagger}f_{Rn}\label{eq:Quasiparticle_Hamiltonian}
\end{align}

\section{\label{sec:Example:-single-band}Example: the extended single-band
Hubbard model}

\subsection{\label{subsec:Main-Hamiltonian}Hamiltonian and setup}

As a further example of the general formalism presented in Section~\ref{sec:Gutzwiller-Lagrangian},
here we consider the single-band extended Hubbard model \cite{Amaricci_2010,Zhang_1989,Yan_1993,Dogen_1994,Dogen_1994-1,Dogen_1996,Chattopadhyay,Nayak_2002,Aichorn_2004,Onari_2004,Schuler_2013,Ayral_2013,Terletska_2017,Rohringer_2018,Schiller_1995,Stanescu_2004,Lichtenstein_2000,biermann_2003,Si_1996,Chitra_2000,Werner_2010}:
\begin{align}
H_{SB}= & -t\sum_{\left\langle R,R'\right\rangle ,\sigma=\pm}\left(c_{R\sigma}^{\dagger}c_{R'\sigma}+h.c.\right)+U\sum_{R}n_{R\uparrow}n_{R\downarrow}\nonumber \\
 & +V\sum_{\left\langle R,R'\right\rangle }n_{R}n_{R'}-J\sum_{\left\langle R,R'\right\rangle }\vec{S}_{R}\cdot\vec{S}_{R'}+\nonumber \\
 & +X\sum_{\left\langle R,R'\right\rangle ,\sigma=\pm}\left(c_{R\sigma}^{\dagger}c_{R'\sigma}+h.c.\right)\left(n_{R-\sigma}+n_{R'-\sigma}\right)\nonumber \\
 & +Y\sum_{\left\langle R,R'\right\rangle }\left(c_{R\uparrow}^{\dagger}c_{R\downarrow}^{\dagger}c_{R'\downarrow}c_{R'\uparrow}+h.c.\right)-\mu\sum_{R\sigma}n_{R\sigma},\label{eq:Extended_Hubbard-1-1}
\end{align}
where $\left\langle R,R'\right\rangle $ denotes nearest neighbors
$R$ and $R'$ with the sum being over both $R$ and $R'$. In the
next section we will write the single-band Hubbard model Gutzwiller
Lagrange function, assuming translational invariance. We summarize
the relevant terms in Table \ref{tab:Symbols}.

\subsection{\label{sec:Gutzwiller-Lagrangian-1}Gutzwiller Lagrange function}
\begin{widetext}
Here we consider the extended single-band Hubbard model, see Eq. (\ref{eq:Extended_Hubbard-1-1}),
for an arbitrary Bravis lattice (e.g., square, cubic, triangular,
body centered cubic (BCC), face centered cubic (FCC)). Following our
derivations in Section \ref{sec:Gutzwiller-Lagrangian}, the single
band Gutzwiller Lagrange function is given by: 

\begin{align}
 & \mathcal{L}_{N}\left(\left\{ D_{\sigma},E_{\sigma},F_{\sigma a},F,G\right\} ,\left\{ D_{\sigma}^{*},E_{\sigma}^{*},F^{*},G^{*}\right\} ,\left\{ \mathcal{R}_{\sigma},\mathcal{S}_{\sigma},\mathcal{T}_{\sigma a},\mathcal{T},\mathcal{U}\right\} ,\left\{ \mathcal{R}_{\sigma}^{*},\mathcal{S}_{\sigma}^{*},\mathcal{T}^{*},\mathcal{U}^{*}\right\} \right.\nonumber \\
 & \left.\lambda_{\sigma},\lambda_{\sigma}^{\left(n.n.\right)},\lambda_{\sigma}^{c},\lambda_{\sigma}^{b},\Delta_{\sigma},\Delta_{\sigma}^{\left(n.n.\right)},o_{\sigma},E,\mu,\left|\Psi_{0}\right\rangle ,\left|\Phi\right\rangle \right)+\nonumber \\
 & =\mathcal{L}_{QP}\left(\left\{ \mathcal{R}_{\sigma},\mathcal{S}_{\sigma}\right\} ,\left\{ \mathcal{R}_{\sigma}^{*},\mathcal{S}_{\sigma}^{*}\right\} ,\lambda_{\sigma},E,\mu,\left|\Psi_{0}\right\rangle \right)=\nonumber \\
 & +\mathcal{L}_{Embed}\left(\left\{ D_{\sigma},E_{\sigma},F_{\sigma a},F,G\right\} ,\left\{ D_{\sigma}^{*},E_{\sigma}^{*},F^{*},G^{*}\right\} ,E^{c},\lambda_{\sigma}^{b},\Delta_{a},\left|\Phi\right\rangle \right)+\nonumber \\
 & +\mathcal{L}_{Mix}\left(\left\{ D_{\sigma},E_{\sigma},F_{\sigma a},F,G\right\} ,\left\{ D_{\sigma}^{*},E_{\sigma}^{*},F^{*},G^{*}\right\} ,\left\{ \mathcal{R}_{\sigma},\mathcal{S}_{\sigma},\mathcal{T}_{\sigma a},\mathcal{T},\mathcal{U}\right\} ,\left\{ \mathcal{R}_{\sigma}^{*},\mathcal{S}_{\sigma}^{*},\mathcal{T}^{*},\mathcal{U}^{*}\right\} ,\right.\nonumber \\
 & \left.\lambda_{\sigma},\lambda_{\sigma}^{\left(n.n.\right)},\lambda_{\sigma}^{c},\lambda_{\sigma}^{b},\Delta_{\sigma},\Delta_{\sigma}^{\left(n.n.\right)},o_{\sigma}\right)+\nonumber \\
 & +\mathcal{L}_{HF}\left(\left\{ \mathcal{T}_{\sigma a},\mathcal{T},\mathcal{U}\right\} ,\left\{ \mathcal{T}^{*},\mathcal{U}^{*}\right\} ,\Delta_{\sigma}^{\left(n.n.\right)},o_{\sigma}\right).\label{eq:Lagrangian-1}
\end{align}
Where: 
\begin{align}
 & \mathcal{L}_{QP}\left(\left\{ \mathcal{R}_{\sigma},\mathcal{S}_{\sigma}\right\} ,\left\{ \mathcal{R}_{\sigma}^{*},\mathcal{S}_{\sigma}^{*}\right\} ,\lambda_{\sigma},E,\mu,\left|\Psi_{0}\right\rangle \right)=\nonumber \\
 & =\left\langle \Psi_{0}\right|-t\sum_{\left\langle RR'\right\rangle \sigma}\mathcal{R}_{\sigma}\mathcal{R}_{\sigma}^{\ast}f_{R\sigma}^{\dagger}f_{R'\sigma}+\left[X\sum_{\left\langle RR'\right\rangle \sigma}\mathcal{S}_{\sigma}\mathcal{R}_{\sigma}^{*}f_{R\sigma}^{\dagger}f_{R'\sigma}+h.c.\right]+\nonumber \\
 & +\sum_{\left\langle RR'\right\rangle \sigma}\lambda_{\sigma}^{\left(n.n.\right)}f_{R\sigma}^{\dagger}f_{R'\sigma}+\sum_{R\sigma}\lambda_{\sigma}f_{R\sigma}^{\dagger}f_{R\sigma}-\mu\sum_{R\sigma}f_{R\sigma}^{\dagger}f_{R\sigma}\left|\Psi_{0}\right\rangle +E\left(1-\left\langle \Psi_{0}\mid\Psi_{0}\right\rangle \right)+\mu N,\label{eq:Quasiparticle_lagrangian-1}
\end{align}

\begin{equation}
\mathcal{L}_{Embed}\left(\left\{ D_{\sigma},E_{\sigma},F_{\sigma a},F,G\right\} ,\left\{ D_{\sigma}^{*},E_{\sigma}^{*},F^{*},G^{*}\right\} ,E^{c},\lambda_{\sigma}^{b},\Delta_{a},\left|\Phi\right\rangle \right)=\left\langle \Phi\right|H_{Embed}\left|\Phi\right\rangle +E^{c}\left(1-\left\langle \Phi\mid\Phi\right\rangle \right),\label{eq:Embed-1}
\end{equation}
\begin{align}
H_{embed} & =U\hat{c}_{\uparrow}^{\dagger}\hat{c}_{\uparrow}\hat{c}_{\downarrow}^{\dagger}\hat{c}_{\downarrow}+\sum_{\sigma}\lambda_{\sigma}^{c}\hat{f}_{\sigma}\hat{f}_{\sigma}^{\dagger}+\left[\sum_{\sigma}D_{\sigma}\hat{c}_{\sigma}^{\dagger}\hat{f}_{\sigma}+h.c.\right]\nonumber \\
 & +\left[\sum_{\sigma}E_{\sigma}\hat{c}_{\sigma}^{\dagger}\hat{c}_{\bar{\sigma}}^{\dagger}\hat{c}_{\bar{\sigma}}\hat{f}_{\sigma}+h.c.\right]+\sum_{\sigma a}F_{\sigma a}\left(\hat{c}_{\sigma}^{\dagger}\hat{c}_{\sigma}\hat{f}_{a}\hat{f}_{a}^{\dagger}-\hat{c}_{\sigma}^{\dagger}\hat{c}_{\sigma}\Delta_{a}\right)+\nonumber \\
 & +\left[F\hat{c}_{\uparrow}^{\dagger}\hat{c}_{\downarrow}\hat{f}_{\downarrow}\hat{f}_{\uparrow}^{\dagger}+h.c.\right]-\left[G\hat{c}_{\uparrow}^{\dagger}\hat{c}_{\downarrow}^{\dagger}\hat{f}_{\downarrow}\hat{f}_{\uparrow}+h.c.\right]+\sum_{\sigma}\lambda_{\sigma}^{b}\hat{c}_{\sigma}^{\dagger}\hat{c}_{\sigma},\label{eq:Embeeding_Hamiltonian}
\end{align}
\begin{align}
 & \mathcal{L}_{Mix}\left(\left\{ D_{\sigma},E_{\sigma},F_{\sigma a},F,G\right\} ,\left\{ D_{\sigma}^{*},E_{\sigma}^{*},F^{*},G^{*}\right\} ,\left\{ \mathcal{R}_{\sigma},\mathcal{S}_{\sigma},\mathcal{T}_{\sigma a},\mathcal{T},\mathcal{U}\right\} ,\left\{ \mathcal{R}_{\sigma}^{*},\mathcal{S}_{\sigma}^{*},\mathcal{T}^{*},\mathcal{U}^{*}\right\} ,\right.\nonumber \\
 & \left.\lambda_{\sigma},\lambda_{\sigma}^{\left(n.n.\right)},\lambda_{\sigma}^{c},\lambda_{\sigma}^{b},\Delta_{\sigma},\Delta_{\sigma}^{\left(n.n.\right)},o_{\sigma}\right)\nonumber \\
 & =-\sum_{\sigma}\left(\lambda_{\sigma}+\lambda_{\sigma}^{c}\right)\Delta_{\sigma}-\sum_{\sigma}\lambda_{\sigma}^{\left(n.n.\right)}\Delta_{\sigma}^{\left(n.n.\right)}-\sum_{\sigma}\lambda_{\sigma}^{b}o_{\sigma}-\sum_{\sigma}\left[D_{\sigma}\left(\mathcal{R}_{\sigma}\sqrt{\left(1-\Delta_{\sigma}\right)\Delta}_{\sigma}\right)+c.c.\right]\nonumber \\
 & -\sum_{\sigma}\left[E_{\sigma}\left(\left[\mathcal{S}_{\sigma}\sqrt{\left(1-\Delta_{\sigma}\right)\Delta}_{\sigma}\right]\right)+c.c.\right]-\sum_{\sigma a}F_{\sigma a}\left(1-\Delta_{a}\right)\Delta_{a}\mathcal{T}_{\sigma a}-\nonumber \\
 & -\left[F\sqrt{\left(1-\Delta_{\uparrow}\right)\Delta_{\uparrow}}\mathcal{T}\sqrt{\left(1-\Delta_{\downarrow}\right)\Delta_{\downarrow}}+c.c.\right]-\left[G\sqrt{\left(1-\Delta_{\uparrow}\right)\Delta_{\uparrow}}\mathcal{U}\sqrt{\left(1-\Delta_{\downarrow}\right)\Delta_{\downarrow}}+c.c.\right],\label{eq:Mixing_Lagrangian_single_band}
\end{align}

\begin{align}
 & \mathcal{L}_{HF}\left(\left\{ \mathcal{T}_{\sigma a},\mathcal{T},\mathcal{U}\right\} ,\left\{ \mathcal{T}^{*},\mathcal{U}^{*}\right\} ,\Delta_{\sigma}^{\left(n.n.\right)},o_{\sigma}\right)=\nonumber \\
 & =\frac{z}{2}V\left(\sum_{\sigma}o_{\sigma}\right)^{2}+\frac{z}{8}J\left(\sum_{\sigma}\sigma o_{\sigma}\right)^{2}+\frac{z}{2}Y\left(\sum_{\sigma}\Delta_{\sigma}^{\left(n.n.\right)}\Delta_{\bar{\sigma}}^{\left(n.n.\right)}\right)\mathcal{U}\cdot\mathcal{U}^{*}\nonumber \\
 & -\frac{z}{2}V\sum_{\sigma\sigma'aa}\Delta_{a}^{\left(n.n.\right)}\Delta_{a}^{\left(n.n\right)}\mathcal{T}_{\sigma a}\mathcal{T}_{\sigma'a}-\frac{z}{8}J\sum_{\sigma\sigma'a}\sigma\sigma'\Delta_{a}^{\left(n.n.\right)}\Delta_{a}^{\left(n.n\right)}\mathcal{T}_{\sigma a}\mathcal{T}_{\sigma'a}-zJ\left[\Delta_{\uparrow}^{\left(n.n.\right)}\Delta_{\downarrow}^{\left(n.n\right)}\mathcal{T}\mathcal{T}^{*}+c.c.\right].\label{eq:Hartree_fock}
\end{align}
In Appendix \ref{sec:Mott-Gap} below we show that, for this single-band
problem, it is possible to reduce the problem of extremizing the Lagrange
function $\mathcal{L}$ to the problem of minimizing the Gutzwiller
energy as a function of the double occupancy $\eta$. 
\begin{table}
\begin{widetext}
\label{tab:Symbols}
\end{widetext}

\begin{tabular}{|c|c|c|c|}
\hline 
Symbol & Meaning & Term & Gutzwiller equivalent (half filling)\tabularnewline
\hline 
\hline 
$t$ & Hopping & $-t\sum_{\left\langle R,R'\right\rangle ,\sigma=\pm}\left(c_{R\sigma}^{\dagger}c_{R'\sigma}+h.c.\right)$ & $-8t\eta\left(1-2\eta\right)\sum_{\left\langle R,R'\right\rangle ,\sigma=\pm}\left(c_{R\sigma}^{\dagger}c_{R'\sigma}+h.c.\right)$\tabularnewline
\hline 
$U$ & Hubbard U & $U\sum_{R}n_{R\uparrow}n_{R\downarrow}$ & $U\eta\sum_{R}1$\tabularnewline
\hline 
$V$ & Density density & $V\sum_{\left\langle R,R'\right\rangle }n_{R}n_{R'}$ & $V\sum_{\left\langle R,R'\right\rangle }\left(4\eta n_{R}+1-4\eta\right)\left(4\eta n_{R'}+1-4\eta\right)$\tabularnewline
\hline 
$X$ & Coulomb assisted hopping & $X\sum_{\left\langle R,R'\right\rangle ,\sigma=\pm}\left(c_{R\sigma}^{\dagger}c_{R'\sigma}+h.c.\right)\left(n_{R-\sigma}+n_{R'-\sigma}\right)$ & $-8X\eta\left(1-2\eta\right)\sum_{\left\langle R,R'\right\rangle ,\sigma=\pm}\left(c_{R\sigma}^{\dagger}c_{R'\sigma}+h.c.\right)$\tabularnewline
\hline 
$Y$ & Pair hopping & $Y\sum_{\left\langle R,R'\right\rangle }\left(c_{R\uparrow}^{\dagger}c_{R\downarrow}^{\dagger}c_{R'\downarrow}c_{R'\uparrow}+h.c.\right)$ & $16\eta^{2}Y\sum_{\left\langle R,R'\right\rangle }\left(c_{R\uparrow}^{\dagger}c_{R\downarrow}^{\dagger}c_{R'\downarrow}c_{R'\uparrow}+h.c.\right)$\tabularnewline
\hline 
$J$ & Exchange interaction & $-J\sum_{\left\langle R,R'\right\rangle }\vec{S}_{R}\cdot\vec{S}_{R'}$ & $-4J\left(1-2\eta\right)^{2}\sum_{\left\langle R,R'\right\rangle }\vec{S}_{R}\cdot\vec{S}_{R'}$\tabularnewline
\hline 
\end{tabular}

\caption{\textbf{The various terms entering the Hamiltonian in Eq. (\ref{eq:Extended_Hubbard-1-1})
as well as their operatorial equivalences, which are explained in
the main text.}}
\end{table}
\end{widetext}

\subsection{\label{subsec:More-detailed-treatment}Extended Hubbard model: \label{subsec:Half-Filling}half
filling}

We express the Gutzwiller operators in the so-called ``mixed-basis
representation'' \citep{Lanata_2015,Lanata_2016,Lanata_2017} defined
as follows: 
\begin{equation}
P_{R}=\sum_{\Gamma,n}[\Lambda_{R}]_{\Gamma n}\left|\Gamma_{R}\right\rangle \left\langle n_{R}\right|\,,\label{eq:Projector-1-1}
\end{equation}
Here the basis set we use is: 
\begin{align}
\left\{ \left|\Gamma_{R}\right\rangle \right\}  & =\left\{ \left|0\right\rangle ,\left|\uparrow\right\rangle ,\left|\downarrow\right\rangle ,\left|\uparrow\downarrow\right\rangle \right\} ,\nonumber \\
\left\{ \left|n_{R}\right\rangle \right\}  & =\left\{ \left|0\right\rangle ,\left|\uparrow\right\rangle ,\left|\downarrow\right\rangle ,\left|\uparrow\downarrow\right\rangle \right\} .\label{eq:Basis-2-1}
\end{align}
For clarity, here we briefly review some definitions in relation with
the notation utilized in previous work \citep{Li2009,Kotliar_1986,Lanata2009}.
We consider the projector \citep{Lanata2012,Lanata2009,Lanata_2015,Lanata_2016,Li2009,Sandri_2014}:
\begin{equation}
\Lambda_{R}=\left(\begin{array}{cccc}
a_{00} & 0 & 0 & 0\\
0 & a_{\uparrow0} & 0 & 0\\
0 & 0 & a_{0\downarrow} & 0\\
0 & 0 & 0 & g
\end{array}\right)\,,\label{eq:Projector-2}
\end{equation}
which we assume does not depend on $R$ as we have assumed that translational
invariance is preserved. We conveniently express the local reduced
density matrix of $|\Psi_{0}\rangle$ as follows: 
\begin{align}
\rho_{R}^{0} & \equiv Tr_{R'\neq R}\left|\Psi_{0}\right\rangle \left\langle \Psi_{0}\right|\nonumber \\
 & =\frac{1}{Z}\exp\left(-\sum_{\sigma}\left[\ln\left(\frac{1-\Delta_{RR\sigma}}{\Delta_{RR\sigma}}\right)\right]c_{R\sigma}^{\dagger}c_{R\sigma}\right)\,,\label{eq:Density_matrix-1-1}
\end{align}
where 
\begin{align}
[\Delta_{R_{1}R_{2}}]_{\sigma}=\left\langle \Psi_{0}\right|c_{R_{1}\sigma}^{\dagger}c_{R_{2}\sigma}\left|\Psi_{0}\right\rangle \label{deltadef-1}
\end{align}
$Z$ is a normalization constant insuring that $Tr\left[\rho_{R}^{0}\right]=1$.
In the same basis as Eq. (\ref{eq:Basis-2-1}) we may write: 
\begin{equation}
[P_{R}^{0}]_{nn'}=\left\langle n_{R}\mid\rho_{R}^{0}\mid n'_{R}\right\rangle \,.\label{eq:precise_for_beginers-1}
\end{equation}
We now assume that $P_{R}^{0}$ (with respect to the same basis set)
is given by \citep{Lanata2012,Lanata2009,Lanata_2015,Lanata_2016,Li2009,Sandri_2014}:

\begin{equation}
P_{R}^{0}=\left(\begin{array}{cccc}
\left(1-n_{\uparrow}\right)\left(1-n_{\downarrow}\right) & 0 & 0 & 0\\
0 & n_{\uparrow}\left(1-n_{\downarrow}\right) & 0 & 0\\
0 & 0 & n_{\downarrow}\left(1-n_{\uparrow}\right) & 0\\
0 & 0 & 0 & n_{\uparrow}n_{\downarrow}
\end{array}\right)\,,\label{eq:Density_matrix-2}
\end{equation}
Following Kotliar and Ruckenstein \citep{Kotliar_1986} we define
the matrix of slave boson amplitudes \citep{Lanata2012,Lanata2009,Lanata_2015,Lanata_2016,Li2009,Sandri_2014,Lanata_2017}:
\begin{equation}
\phi_{R}=\left(\begin{array}{cccc}
e & 0 & 0 & 0\\
0 & p_{\uparrow} & 0 & 0\\
0 & 0 & p_{\downarrow} & 0\\
0 & 0 & 0 & d
\end{array}\right)=\Lambda_{R}\sqrt{P_{R}^{0}}\,.\label{eq:density_matrix-1}
\end{equation}
We will assume that:

\begin{equation}
p_{\uparrow}=p_{\downarrow}=p\label{eq:No_magnetization-1}
\end{equation}
that is paramagnetism. We will solve this problem in the limit of
large co-ordinations number and using the $P_{R}^{\dagger}P_{R}-I$
expansion (see Appendices \ref{sec:Weak-coupling-expansion} and \ref{sec:The--scaling}).

The Gutzwiller constraints are that: 
\begin{align}
p_{\uparrow}^{2}+d^{2} & =\Delta_{\uparrow}\equiv\left\langle \Psi_{0}\right|f_{R\uparrow}^{\dagger}f_{R\uparrow}\left|\Psi_{0}\right\rangle \nonumber \\
p_{\downarrow}^{2}+d^{2} & =\Delta_{\downarrow}\equiv\left\langle \Psi_{0}\right|f_{R\downarrow}^{\dagger}f_{R\downarrow}\left|\Psi_{0}\right\rangle \nonumber \\
e^{2}+p_{\uparrow}^{2}+p_{\downarrow}^{2}+d^{2} & =1\label{eq:Constraints-1-1}
\end{align}
Furthermore as $\left[P_{R},\rho_{R0}P_{R}^{\dagger}\right]=0$ (this
is not the case in general) we have that: 
\begin{align}
\left\langle \Phi_{R}\right|c_{R\uparrow}^{\dagger}c_{R\uparrow}\left|\Phi_{R}\right\rangle \equiv n_{\uparrow} & =\Delta_{\uparrow}\nonumber \\
\left\langle \Phi_{R}\right|c_{R\downarrow}^{\dagger}c_{R\downarrow}\left|\Phi_{R}\right\rangle \equiv n_{\downarrow} & =\Delta_{\downarrow}\label{eq:Natural_basis-1-1}
\end{align}
Indeed we have that: 
\begin{align}
 & \left\langle \Phi_{R}\right|c_{R\alpha}^{\dagger}c_{R\beta}\left|\Phi_{R}\right\rangle \nonumber \\
 & =Tr\left[\rho_{R0}P_{R}^{\dagger}c_{R\alpha}^{\dagger}c_{R\beta}P_{R}\right]\nonumber \\
 & =Tr\left[\rho_{R0}P_{R}^{\dagger}P_{R}c_{R\alpha}^{\dagger}c_{R\beta}\right]+Tr\left[\left[P_{R},\rho_{R0}P_{R}^{\dagger}\right]c_{R\alpha}^{\dagger}c_{R\beta}\right]\nonumber \\
 & =Tr\left[\rho_{R0}c_{R\alpha}^{\dagger}c_{R\beta}\right]\label{eq:commutators}
\end{align}
We note that for general use we phrased the proof in a general language
applicable to multiband models. In particular we may now set $c_{\sigma}=f_{\sigma}$.
We note that the solutions to the Gutzwiller equivalences depend only
on $e,\,p_{\uparrow},\,p_{\downarrow},\,d$ as well as $\Delta_{\uparrow}$
and $\Delta_{\downarrow}$. Now because of Eq. (\ref{eq:Natural_basis-1-1})
we have that $\Delta_{\uparrow}$ and $\Delta_{\downarrow}$ are fixed
for all states with a fixed magnetization and occupation and because
we are considering only the paramagnetic case $\Delta_{\uparrow}=\Delta_{\downarrow}$.
Furthermore because of the Gutzwiller constraints we have that $e,\,p_{\uparrow}=\,p_{\downarrow}$
are functions of $d$ and the occupation and magnetization (which
we will ignore in the paramagnetic case). As such the renormalization
coefficients $\left\{ \mathcal{R}_{\sigma},\mathcal{S}_{\sigma},\mathcal{T}_{\sigma a},\mathcal{T},\mathcal{U}\right\} $
depend only on $d$ and on the occupation and magnetization (which
are fixed, with the magnetization being zero in this case) and not
explicitly on the wavefunction $\left|\Psi_{0}\right\rangle $. This
greatly facilitates analytic calculations in this special case. 

To assess the influence of inter-site interactions on the Mott physics,
in this subsection we study the Hamiltonian in Eq. (\ref{eq:Extended_Hubbard-1-1})
at half filling. Note that, because we are at half filling in the
normal phase, we have that $n_{\uparrow}=n_{\downarrow}=\frac{1}{2}$.

As shown in Appendix~\ref{sec:Mott-Gap}, for this system we obtain
following operator equivalences: 
\begin{align}
P_{R}^{\dagger}c_{R\sigma}P_{R} & \sim2\sqrt{2\eta\left(1-2\eta\right)}c_{R\sigma}\nonumber \\
P_{R}^{\dagger}n_{R}P_{R} & \sim4\eta n_{R}+\left(1-4\eta\right)I\nonumber \\
P_{R}^{\dagger}c_{R\uparrow}^{\dagger}c_{R\downarrow}^{\dagger}P_{R} & \sim4\eta c_{R\uparrow}^{\dagger}c_{R\downarrow}^{\dagger}\nonumber \\
P_{R}^{\dagger}\vec{S}_{R}P_{R} & \sim2\left(1-2\eta\right)\vec{S}_{R}\nonumber \\
P_{R}^{\dagger}c_{R\sigma}^{\dagger}n_{R-\sigma}P_{R} & \sim\sqrt{2\eta\left(1-2\eta\right)}c_{R\sigma}^{\dagger}.\label{eq:Equivalences-1}
\end{align}
This leads to the result that extremizing Eq.~(\ref{eq:Lagrangian-1})
amounts to minimize the following energy function of the local double
occupancy $\eta\equiv d^{2}$: 
\begin{align}
\left\langle H\right\rangle  & =\eta^{2}\left[32\left(t-X\right)z\chi-16\left(V-Y\right)z\chi^{2}-12zJ\chi^{2}\right]\nonumber \\
 & +\eta\left[-16\left(t-X\right)z\chi+12zJ\chi^{2}+U\right]\nonumber \\
 & +\frac{1}{2}Vz-3zJ\chi^{2}-\mu_{1/2}.\label{eq:Polynomial}
\end{align}
Here $\mu_{1/2}$ is the chemical potential needed to enforce that
the system is at half filling. For simplicity, from now on we will
assume that $J=0$, as typically $J\ll t-X,V-Y,U$~\citep{Fazekas_1999,Hubbard_1963}.
It is convenient to express our variables in terms of the following
dimensionless quantities: 
\begin{align}
u & =\frac{U}{16\left(t-X\right)z\chi}\nonumber \\
v & =\frac{16\left(V-Y\right)z\chi^{2}}{16\left(t-X\right)z\chi}.\label{eq:Rescaled}
\end{align}
The results are identical to Section \ref{sec:Example:-single-band-1}
except for rescaling of the variables and will be repeated for clarity.

Consistently with the fact that our theory reduces to the ordinary
GA in the limit of vanishing intersite interactions, at $v=0$ we
recover the Brinkman Rice transition \citep{Brinkman_1970}, where
$\eta=0$ (i.e., the charge fluctuations are frozen) for all $u\geq1$.
More generally the Brinkman Rice phase occurs when $u>1$ and $v<2u$.

\subsubsection{\label{subsec:Goldstein-Kotliar-Corssover-1}Metallic phase: enhanced-valence
crossover}

Minimizing the energy function {[}Eq.~(\ref{eq:Polynomial}){]} it
can be readily shown that for $u<1$ and $v<v_{c}=1+u$ the system
remains metallic and that, in this phase, the double occupancy is
given by: 
\begin{equation}
\eta=\frac{1-u}{4\left(1-v/2\right)}\,.\label{eq:sSoluton}
\end{equation}
Eq.~(\ref{eq:sSoluton}) shows that the intersite Coulomb interaction
can enhance dramatically charge fluctuations. In particular, we note
that $\eta$ can even exceed $\frac{1}{4}$ for $v_{c}>v>2u$ while
this is impossible in the half-filled single-band Hubbard model with
only local Hubbard repulsion. The line where $\eta=\frac{1}{4}$ we
refer to as the "enhanced-valence crossover,"

\subsubsection{\label{subsec:Goldstein-kotliar-Transition-1}Valence-skipping phase}

Remarkably, we find that the non-local Coulomb interaction can induce
a phase transition into a phase with double occupancy $\eta=\frac{1}{2}$,
which is stable for $v>1+u$ and $v>2u$. 

In this work we refer to this region as the "valence-skipping phase".
We note that we call this a valence skipping phase because the single
site density matrix in this phase is given by:
\begin{align}
\rho_{R} & =Tr_{R'\neq R}\left|\Psi\right\rangle \left\langle \Psi\right|=P_{R}\rho_{R}^{0}P_{R}^{\dagger}+O\left(\frac{1}{z}\right),\nonumber \\
P_{R}\rho_{R}^{0}P_{R}^{\dagger} & =\left(\begin{array}{cccc}
1/2 & 0 & 0 & 0\\
0 & 0 & 0 & 0\\
0 & 0 & 0 & 0\\
0 & 0 & 0 & 1/2
\end{array}\right)\label{eq:Valence_skipping}
\end{align}
For $v>2$ there is a first order phase transition between the Brinkman
Rice phase \citep{Brinkman_1970} and the valence skipping phase while
for $v<2$ there is a second order phase transition between the metallic
phase and the valence skipping phase. The order of the transition
is signified by the continuity or discontinuity of $\eta$ across
these phase transitions. $\left(u,v\right)=\left(1,2\right)$ is a
multiciritical point.

\subsection{\label{subsec:General-dispersion}Quasiparticle dispersion}

In this Section we restore $J\neq0$. We have that the quasiparticle
dispersion is given by (see Appendix \ref{sec:Quasiparticle-energies}):
\begin{align}
H_{QP}= & \sum_{\left\langle R,R'\right\rangle ,\sigma}c_{R\sigma}^{\dagger}c_{R'\sigma}[-8\eta\left(1-2\eta\right)\left(t-X\right)\nonumber \\
 & -(3\left(1-2\eta\right)^{2}J+16\eta^{2}\left(V-Y\right))\chi]\nonumber \\
 & \equiv-Zt\sum_{\left\langle R,R'\right\rangle ,\sigma}c_{R\sigma}^{\dagger}c_{R'\sigma}+\mathscr{\varDelta}t\sum_{\left\langle R,R'\right\rangle ,\sigma}c_{R\sigma}^{\dagger}c_{R'\sigma},\label{eq:dispersion}
\end{align}
where 
\begin{align}
Z & =8\eta\left(1-2\eta\right)\nonumber \\
\varDelta & =-8\eta\left(1-2\eta\right)\frac{X}{t}\nonumber \\
 & -\left[3\left(1-2\eta\right)^{2}\frac{J}{t}+16\eta^{2}\frac{\left(V-Y\right)}{t}\right]\chi.\label{eq:Definitions}
\end{align}
Now we follow \citep{Lanata_2015,Bunemann_2003} and write

\begin{align}
G^{coh} & =\frac{Z}{i\omega_{n}-Z\varepsilon\left(k\right)+\varDelta\varepsilon\left(k\right)+\bar{\mu}_{1/2}}\nonumber \\
 & =\frac{1}{i\omega_{n}-\varepsilon\left(k\right)+\mu_{1/2}-\Sigma\left(k,\omega\right)},\label{eq:Self_energy-1}
\end{align}
where $\bar{\mu}_{1/2}$ is the chemical potential at half filling
for the quasiparticle Hamiltonian, so that: 
\begin{equation}
\Sigma\left(k,\omega_{n}\right)=-\frac{\varDelta}{Z}\varepsilon\left(k\right)+\mu_{1/2}+i\omega_{n}\left[1-\frac{1}{Z}\right]-\frac{\bar{\mu}_{1/2}}{Z}.\label{eq:Chemical_simplified}
\end{equation}
We note that, consistently with previous GW+DMFT studies of the two
dimensional extended Hubbard model \citep{Ayral_2013,Terletska_2017},
the self energy is a function of both momentum and frequency. Furthermore
we note that $G^{coh}$ is the coherent part of the Green's functions
\cite{Goldstein_2020}. Indeed the coherent part of the Greens function
satisfies:
\begin{align}
G^{coh}\left(k,\omega\right) & =\int\frac{A^{coh}\left(k,\nu\right)}{\omega-\nu}d\nu\nonumber \\
A^{coh}\left(k,\omega\right) & =Z\delta\left(\omega-\left[Z\varepsilon\left(k\right)-\varDelta\varepsilon\left(k\right)-\bar{\mu}_{1/2}\right]\right)\nonumber \\
\int A^{coh}\left(k,\omega\right)d\omega & =Z<1\label{eq:Z_coherent}
\end{align}
 The missing quasiparticle weight $1-Z$ is provided by the incoherent
piece \cite{Goldstein_2020}.

\section{\label{sec:Mott-Gap}Mott Gap}

\subsection{\label{Mott-Gap:-Beyond}Mott gap}

As a benchmark of our theory, in this subsection we calculate the
Mott gap (defined as the jump in chemical potential across the Mott
transition) and compare our results with Ref.~\citep{Lavagna_1990}.
As shown below, extremizing Eq.~(\ref{eq:Lagrangian-1}), we obtain
that the Mott gap is given by: 
\begin{equation}
\Delta\mu=\bar{U}\sqrt{1-\frac{\bar{U}_{c}}{\bar{U}}}\,,\label{eq:Mott_Jump}
\end{equation}
where 
\begin{align}
\bar{U} & =U+12Jz\chi^{2}\nonumber \\
\bar{U}_{c} & =16z\chi(t-X)
\,.\label{eq:redefinitions-1}
\end{align}
We note that Eq.~(\ref{eq:Mott_Jump}) reduces to the result of Ref.~\citep{Lavagna_1990}
for the standard Hubbard model when we set the intersite interactions
to zero, where, by definition, $\bar{U}=U$ and $\bar{U}_{c}={U}_{c}$.
The details of the calculations are given below.

\subsection{\label{subsec:General-Setup}Expectation values and equivalence relationships}

We want to solve for the Mott gap in the Brinkman Rice transition.
We will consider the generic single band Hubbard Hamiltonian with
nearest neighbor interactions given in Eq. (\ref{eq:Extended_Hubbard-1-1}).
We will assume that

\begin{equation}
p_{\uparrow}=p_{\downarrow}=p\label{eq:No_magnetization}
\end{equation}
In this case the Gutzwiller equivalences (which may be obtained from
the Lagrange multiplier terms in the Lagrangian in Eq. (\ref{eq:Lagrangian-1}))
are given by: 
\begin{align}
P_{R}^{\dagger}c_{\alpha}P_{R} & \sim\frac{\left(e+d\right)p}{\sqrt{1-d^{2}-p^{2}}\sqrt{1-e^{2}-p^{2}}}c_{\alpha}\nonumber \\
P_{R}^{\dagger}nP_{R} & \sim\frac{\left[e^{2}-d^{2}+\frac{d^{2}}{p^{2}+d^{2}}\right]}{\left(1-p^{2}-d^{2}\right)}n+\nonumber \\
 & \left[+2p^{2}+2d^{2}-\right.\nonumber \\
 & \left.-2\left[\frac{\left[e^{2}-d^{2}+\frac{d^{2}}{p^{2}+d^{2}}\right]}{\left(1-p^{2}-d^{2}\right)}\right]\left(p^{2}+d^{2}\right)\right]I\nonumber \\
P_{R}^{\dagger}c_{\uparrow}^{\dagger}c_{\downarrow}^{\dagger}P_{R} & \sim\frac{de}{\left(1-p^{2}-d^{2}\right)\left(p^{2}+d^{2}\right)}c_{\uparrow}^{\dagger}c_{\downarrow}^{\dagger}\nonumber \\
P_{R}^{\dagger}\vec{S}P_{R} & \sim\frac{p^{2}}{\left(1-p^{2}-d^{2}\right)\left(p^{2}+d^{2}\right)}\vec{S}\nonumber \\
P_{R}^{\dagger}c_{\alpha}^{\dagger}n_{\bar{\alpha}}P_{R} & \sim\frac{dp}{\sqrt{1-d^{2}-p^{2}}\sqrt{1-e^{2}-p^{2}}}c_{\alpha}^{\dagger}\nonumber \\
P_{R}^{\dagger}n_{\uparrow}n_{\downarrow}P_{R} & \sim\left(\frac{d^{2}}{p^{2}+d^{2}}\right)n-d^{2}I\label{eq:Equivalences}
\end{align}
Indeed one can take derivatives with respect to $D_{\sigma},\,E_{\sigma},\,F_{\sigma\sigma'},\,F$
and set them to zero to obtain Eq. (\ref{eq:Equivalences}).

\subsection{\label{subsec:Gutzwiller-Lagrange-function}Gutzwiller Lagrange function}
\begin{widetext}
We now calculate the Gutzwiller Lagrange function, we follow the notation
in \citep{Lavagna_1990} which facilities comparison. In this notation
using the Gutzwiller equivalences in Eq. (\ref{eq:Equivalences})
the Gutzwiller energy functional is given by: 
\begin{align}
\mathcal{L} & =-2\frac{\left(e+d\right)^{2}p^{2}}{\left(1-d^{2}-p^{2}\right)\left(1-e^{2}-p^{2}\right)}tz\chi+Ud^{2}\nonumber \\
 & +2V\left(\left\langle c^{\dagger}c\right\rangle \right)^{2}-\frac{3}{4}\left[\frac{p^{2}}{\left(1-p^{2}-d^{2}\right)\left(p^{2}+d^{2}\right)}\right]^{2}zJ\chi^{2}-\left(\frac{\left[e^{2}-d^{2}+\frac{d^{2}}{p^{2}+d^{2}}\right]}{\left(1-p^{2}-d^{2}\right)}\right)^{2}Vz\chi^{2}\nonumber \\
 & +Yz\frac{d^{2}e^{2}}{\left(1-p^{2}-d^{2}\right)^{2}\left(p^{2}+d^{2}\right)^{2}}\chi^{2}+4\frac{d\left(e+d\right)p^{2}}{\left(1-d^{2}-p^{2}\right)\left(1-e^{2}-p^{2}\right)}z\chi X\nonumber \\
 & +\lambda^{\left(1\right)}\left(e^{2}+2p^{2}+d^{2}-1\right)+\lambda^{\left(2\right)}\left(2p^{2}+2d^{2}-2\left\langle f^{\dagger}f\right\rangle \right)-\mu\left(n-2\left\langle f^{\dagger}f\right\rangle \right)\label{eq:Ground_state-1}
\end{align}
We now replace $\left\langle c_{\sigma}^{\dagger}c_{\sigma}\right\rangle =\left\langle f^{\dagger}f\right\rangle $
(see Eqs. (\ref{eq:Natural_basis-1-1})) which is special to the case
at hand. Here we have introduced the same notation as \citep{Lavagna_1990}
and have gotten rid of the terms enforcing the Gutzwiller constraints
as they are of Lagrange multiplier form and do not contribute to the
total energy and replaced the coefficients $\mathcal{R}_{\mu i}$
with their saddle point values. Furthermore since there is no magnetization
we have simplified our calculations to the case of spinless fermions
(by doubling certain terms). Using the Gutzwiller constraints we have
that 
\begin{align}
p^{2} & =\frac{n}{2}-d^{2}\nonumber \\
e^{2} & =1-n+d^{2}\label{eq:Solutions}
\end{align}
We now have that the energy functional is given by: 
\begin{align}
\left\langle H\right\rangle  & =-2\frac{\left(\sqrt{1-n+d^{2}}+d\right)^{2}\left(\frac{n}{2}-d^{2}\right)}{\left(1-\frac{n}{2}\right)\left(\frac{n}{2}\right)}tz\chi+Ud^{2}\nonumber \\
 & +\frac{1}{2}Vn^{2}z-\frac{3}{4}\left[\frac{\left(\frac{n}{2}-d^{2}\right)}{\left(1-\frac{n}{2}\right)\left(\frac{n}{2}\right)}\right]^{2}zJ\chi^{2}-\left(\frac{\left[\left(1-n\right)+\frac{d^{2}}{\frac{n}{2}}\right]}{\left(1-\frac{n}{2}\right)}\right)^{2}Vz\chi^{2}\nonumber \\
 & +Yz\frac{d^{2}\left(1-n+d^{2}\right)}{\left(1-\frac{n}{2}\right)^{2}\left(\frac{n}{2}\right)^{2}}\chi^{2}+4\frac{d\left(\sqrt{1-n+d^{2}}+d\right)\left(\frac{n}{2}-d^{2}\right)}{\left(1-\frac{n}{2}\right)\left(\frac{n}{2}\right)}z\chi X\label{eq:Energy-2}
\end{align}
We now assume that $n=1+\delta$ with $\delta,d\ll1$ then we have
that 
\begin{align}
\left\langle H\right\rangle  & =-8\frac{\left(\sqrt{d^{2}-\delta}+d\right)^{2}\left(\frac{1+\delta}{2}-d^{2}\right)}{1-\delta^{2}}tz\chi+Ud^{2}\nonumber \\
 & +\frac{1}{2}Vn^{2}z-12\left[\frac{\left(\frac{1+\delta-2d^{2}}{2}\right)}{1-\delta^{2}}\right]^{2}zJ\chi^{2}-2\left(\frac{\left[-\delta+2\frac{d^{2}}{1+\delta}\right]}{\left(1-\delta\right)}\right)^{2}Vz\chi^{2}\nonumber \\
 & +16Yz\frac{d^{2}\left(d^{2}-\delta\right)}{\left(1-\delta^{2}\right)^{2}}\chi^{2}+16\frac{d\left(\sqrt{d^{2}-\delta}+d\right)\left(\frac{1+\delta}{2}-d^{2}\right)}{1-\delta^{2}}z\chi X\label{eq:Energy_d_delta}
\end{align}
Now we introduce the variable $x=e+d$ then we have that \citep{Vollhardt_1987}:
\begin{align}
d^{2} & =\left(\frac{x^{2}+\delta}{2x}\right)^{2}\label{eq:Relation_x_d}\\
p^{2} & =\frac{1+\delta}{2}-\left(\frac{x^{2}+\delta}{2x}\right)^{2}\\
e^{2} & =\left(\frac{x^{2}+\delta}{2x}\right)^{2}-\delta
\end{align}
Then we have that \citep{Vollhardt_1987}: 
\begin{align}
\left\langle H\right\rangle  & =-2\left[1-\frac{\left(1-x^{2}\right)^{2}}{1-\delta^{2}}\right]tz\chi+U\left(\frac{x^{2}+\delta}{2x}\right)^{2}\nonumber \\
 & +\frac{1}{2}V\left(1+\delta\right)^{2}z-\frac{3}{4}\left[\frac{2x^{2}-\left(x^{4}+\delta^{2}\right)}{x^{2}\left[1-\delta^{2}\right]}\right]^{2}zJ\chi^{2}-\left(\frac{\left[-\delta^{2}2x^{2}+\left(x^{4}+\delta^{2}\right)\right]}{\left(1-\delta^{2}\right)x^{2}}\right)^{2}Vz\chi^{2}\nonumber \\
 & +Yz\frac{\left(x^{4}-\delta^{2}\right)^{2}}{\left(1-\delta^{2}\right)^{2}x^{4}}\chi^{2}+2\frac{\left(x^{2}+\delta\right)\left(2x^{2}-\left(x^{4}+\delta^{2}\right)\right)}{\left[1-\delta^{2}\right]x^{2}}z\chi X\label{eq:Energy_function}
\end{align}
We now introduce $y=x^{2}$ and write: 
\begin{align}
\left\langle H\right\rangle  & =-2\left[1-\frac{\left(1-y\right)^{2}}{1-\delta^{2}}\right]tz\chi+U\left(\frac{y}{4}+\frac{\delta}{2}+\frac{\delta^{2}}{4y}\right)\nonumber \\
 & +\frac{1}{2}V\left(1+\delta\right)^{2}z-\frac{3}{4}\left[\frac{2y-\left(y^{2}+\delta^{2}\right)}{y\left[1-\delta^{2}\right]}\right]^{2}zJ\chi^{2}-\left(\frac{\left[-\delta^{2}2y+\left(y^{2}+\delta^{2}\right)\right]}{\left(1-\delta^{2}\right)y}\right)^{2}Vz\chi^{2}\nonumber \\
 & +Yz\frac{\left(y^{2}-\delta^{2}\right)^{2}}{\left(1-\delta^{2}\right)^{2}y^{2}}\chi^{2}+2\left[1-\frac{\left(1-y\right)^{2}}{1-\delta^{2}}+\frac{\delta}{y}\left[1-\frac{\left(1-y\right)^{2}}{1-\delta^{2}}\right]\right]z\chi X\label{eq:Energy_function_y}
\end{align}
Now minimizing the energy with respect to $y$ we get that: 
\begin{align}
\frac{\partial H}{\partial y} & =-4\left[\frac{\left(1-y\right)}{1-\delta^{2}}\right]\left(t-X\right)z\chi+U\left(\frac{1}{4}-\frac{\delta^{2}}{4y^{2}}\right)-\frac{3}{2}\left[\frac{2y-\left(y^{2}+\delta^{2}\right)}{y\left[1-\delta^{2}\right]}\right]\left[\frac{2-2y}{y\left[1-\delta^{2}\right]}-\frac{2y-\left(y^{2}+\delta^{2}\right)}{y^{2}\left[1-\delta^{2}\right]}\right]zJ\chi^{2}\nonumber \\
 & +2\left(\frac{\left[-\delta^{2}2y+\left(y^{2}+\delta^{2}\right)\right]}{\left(1-\delta^{2}\right)y}\right)\left[\frac{2}{y}-\frac{\left[-\delta^{2}2y+\left(y^{2}+\delta^{2}\right)\right]}{\left(1-\delta^{2}\right)y^{2}}\right]Vz\chi^{2}+Yz\left[4\frac{\left(y^{2}-\delta^{2}\right)}{\left(1-\delta^{2}\right)^{2}y}-2\frac{\left(y^{2}-\delta^{2}\right)^{2}}{\left(1-\delta^{2}\right)^{2}y^{3}}\right]\chi^{2}\nonumber \\
 & \left[-2\frac{\delta}{y^{2}}\left[1-\frac{\left(1-y\right)^{2}}{1-\delta^{2}}\right]+4\frac{\delta}{y}\left[\frac{\left(1-y\right)}{1-\delta^{2}}\right]\right]z\chi X=0\label{eq:Minimum_equation}
\end{align}
We now take $y,\delta\ll1$, with this $\frac{\partial H}{\partial y}$
is approximately given by: 
\begin{align}
\frac{\partial H}{\partial y} & =-4\left(t-X\right)z\chi+U\left(\frac{1}{4}-\frac{\delta^{2}}{4y^{2}}\right)\nonumber \\
 & -3\left[\frac{\left(-y^{2}+\delta^{2}\right)}{y^{2}}\right]zJ\chi^{2}=0\label{eq:Derivative}
\end{align}
Where we have taken: 
\begin{equation}
y=O\left(\delta\right),\label{eq:Order}
\end{equation}
and have kept only the order one terms, the rest will not contribute
to the Mott gap. Where we used approximations like: 
\begin{equation}
-2\frac{\delta}{y^{2}}\left[1-\frac{\left(1-y\right)^{2}}{1-\delta^{2}}\right]\cong-4\frac{\delta}{y}\label{eq:Approximation}
\end{equation}
We now rewrite Eq. (\ref{eq:Derivative}) as: 
\begin{equation}
y^{2}\left[-4\left(t-X\right)z\chi+\frac{U}{4}+3zJ\chi^{2}\right]=\delta^{2}\left[\frac{U}{4}+3zJ\chi^{2}\right]\label{eq:Final_equation-1}
\end{equation}
In this case: 
\begin{equation}
y=\left|\delta\right|\sqrt{\frac{\left[U+12zJ\chi^{2}\right]}{\left[U+12zJ\chi^{2}-16\left(t-X\right)z\chi\right]}}\equiv\frac{\left|\delta\right|}{\varsigma}\label{eq:Solution}
\end{equation}
From this we see that 
\begin{equation}
U_{c}=16\left(t-X\right)z\chi-12zJ\chi^{2}\label{eq:Critical}
\end{equation}
is the critical onsite Hubbard interaction where the Mott gap opens.
This matches with Eq. (\ref{eq:redefinitions-1}) from the main text.
We now write \citep{Lavagna_1990} 
\begin{align}
y_{\pm} & =\frac{\left|\delta\right|}{\varsigma}\nonumber \\
e^{2} & =\left(\frac{y}{4}+\frac{\delta}{2}+\frac{\delta^{2}}{4y}\right)-\delta=\left(\frac{\left|\delta\right|}{4\varsigma}+\frac{\left|\delta\right|\varsigma}{4}-\frac{\delta}{2}\right)=\varsigma_{\pm e}\left|\delta\right|\nonumber \\
d^{2} & =\left(\frac{y}{4}+\frac{\delta}{2}+\frac{\delta^{2}}{4y}\right)=\left(\frac{\left|\delta\right|}{4\varsigma}+\frac{\left|\delta\right|\varsigma}{4}+\frac{\delta}{2}\right)=\varsigma_{\pm d}\left|\delta\right|\nonumber \\
p^{2} & =\frac{1}{2}-\left(\frac{\left|\delta\right|}{4\varsigma}+\frac{\left|\delta\right|\varsigma}{4}\right),\label{eq:Relations-2}
\end{align}
Where 
\begin{align}
\sqrt{\varsigma_{\pm d}} & =\sqrt{\frac{1}{4\varsigma}+\frac{\varsigma}{4}\pm\frac{1}{2}}\nonumber \\
\sqrt{\varsigma_{\pm e}} & =\sqrt{\frac{1}{4\varsigma}+\frac{\varsigma}{4}\mp\frac{1}{2}}.\label{eq:Equation-1}
\end{align}
\end{widetext}

\subsection{\label{subsec:The-Mott-gap}The Mott gap}

Now going back to Eq. (\ref{eq:Ground_state-1}) we get that: 
\begin{align*}
\left\langle H\right\rangle  & =\left\langle H_{0}\right\rangle +\lambda^{\left(1\right)}\left(e^{2}+2p^{2}+d^{2}-1\right)\\
 & +\lambda^{\left(2\right)}\left(2p^{2}+2d^{2}-2\left\langle f^{\dagger}f\right\rangle \right)-\mu\left(n-2\left\langle f^{\dagger}f\right\rangle \right)
\end{align*}
We then have that

\begin{align}
\left\langle H_{0}\right\rangle  & =-2\frac{\left(e+d\right)^{2}p^{2}}{\left(1-d^{2}-p^{2}\right)\left(1-e^{2}-p^{2}\right)}tz\chi+Ud^{2}\nonumber \\
 & +2V\left(\left\langle c^{\dagger}c\right\rangle \right)^{2}-\frac{3}{4}\left[\frac{p^{2}}{\left(1-p^{2}-d^{2}\right)\left(p^{2}+d^{2}\right)}\right]^{2}zJ\chi^{2}\nonumber \\
 & -\left(\frac{\left[e^{2}-d^{2}+\frac{d^{2}}{p^{2}+d^{2}}\right]}{\left(1-p^{2}-d^{2}\right)}\right)^{2}Vz\chi^{2}\nonumber \\
 & +Yz\frac{d^{2}e^{2}}{\left(1-p^{2}-d^{2}\right)^{2}\left(p^{2}+d^{2}\right)^{2}}\chi^{2}\nonumber \\
 & +4\frac{d\left(e+d\right)p^{2}}{\left(1-d^{2}-p^{2}\right)\left(1-e^{2}-p^{2}\right)}z\chi X\label{eq:Hamiltonian_energy}
\end{align}
Furthermore: 
\begin{equation}
0=\frac{\partial\left\langle H\right\rangle }{\partial\left\langle f^{\dagger}f\right\rangle }\Rightarrow\mu=\lambda^{\left(2\right)}\label{eq:Equation}
\end{equation}
\begin{align}
0 & =\frac{\partial\left\langle H\right\rangle }{\partial e}\Rightarrow\lambda^{\left(1\right)}=-\frac{\frac{\partial\left\langle H_{0}\right\rangle }{\partial e}}{2e}\nonumber \\
0 & =\frac{\partial\left\langle H\right\rangle }{\partial d}\Rightarrow2d\left(\lambda^{\left(1\right)}+2\lambda^{\left(2\right)}\right)=-\frac{\partial\left\langle H_{0}\right\rangle }{\partial d}\nonumber \\
 & \Rightarrow\lambda^{\left(2\right)}=-\frac{\frac{\partial\left\langle H_{0}\right\rangle }{\partial d}}{4d}+\frac{\frac{\partial\left\langle H_{0}\right\rangle }{\partial e}}{4e}\label{eq:Gap-2}
\end{align}
Now we take: 
\begin{align}
\frac{\partial\left\langle H_{0}\right\rangle }{\partial d} & =-8\left(e+d\right)tz\chi+2Ud+8\left[e+2d\right]z\chi X+...\nonumber \\
\frac{\partial\left\langle H_{0}\right\rangle }{\partial e} & =-8\left(e+d\right)tz\chi+8dz\chi X+......\label{eq:Derivatives}
\end{align}
Therefore: 
\begin{align}
\lambda_{\pm}^{\left(2\right)} & =const-2\left(t-X\right)z\chi\left[\frac{e}{d}-\frac{d}{e}\right]+.......\nonumber \\
 & =const-2\left(t-X\right)z\chi\left[\frac{\sqrt{\varsigma_{\pm e}}}{\sqrt{\varsigma_{\pm d}}}-\frac{\sqrt{\varsigma_{\pm d}}}{\sqrt{\varsigma_{\pm e}}}\right]\label{eq:Lanbda+-}
\end{align}
Form this we see that: 
\begin{equation}
\Delta\mu=\lambda_{+}^{\left(2\right)}-\lambda_{-}^{\left(2\right)}\label{eq:Jump}
\end{equation}
This means that 
\begin{align}
\Delta\mu & =4\left(\left(t-X\right)z\chi\right)\left[\frac{\sqrt{\varsigma_{+d}}}{\sqrt{\varsigma_{+e}}}-\frac{\sqrt{\varsigma_{+e}}}{\sqrt{\varsigma_{+d}}}\right]\nonumber \\
 & =\frac{16\left(\left(t-X\right)z\chi\right)}{\sqrt{\frac{\left[U+12zJ\chi^{2}\right]}{\left[U+12zJ\chi^{2}-16\left(t-X\right)z\chi\right]}}-\sqrt{\frac{\left[U+12zJ\chi^{2}-16\left(t-X\right)z\chi\right]}{\left[U+12zJ\chi^{2}\right]}}}\label{eq:Gap}
\end{align}

\subsubsection{\label{subsec:Simplifying-the-expressions}Simplifying the expressions}

We write: 
\begin{align}
\bar{U} & =U+12Jz\chi^{2}\nonumber \\
\bar{t} & =t-X\label{eq:redefinitions-1-1}
\end{align}
Then 
\begin{equation}
\Delta\mu=\frac{16z\chi\bar{t}}{\sqrt{\frac{\bar{U}}{\bar{U}-16z\chi\bar{t}}}-\sqrt{\frac{\bar{U}-16z\chi\bar{t}}{\bar{U}}}}\label{eq:Gap-1}
\end{equation}
We now introduce 
\begin{equation}
\bar{U}_{c}=16z\chi\bar{t}\label{eq:Critical-1}
\end{equation}
Then 
\begin{equation}
\Delta\mu=\sqrt{\bar{U}\left(\bar{U}-\bar{U}_{c}\right)}\label{eq:Gap-3}
\end{equation}
Then there are two cases to consider $\delta\bar{U}\equiv\bar{U}-\bar{U}_{c}\ll\bar{U}_{c}$
and $\bar{U}\gg\bar{U}_{c}$

\paragraph{\label{subsec:Case}Case $\delta\bar{U}\ll\bar{U}_{c}$}

Then

\begin{equation}
\Delta\mu\cong\sqrt{U_{c}}\sqrt{\delta U}\label{eq:Square_root}
\end{equation}

\paragraph{\label{subsec:Case-1}Case $\bar{U}\gg\bar{U}_{c}$}

Then

\begin{equation}
\Delta\mu\cong\bar{U}-\frac{1}{2}\bar{U}_{c}\label{eq:Square_root-1}
\end{equation}

\section{\label{sec:Calculating-the-embedding}Calculating the embedding Hamiltonian}

We would like to qualitatively motivate the results of the previous
section that the gap does not depend on $V$ and $Y$ by studying
the embedding Hamiltonian \citep{Lanata_2015,Lanata_2016,Lanata_2017,Lanata_2017(2)}.
To do so we note that for cases when some of the variables $\left\{ \mathcal{R}_{\sigma},\mathcal{S}_{\sigma},\mathcal{T}_{\sigma a},\mathcal{T},\mathcal{U}\right\} ,\left\{ \mathcal{R}_{\sigma}^{*},\mathcal{S}_{\sigma}^{*},\mathcal{T}^{*},\mathcal{U}^{*}\right\} $
become zero the corresponding piece of the embedding Hamiltonian becomes
zero. Indeed taking derivatives we get that: 
\begin{widetext}
\begin{align}
\frac{\partial\mathcal{L}}{\partial\mathcal{R}_{\sigma}} & =0\Rightarrow0=\left\langle \Psi_{0}\right|-t\sum_{\left\langle RR'\right\rangle \sigma}\mathcal{R}_{\sigma}^{\ast}f_{R\sigma}^{\dagger}f_{R'\sigma}+X\sum_{\left\langle RR'\right\rangle \sigma}\mathcal{S}_{\sigma}^{*}f_{R\sigma}^{\dagger}f_{R'\sigma}\left|\Psi_{0}\right\rangle =D_{\sigma}\sqrt{\left(1-\Delta_{\sigma}\right)\Delta}_{\sigma}\nonumber \\
\frac{\partial\mathcal{L}}{\partial\mathcal{S}_{\sigma}} & =0\Rightarrow0=\left\langle \Psi_{0}\right|X\sum_{\left\langle RR'\right\rangle \sigma}\mathcal{R}_{\sigma}^{*}f_{R\sigma}^{\dagger}f_{R'\sigma}\left|\Psi_{0}\right\rangle =E_{\sigma}\sqrt{\left(1-\Delta_{\sigma}\right)\Delta}_{\sigma}\nonumber \\
\frac{\partial\mathcal{L}}{\partial\mathcal{T}} & =0\Rightarrow0=-zJ\Delta_{\uparrow}^{\left(n.n.\right)}\Delta_{\downarrow}^{\left(n.n\right)}\mathcal{T}^{*}=F\sqrt{\left(1-\Delta_{\uparrow}\right)\Delta_{\uparrow}}\sqrt{\left(1-\Delta_{\downarrow}\right)\Delta_{\downarrow}}\nonumber \\
\frac{\partial\mathcal{L}}{\partial\mathcal{U}} & =0\Rightarrow0=\frac{z}{2}Y\left(\sum_{\sigma}\Delta_{\sigma}^{\left(n.n.\right)}\Delta_{\bar{\sigma}}^{\left(n.n.\right)}\right)\mathcal{U}=G\sqrt{\left(1-\Delta_{\uparrow}\right)\Delta_{\uparrow}}\sqrt{\left(1-\Delta_{\downarrow}\right)\Delta_{\downarrow}}\label{eq:Lagrange_equations_simple}
\end{align}

From this we see that $G,\,F,\,E_{\sigma},\,D_{\sigma}$ vanish whenever
$\mathcal{U},\,\mathcal{T},\,\mathcal{S}_{\sigma},\,\mathcal{R}_{\sigma}$
vanish. The situation with $F_{\sigma a}$ is more complex. The coefficients
$\mathcal{T}_{\sigma a}$ do not vanish simultaneously anywhere in
the phase diagram but instead satisfy $\mathcal{T}_{\uparrow\uparrow}=-\mathcal{T}_{\uparrow\downarrow}=\mathcal{T}_{\downarrow\downarrow}=-\mathcal{T}_{\downarrow\uparrow}$
when $\eta=0$, $\Delta_{\uparrow}^{\left(n.n.\right)}=\Delta_{\downarrow}^{\left(n.n\right)}=\Delta^{\left(n.n.\right)}=\chi$
and $\Delta_{a}=\Delta_{\bar{a}}=\Delta=n_{a}=n_{\bar{a}}=n=\frac{1}{2}$
(we note that $\Delta_{a}=n_{a}$ is a direct consequence of our assumption
that $\left[\Lambda_{R},P_{R}^{0}\right]=0$). We will now show that
this implies the following relations between the coefficients $F_{\sigma a}$:
\begin{equation}
F_{\uparrow\uparrow}=-F_{\uparrow\downarrow}=F_{\downarrow\downarrow}=-F_{\downarrow\uparrow}\equiv\tilde{F}\label{eq:F_coeeficients}
\end{equation}
Indeed taking appropriate combinations of derivatives of the Lagrange
function we get that: 
\begin{align}
\frac{\partial\mathcal{L}}{\partial\mathcal{T}_{\sigma a}}+\frac{\partial\mathcal{L}}{\partial\mathcal{T}_{\bar{\sigma}a}}=0\Rightarrow & 0=-2zV\sum_{\sigma'}\Delta_{a}^{\left(n.n.\right)}\Delta_{a}^{\left(n.n\right)}\mathcal{T}_{\sigma'a}+\frac{z}{4}J\sum_{\sigma'a}\left(\sigma+\bar{\sigma}\right)\sigma'\Delta_{a}^{\left(n.n.\right)}\Delta_{a}^{\left(n.n\right)}\mathcal{T}_{\sigma'a}\nonumber \\
 & =F_{\sigma a}\left(1-\Delta_{a}\right)\Delta_{a}+F_{\bar{\sigma}a}\left(1-\Delta_{a}\right)\Delta_{a}\nonumber \\
\frac{\partial\mathcal{L}}{\partial\mathcal{T}_{\sigma a}}+\frac{\partial\mathcal{L}}{\partial\mathcal{T}_{\sigma\bar{a}}}=0\Rightarrow & 0=-zV\sum_{\sigma'}\Delta_{a}^{\left(n.n.\right)}\Delta_{a}^{\left(n.n\right)}\mathcal{T}_{\sigma'a}-zV\sum_{\sigma'}\Delta_{\bar{a}}^{\left(n.n.\right)}\Delta_{\bar{a}}^{\left(n.n\right)}\mathcal{T}_{\sigma'\bar{a}}+\nonumber \\
 & +\frac{z}{4}J\sum_{\sigma'}\sigma\sigma'\Delta_{a}^{\left(n.n.\right)}\Delta_{a}^{\left(n.n\right)}\mathcal{T}_{\sigma a}\mathcal{T}_{\sigma'a}+zJ\sum_{\sigma'}\sigma\sigma'\Delta_{\bar{a}}^{\left(n.n.\right)}\Delta_{\bar{a}}^{\left(n.n\right)}\mathcal{T}_{\sigma'a}\nonumber \\
 & =F_{\sigma a}\left(1-\Delta_{a}\right)\Delta_{a}+F_{\sigma\bar{a}}\left(1-\Delta_{\bar{a}}\right)\Delta_{\bar{a}}\label{eq:Lagrange_equations_hard}
\end{align}
From this we get that Eq. (\ref{eq:F_coeeficients}) follows. In particular
at half filling for $\eta=0$ we have that 
\begin{equation}
H_{embed}=U\hat{c}_{\uparrow}^{\dagger}\hat{c}_{\uparrow}\hat{c}_{\downarrow}^{\dagger}\hat{c}_{\downarrow}+\sum_{\sigma}\lambda_{\sigma}^{c}\hat{f}_{\sigma}\hat{f}_{\sigma}^{\dagger}+\left[F\hat{c}_{\uparrow}^{\dagger}\hat{c}_{\downarrow}\hat{f}_{\downarrow}\hat{f}_{\uparrow}^{\dagger}+h.c.\right]+\tilde{F}\sum_{\sigma a}\sigma a\left(\hat{c}_{\sigma}^{\dagger}\hat{c}_{\sigma}\hat{f}_{a}\hat{f}_{a}^{\dagger}\right)+\sum_{\sigma}\lambda_{\sigma}^{b}\hat{c}_{\sigma}^{\dagger}\hat{c}_{\sigma}\label{eq:Embedding_mott}
\end{equation}
We see that there is no density density type of interaction or pair
hopping. We note that we can get a relation between $F$ and $\tilde{F}$
via: 
\begin{align}
F & =\frac{-zJ\Delta_{\uparrow}^{\left(n.n.\right)}\Delta_{\downarrow}^{\left(n.n\right)}\mathcal{T}^{*}}{\sqrt{\left(1-\Delta_{\uparrow}\right)\Delta_{\uparrow}}\sqrt{\left(1-\Delta_{\downarrow}\right)\Delta_{\downarrow}}}=\frac{-zJ\chi^{2}\mathcal{T}^{*}}{\left(1-\Delta\right)\Delta}\nonumber \\
\frac{\delta\mathcal{L}}{\delta\mathcal{T}_{\uparrow\uparrow}}=0\Rightarrow & -\frac{z}{4}J\sum_{\sigma\sigma'ab}\sigma'\Delta_{\uparrow}^{\left(n.n.\right)}\Delta_{\uparrow}^{\left(n.n\right)}\mathcal{T}_{\sigma'\uparrow}=\tilde{F}\left(1-\Delta_{\uparrow}\right)\Delta_{\uparrow}\Rightarrow\tilde{F}=\frac{-zJ\chi^{2}\mathcal{T}_{\uparrow\uparrow}^{*}}{2\left(1-\Delta\right)\Delta}\label{eq:Relations}
\end{align}
We now use that $\mathcal{T}_{\uparrow\uparrow}=\frac{\mathcal{T}}{2}=1$.
Therefore we have that: 
\begin{equation}
H_{embed}=U\hat{c}_{\uparrow}^{\dagger}\hat{c}_{\uparrow}\hat{c}_{\downarrow}^{\dagger}\hat{c}_{\downarrow}+\sum_{\sigma}\lambda_{\sigma}^{c}\hat{f}_{\sigma}\hat{f}_{\sigma}^{\dagger}-8zJ\chi^{2}\left(\vec{S}_{c}\cdot\vec{S}_{f}\right)+\sum_{\sigma}\lambda_{\sigma}^{b}\hat{c}_{\sigma}^{\dagger}\hat{c}_{\sigma}\label{eq:Embedding_mott-1}
\end{equation}
Here $\vec{S}_{c}$ is the spin operator for the impurity and $\vec{S}_{f}$
is the spin operator for the bath. We note that the term: 
\begin{equation}
-\sum_{\sigma a}F_{\sigma a}\hat{c}_{\sigma}^{\dagger}\hat{c}_{\sigma}\Delta_{a}=0\label{eq:term}
\end{equation}
due to Eq. (\ref{eq:F_coeeficients}). We note that the parameters
$V$ and $Y$ do not enter the embedding Hamiltonian at half filling
in the Brinkman Rice phase. 
\end{widetext}

\end{document}